\documentclass{article}
\usepackage{graphicx}
\usepackage{amsmath}
\begin{document}

\begin{center} 
{\Large{\bf The Physics of Ghost imaging}} \\  
\vspace*{3mm}
Yanhua Shih \\
Department of Physics \\
University of Maryland, Baltimore County, \\ Baltimore, MD 21250, U.S.A. 

\end{center}

\begin{abstract}
One of the most surprising consequences of 
quantum mechanics is the nonlocal correlation of a multi-particle system observable 
in joint-detection of distant particle-detectors. Ghost imaging is one of such phenomena.  
Taking a photograph of an object, traditionally, we need to face a camera 
to the object.  But with ghost imaging, we can image the object by pointing a CCD 
camera towards the light source, rather than towards the object.  Ghost imaging is
reproduced at quantum level by a non-factorizable point-to-point image-forming 
correlation between two photons. Two types of ghost imaging have been experimentally 
demonstrated since 1995.  Type-one ghost imaging uses entangled photon pairs as the 
light source. The non-factorizable image-forming correlation is the result of a nonlocal 
constructive-destructive interference among a large number of biphoton amplitudes, 
a nonclassical entity corresponding to different yet indistinguishable alternative ways 
for the photon pair to produce a joint-detction event between distant photodetectors. 
Type-two ghost imaging uses chaotic light. The type-two non-factorizable image-forming 
correlation is caused by the superposition between paired two-photon amplitudes, 
or the symmetrized effective two-photon wavefunction,
corresponding to two different yet indistinguishable alternative ways of triggering a 
join-detection event by two independent photons.   The multi-photon interference nature 
of ghost imaging determines its peculiar features: (1) it is nonlocal; (2) its imaging resolution 
differs from that of classical; and (3) the type-two ghost image is turbulence-free.\footnote{
For instance, any fluctuation of the refraction index or phase disturbance in the optical 
path has no influence to the type-two ghost image.}  
Ghost imaging has attracted a great deal of attention, perhaps due to these features 
for certain applications.  Achieving these features, the realization of nonlocal multi-photon 
interference is a necessary condition.  Classical simulations, such as the man-made 
factorizable speckle-speckle correlation, can never have such features.  
\end{abstract}

\section{Introduction} 
   
\begin{figure}[hbt]
\centering
    \includegraphics[width=85mm]{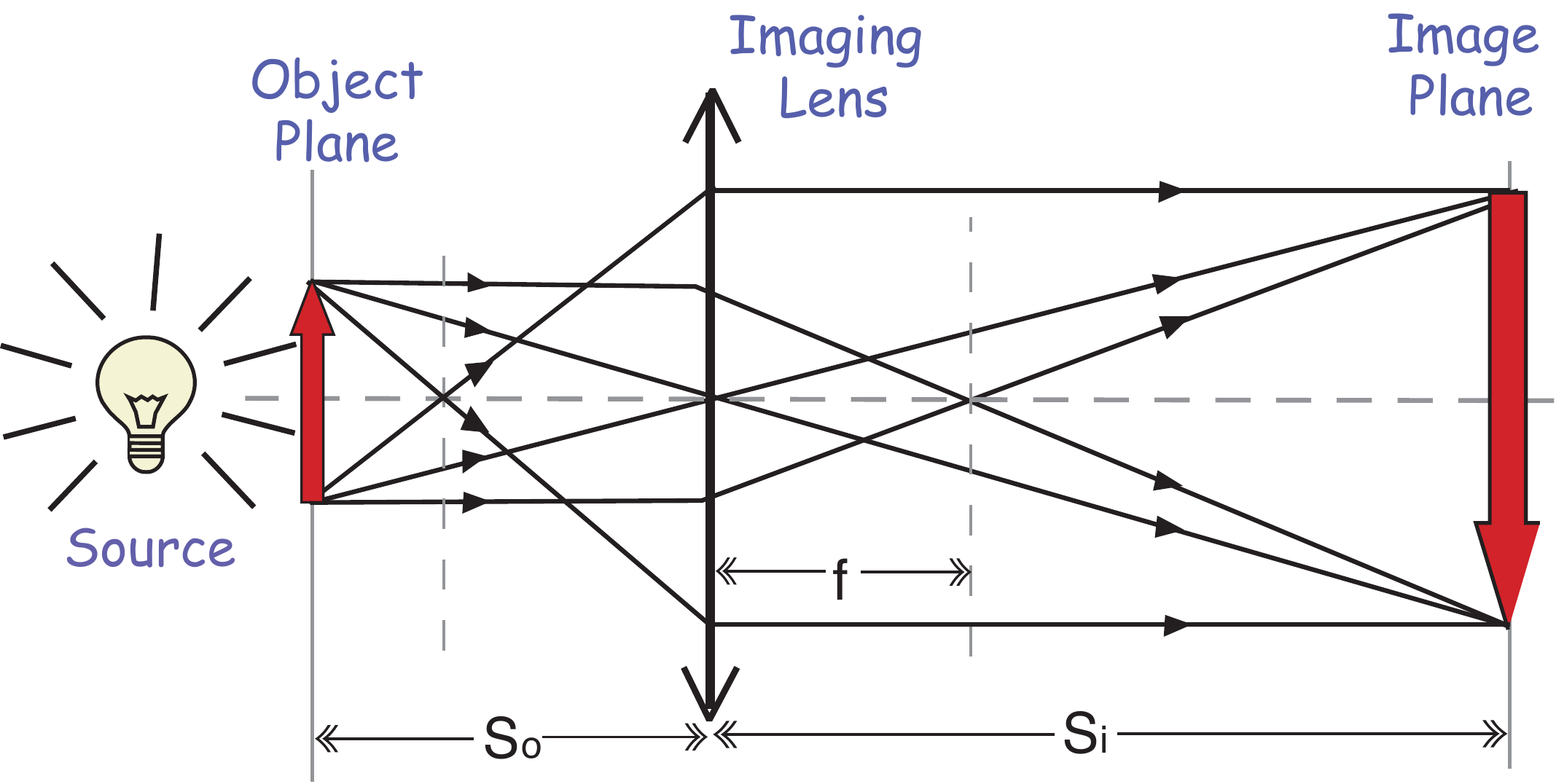}
     \parbox{14.25cm}{\caption{Optical imaging: a lens produces an
    \textit{image} of an object in the plane defined by the Gaussian 
    thin-lens equation $1/s_i+1/s_o=1/f$.  Image formation is
    based on a point-to-point relationship between the object plane 
    and the image plane.  All radiations emitted from a point on the object
    plane will ``collapse" to a unique point on the image plane.  
    }\label{fig:Projection-1}}
\end{figure}

\hspace{6.5mm}Assuming an object that is either self-luminous or externally 
illuminated, imagining each point on the object surface as a point radiation sub-source, 
each point sub-source will emit spherical waves to all possible directions.  
How much chance do we expect to have a spherical wave collapsing into a point or 
a ``speckle" by free propagation?  Obviously, the chance is zero unless 
an imaging system is applied.  The concept of optical imaging was well developed in 
classical optics for this purpose. Figure~\ref{fig:Projection-1} schematically illustrates 
a standard imaging setup.   
In this setup an object is illuminated by a radiation source,  
an imaging lens is used to focus the scattered and reflected light from the object 
onto an image plane which is defined by the ``Gaussian thin lens equation"
\begin{equation}\label{Lens-Eq}
\frac{1}{s_i}+\frac{1}{s_o}=\frac{1}{f},
\end{equation}
where $s_o$ is the distance between the object and the imaging lens, $s_i$ the 
distance between the imaging lens and the image plane, and $f$ the
focal length of the imaging lens.  Basically this equation
defines a point-to-point relationship between the object plane 
and the image plane: any radiation starting from a point on the object
plane will ``collapse" to a unique point on the image plane.   It is not difficult 
to see from Fig.~\ref{fig:Projection-1} that the point-to-point relationship 
is the result of \emph{constructive-destructive interference}.  
The radiation fields coming from a point on 
the object plane will experience equal distance propagation to
superpose constructively at one unique point on the image plane,
and experience unequal distance propagations to superpose
destructively at all other points on the image plane.   
The use of the imaging lens makes this constructive-destructive 
interference possible.  

A perfect point-to-point image-forming relationship between the object and 
image planes produces a perfect image.  The observed image is a reproduction, 
either magnified or demagnified, of the illuminated object, mathematically
corresponding to a convolution between the object distribution function 
$|A(\vec{\rho}_{o})|^2$ (aperture function) and a 
$\delta$-function which characterizes the perfect point-to-point 
relationship between the object and image planes:
\begin{equation}\label{Image-Eq}
I(\vec{\rho}_i) =
\int_{obj} d\vec{\rho}_{o} \, \big{|} A(\vec{\rho}_{o}) \big{|}^2\,
\delta(\vec{\rho}_{o} + \frac{\vec{\rho}_{i}}{m})
\end{equation}
where $I(\vec{\rho}_i)$ is the intensity in the image plane, $\vec{\rho}_{o}$ and 
$\vec{\rho}_{i}$ are 2-D vectors of the 
transverse coordinates in the object and image planes,
respectively, and $m = s_i / s_o$ is the image magnification factor.

In reality, limited by  the finite size of the imaging system, 
we may never obtain a perfect point-to-point correspondence.   
The incomplete constructive-destructive interference turns the
point-to-point correspondence into a point-to-``spot"
relationship.  The $\delta$-function in the convolution of 
Eq.~(\ref{Image-Eq}) will be replaced by a point-spread function:
\begin{equation}\label{Image-Eq-2}
I(\vec{\rho}_i) =
\int_{obj} d\vec{\rho}_{o} \, \big{|} A(\vec{\rho}_{o}) \big{|}^2 \,
somb^2\big{[} \frac{R}{s_o}\,\frac{\omega}{c} \big{|}\vec{\rho}_{o} + 
\frac{\vec{\rho}_{i}}{m} \big{|} \big{]},
\end{equation}
where the sombrero-like function, or the Airy disk, is defined as
$$
somb(x) = \frac{2J_1(x)}{x},
$$
and $J_1(x)$ is the first-order Bessel function, and $R$ the radius of the imaging lens, 
and $R/s_o$ is known as the numerical aperture of the imaging system.  The sombrero-like 
point-spread function, or the Airy disk, defines the spot size on the image plane that is 
produced by the radiation coming from point $\vec{\rho}_{o}$.  It is clear from 
Eq.~(\ref{Image-Eq-2}) that a larger imaging lens and shorter wavelength will result in 
a narrower point-spread function, and thus a higher spatial resolution of the image. 
The finite size of the spot determines the spatial resolution of the imaging system.  

Type-one and type-two ghost imaging, in certain aspects, exhibit a similar point-to-point 
imaging-forming function as that of classical except the ghost image is reproducible only 
in the joint-detection between two independent photodetectors, and the point-to-point 
imaging-forming function is in the form of second-order correlation,
\begin{equation}
R_{12}(\vec{\rho}_i) = \int_{obj} d\vec{\rho}_{o} \, \big{|} A(\vec{\rho}_{o}) \big{|}^2 \,
G^{(2)}(\vec{\rho}_{o}, \vec{\rho}_{i}), 
\end{equation} 
where $R_{12}(\vec{\rho}_i)$ is the joint-detection counting rate between photodetectors 
$D_1$ and $D_2$.  Mathematically, the convolution is taken between the aperture function
of the object $\big{|}A(\vec{\rho}_{o})\big{|}^2$ and a nontrivial poin-to-point second-order correlation 
function $G^{(2)}(\vec{\rho}_{o}, \vec{\rho}_{i})$, corresponding to the probability of observing 
a joint photo-detection event at coordinates $\vec{\rho}_{o}$ and $\vec{\rho}_{i}$.   It is the 
special physics behind $G^{(2)}(\vec{\rho}_{o}, \vec{\rho}_{i})$ made ghost imaging so special. 

\begin{figure}[htb]
    \centering
    \includegraphics[width=95mm]{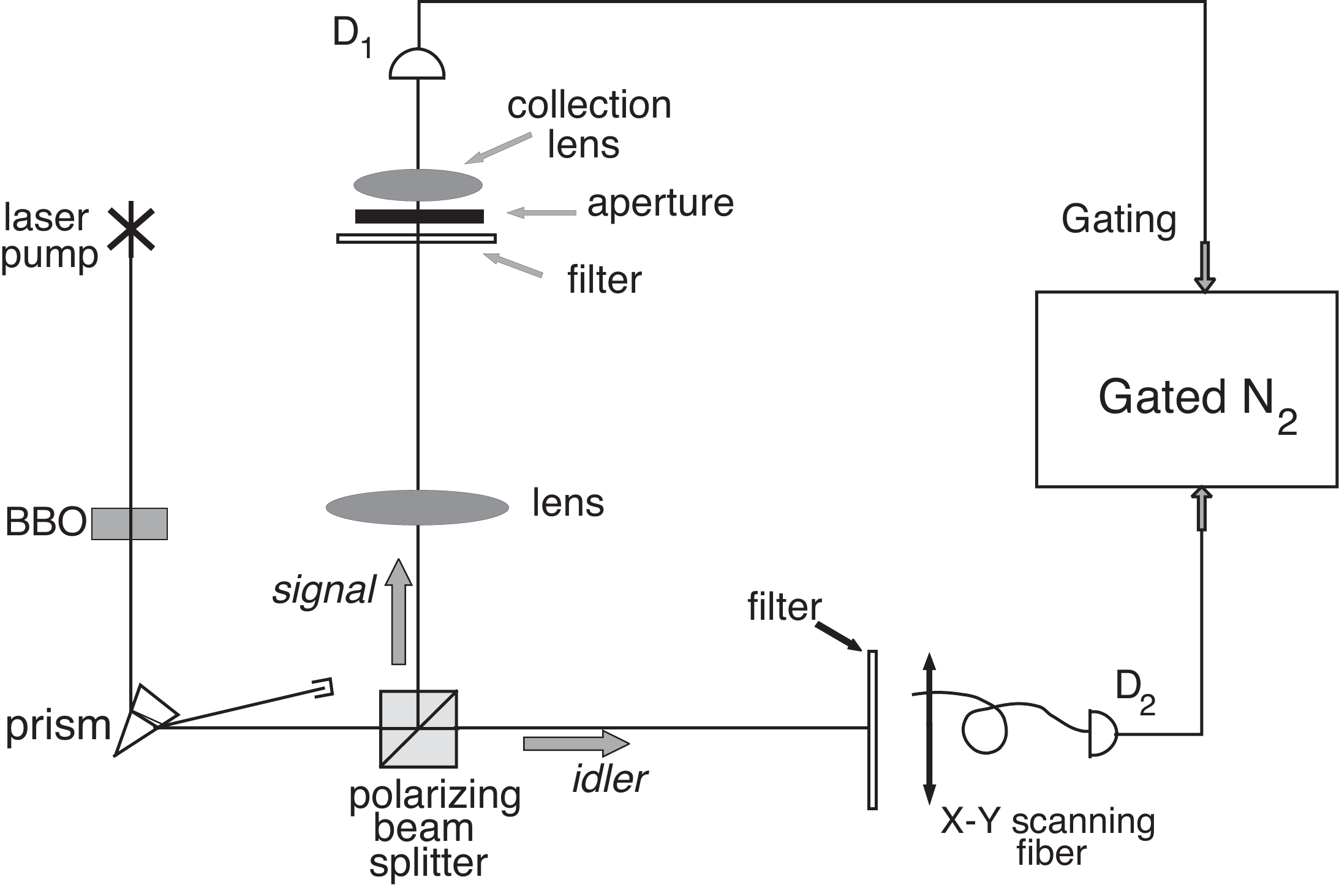}
    \parbox{14.25cm}{\caption{Schematic set-up of the first ``ghost'' image 
    experiment. The experimental demonstrations of ghost imaging and 
    ghost interference \cite{GhostInt} in 1995 together stimulated the foundation 
    of quantum imaging in terms of geometrical and physical optics.}
    \label{Imageset}}
\end{figure}
\begin{figure}
    \centering
    \includegraphics[width=87mm]{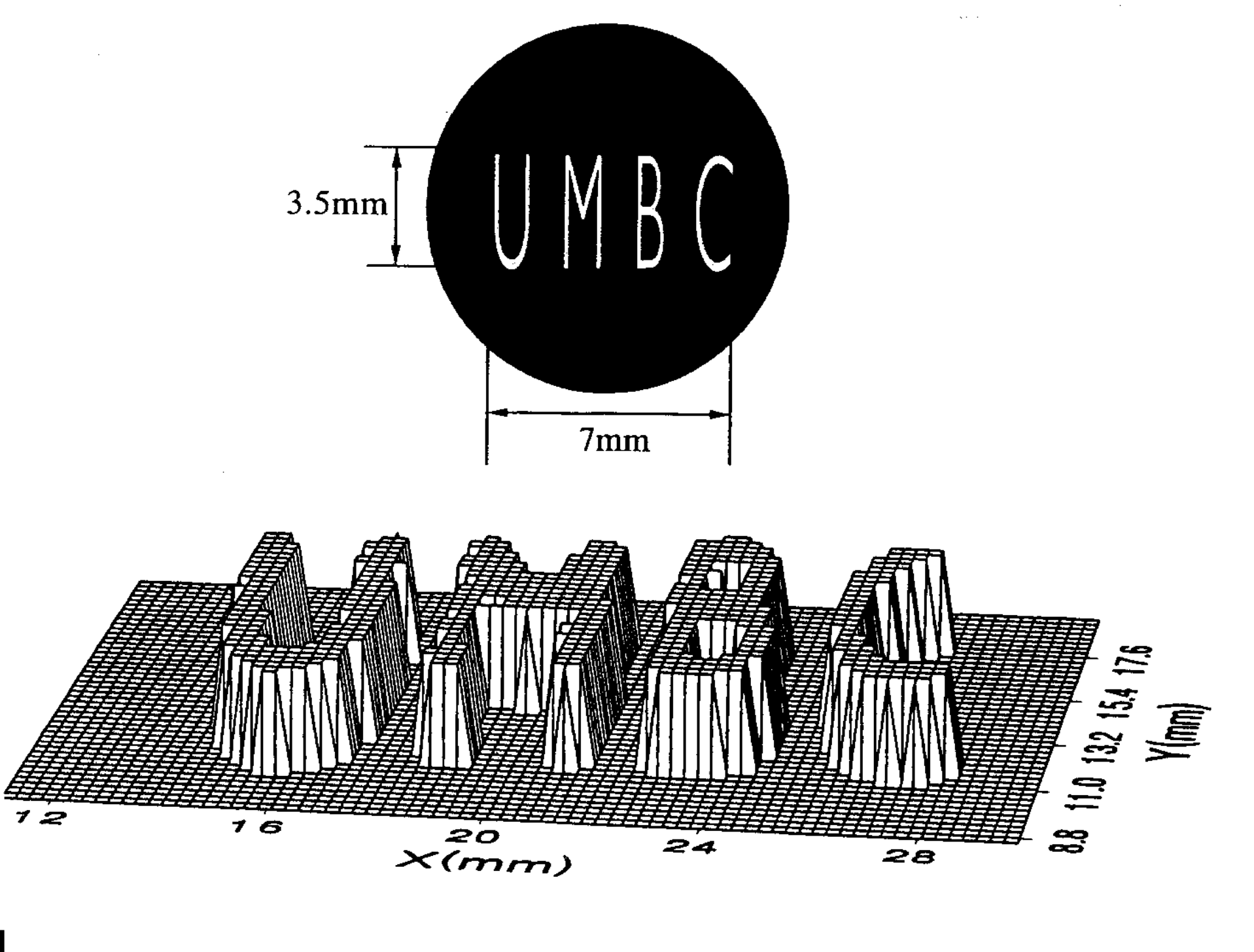}
     \parbox{14.25cm}{\caption{Upper: A reproduction of the
    actual aperture ``UMBC" placed in the signal beam. Lower: The image
    of ``UMBC'': coincidence counts as a function of the fiber tip's
    transverse coordinates in the image plane. The step size is 0.25mm. The 
    image shown is a ``slice'' at the half maximum value.}
    \label{UMBC}}
\end{figure}
The first type-one ghost imaging experiment was demonstrated by Pittman 
\emph{et al}. in 1995 \cite{GhostImage} enlightened by the theoretical work of 
Klyshko \cite{KlyshkoImg}.   
The schematic setup of the experiment is shown in Fig.~\ref{Imageset}.  
A continuous wave (CW) laser is used to pump a nonlinear
crystal to produce an entangled pair of orthogonally polarized signal 
(e-ray of the crystal) and idler (o-ray of the crystal) 
photons in the nonlinear optical process of spontaneous parametric 
down-conversion (SPDC).  The pair emerges from the
crystal collinearly with $\omega _{s}\cong \omega _{i}\cong \omega
_{p}/2$ (degenerate SPDC). The pump is then separated from the signal-idler
pair by a dispersion prism, and the
signal and idler are sent in different directions by a polarization
beam splitting Thompson prism. The signal photon passes through a convex lens
of 400mm focal length and illuminates a chosen aperture (mask). As an
example, one of the demonstrations used the letters ``UMBC'' for the object mask. 
Behind the aperture is the ``bucket'' detector package $D_{1}$, which is made by 
an avalanche photodiode placed at the focus of a short focal length collection lens.  
During the experiment $D_{1}$ is kept in a fixed position.  The idler photon is captured 
by detector package $D_{2} $, which consists of an optical fiber coupled to
another avalanche photodiode. The input tip of the fiber is scannable in the transverse 
plane by two step motors (along orthogonal directions). The output pulses of $D_1$ 
and $D_2$, both operate in the photon counting mode, are independently counted as 
the counting rate of $D_1$ and $D_2$, respectively, and simultaneously, sent to 
a coincidence circuit for counting the joint-detection events of the  
pair.  The single detector counting rates of $D_1$ and $D_2$ are both monitored to be 
constants during the measurement.  Surprisingly, a ghost image of the chosen aperture 
is observed in coincidences during the scanning of the fiber tip, 
when the following two experimental conditions are 
satisfactory: (1) $D_1$ and $D_2$ always measure 
a pair; (2) the distances $s_{o}$, which is the optical distance between the aperture 
to the lens, $s_{i}$, which is the optical distance from the imaging lens going 
backward along the signal photon path to the two-photon source of SPDC then 
going forward along the idler photon path to the fiber tip, and the focal length 
of the imaging lens $f$ satisfy the Gaussian thin lens equation of Eq.~(\ref{Lens-Eq}).

Figure~\ref{UMBC} shows a typical measured ghost image.   It is interesting to
note that while the size of the ``UMBC'' aperture inserted in the signal path is only about 
3.5mm$\times$7mm, the observed image measures 7mm$\times$14mm.  The image is 
therefore magnified by a factor of 2 which equals the expected magnification 
$m = s_{i}/s_{o}$.  In this measurement $s_{o}=$~600mm and $s_{i}=$~1200mm.  
When $D_2$ was scanned on transverse planes other than 
the ghost image plane the images blurred out.

The experiment was immediately given the name ``ghost imaging" by the
physics community due to its nonlocal feature.  In the language of 
Einstein-Podolsky-Rosen (EPR) \cite{EPR}, the non-factorizable\footnote{
Statistically, a factorizable correlation function
$
G^{(2)}(\mathbf{r}_1, t_1; \mathbf{r}_2, t_2) 
= G^{(1)}(\mathbf{r}_1, t_1) \, G^{(1)}(\mathbf{r}_2, t_2)
$
characters independent 
radiations at space-time $(\mathbf{r}_1, t_1)$ and $(\mathbf{r}_2, t_2)$.
In ghost imaging, the light on the object plane and the light at the CCD array is
described by a non-factorizeable point-to-point image-forming function, indicating 
nontrivial statistical correlation between the two measured intensities.} 
point-to-point image-forming correlation 
\begin{equation}
G^{(2)}(\vec{\rho}_{o}, \vec{\rho}_{i}) \sim \delta(\vec{\rho}_{o} + \vec{\rho}_{i}/m)
\end{equation} 
observed in this experiment represents a nonlocal behavior of a measured pair of 
photons: neither the signal photon nor idler photon ``knows" precisely where to go 
when the pair is created at the source.  However, 
if one of them is observed at a point on the object plane, the other one must arrive 
at a unique corresponding point on the image plane.\footnote{The ghost imaging 
experiment is thus considered a demonstration of the historical 
Einstein-Podolsky-Rosen (EPR) experiment. }
Although questions regarding fundamental issues of quantum theory still exist, 
the experimental demonstration of ghost imaging \cite{GhostImage} and 
ghost interference \cite{GhostInt} in 1995 together stimulated the foundation of 
quantum imaging in terms of geometrical and physical optics. 

Type-two ghost imaging uses chaotic radiation sources. Different from type-one, 
the non-factorizable point-to-point image-forming correlation between the object 
and image planes is only partial with at least 50\% constant background, 
\begin{equation}
G^{(2)}(\vec{\rho}_{o}, \vec{\rho}_{i}) \sim 1+ \delta(\vec{\rho}_{o} - \vec{\rho}_{i}). 
\end{equation}
\begin{figure}[hbt]
    \centering
    \includegraphics[width=90mm]{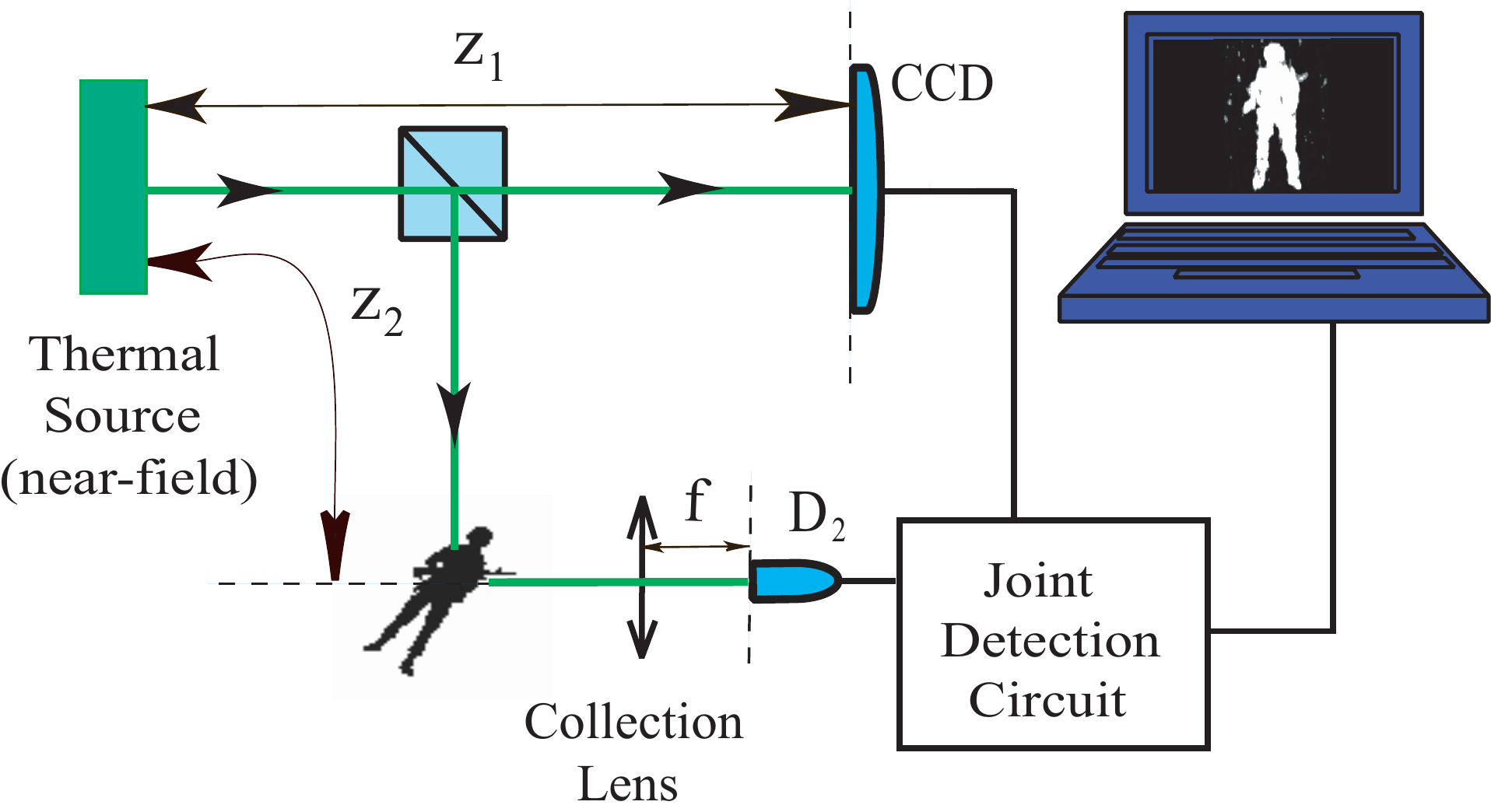}  
    \parbox{14.25cm}{\caption{Near-field lensless ghost imaging of chaotic light 
    demonstrated by Meyers \emph{et al.}.  $D_2$ is a ``bucket" photon counting
    detector that is used to collect and count all random scattered and reflected 
    photons from the object. The joint-detection between $D_2$ and 
    the CCD array is realized by a photon-counting-coincidence circuit. $D_2$ is 
    fixed in space.  The counting rate of $D_2$ and the un-gated output of the CCD 
    are both monitored to be constants during the measurement.  Surprisingly, a 
    1:1 ghost image of the object is captured in joint-detection between $D_2$ and the 
    CCD, when taking $z_1 = z_2$.  The images  ``blurred out" when the CCD is 
    moved away from $z_1 = z_2$, either in the direction of $z_1 > z_2$ or 
    $z_1 < z_2$. } 
    \label{fig:Meyers}}
\end{figure}
\begin{figure}[hbt]
    \centering 
    \includegraphics[height=55mm, width=44mm]{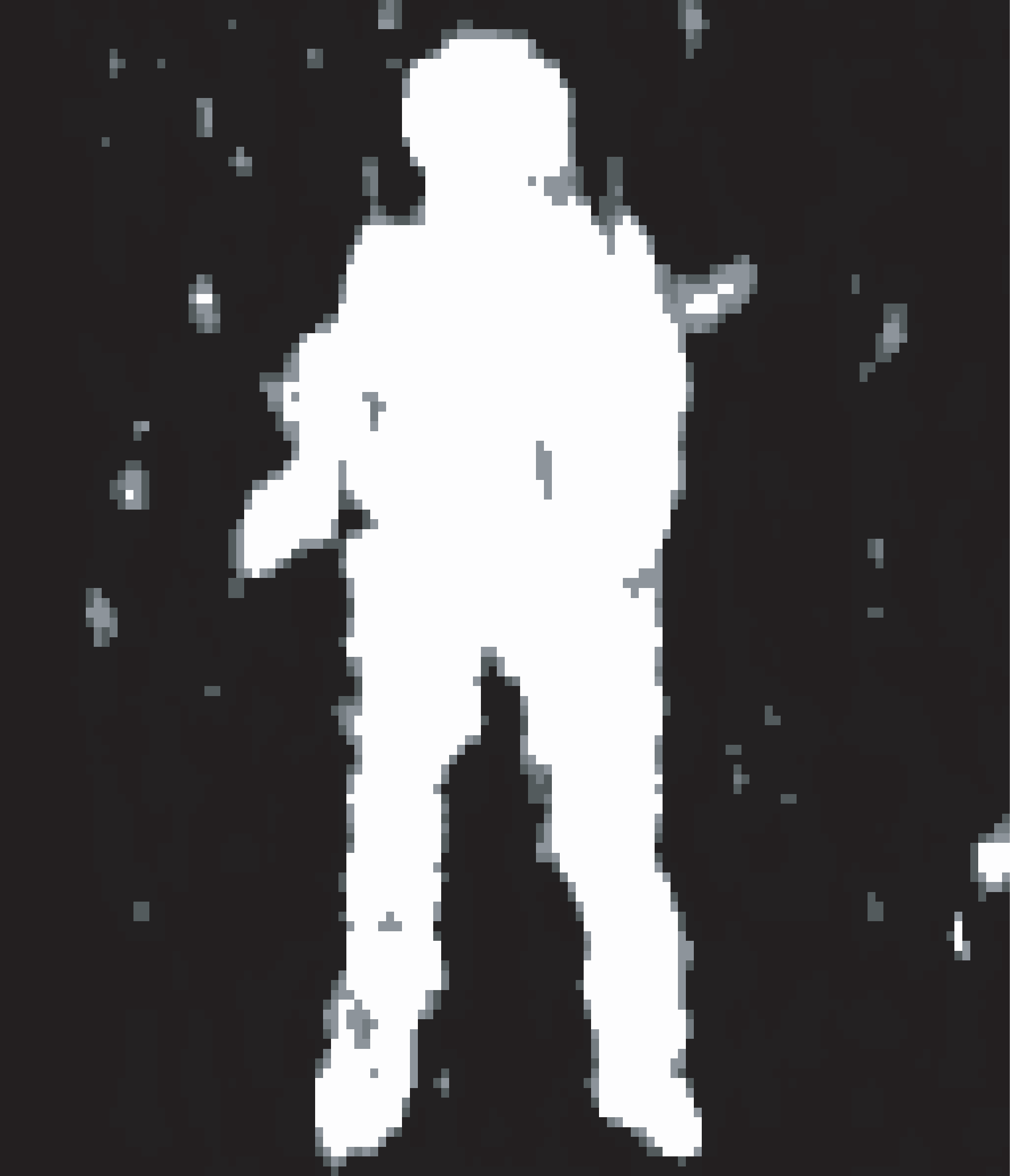}
     \parbox{14.25cm}{\caption{Ghost image of a toy soldier model.} 
    \label{Fig2}}
\end{figure}
The first near-field lensless ghost imaging experiment was demonstrated by 
Scarcelli \emph{et al} in 2005 and 2006 \cite{prl2}\cite{prl1} after their experimental 
demonstration of two-photon interference of chaotic light in 2004 \cite{Europhys}.  
Figure~\ref{fig:Meyers} illustrates an improved setup of the
type-two ghost imaging experiment by Meyers \emph{et al}. \cite{Meyers}. 
The thermal radiation of a chaotic source, which has a fairly large size in 
the transverse dimension, is split into two by a $50\% - 50\%$ beamsplitter. 
One of the beams illuminates a toy soldier as shown in Fig.~\ref{fig:Meyers}.  
The scattered and reflected photons from the solider (object) are collected and counted 
by  a ``bucket" detector $D_2$.  In the other beam a high resolution CCD array, 
operating at the photon counting regime, is placed toward the radiation source for  
joint-detection with the ``bucket"  detector $D_2$.  The counting rate of 
$D_2$ and the un-gated output of the CCD are both monitored to be constants during 
the measurement.  Surprisingly, a 1:1 ghost image of the toy soldier is captured
in the joint-detection between $D_2$ and the CCD, when taking $z_1 = z_2$.  
The 1:1 ghost image of the toy soldier is shown in Fig.~\ref{Fig2}.
The images ``blurred out" when the CCD is moved away from $z_1 = z_2$, 
either to the side of $z_1 > z_2$ or $z_1 < z_2$.   

There is no doubt that chaotic radiations propagate to any transverse plane 
in a random and chaotic manner.   A brief discussion for Fresnel free-propagation 
is given in the appendix.  In the lensless ghost imaging experiment, a large transverse 
sized chaotic light source, as shown in Fig.~\ref{fig:Meyers}, is usually used for 
achieving better spatial resolution.  The source consists a large number of independent 
point sub-sources randomly distributed on the source plane.   Each point sub-source
may randomly radiate independent spherical waves to the object and image planes.  
Due to the chaotic nature of the source there is no 
interference between these sub-fields.  These independent sub-intensities simply 
add together, yielding a constant total intensity in space and in time on any 
transverse plane.   In the lensless ghost imaging setup, there is no lens applied to force 
these spherical waves collapsing to a point or a ``speckle", and there is no chance to 
have two identical copies of any ``speckle" of the source onto the object and image 
planes.  What is the physical cause of the point-to-point image-forming correlation?  Although 
the non-factorizable point-to-point correlation between the object and image planes is only 
partial, the type-two ghost imaging looks more surprising than type-one because of 
the nature of the light source.  Unlike the signal-idler photon pair, the jointly measured 
photons in type-two ghost imaging are just two independent photons that fall into the 
coincidence time window by chance only.   Nevertheless, analogous to EPR, the 
non-factorizable partial point-to-point correlation represents a nonlocal behavior of a measured pair 
of independent photons: neither photon-one nor photon-two ``knows" precisely 
where to go when they are created at each independent sub-sources; 
however, if one of them is observed at a point on the object plane, the other one 
has twice greater probability of arriving at a unique corresponding point on 
the image plane.\footnote{Similar to the HBT correlation, the contrast of the near-field
partial point-to-point image-forming function is $50\%$, i.e., two to one ratio between 
the maximum value and the constant background, see Eq.~(\ref{HBT-1}).}   

We have concluded and will show that the partial point-to-point 
correlation between the object and image planes in type-two ghost imaging is 
the result of  \emph{two-photon interference}.  Similar to that of type-one, 
it involves the nonlocal superposition of two-photon amplitudes, a nonclassical 
entity corresponding to different yet indistinguishable alternative ways of 
triggering a joint-detection event \cite{IEEE-03}. Different from that of type-one,
the joint-detection events observed in type-two ghost imaging are triggered by 
two randomly distributed independent photons.  
It is interesting to see that the quantum mechanical concept of \emph{two-photon 
interference} is applicable to ``classical" thermal light.\footnote{There exist a number 
of definitions for classical light and for quantum light.   One of the commonly 
accepted definitions considers thermal light classical because its positive $P$-function. }
In fact, this is not the first time in the history of physics we apply
quantum mechanical concepts to thermal light.  We should 
not forget Planck's theory of blackbody radiation originated the quantum physics.  
The radiation Planck dealt with was thermal radiation.   
Although the concept of ``two-photon interference" 
comes from the study of entangled biphoton states \cite{IEEE-03}, 
the concept should not be restricted to entangled systems.  The concept is generally 
true and applicable to any radiation, including ``classical" thermal light.  
The partial point-to-point correlation of thermal radiation is not a new discovery either.  
The first set of temporal and spatial far-field intensity-intensity correlations of thermal 
light was demonstrated by Hanbury Brown and Twiss (HBT) in 1956 
\cite{hbt}\cite{hbt-book}.  The HBT experiment created quite a surprise in the physics 
community and lead to a debate about the classical or quantum nature of the 
phenomenon \cite{hbt-book}\cite{Scully-book}.  Although the discovery of HBT initiated 
a number of key concepts of modern quantum optics, the HBT phenomenon itself was 
finally interpreted as statistical correlation of intensity fluctuations and considered as a 
classical effect.    It is then reasonable to ask: 
Is the near-field type-two ghost imaging with thermal light a simple classical 
effect similar to that of HBT? Is it possible that the ghost imaging phenomenon
itself, including the type-one ghost imaging of 1995, is merely a simple classical 
effect of intensity fluctuation correlation?\cite{gatti}\cite{Wang}\cite{Zhu}\cite{MIT}  
This article will address these important questions and explore the multi-photon
interference nature of ghost imaging.

To explore the two-photon interference nature,  
we will analyze the physics of type-one and type-two ghost imaging in five steps.  (1)
Review the physics of coherent and incoherent light propagation; (2)
Review classical imaging as the result of constructive-destructive interference
among electromagnetic waves; (3) analyze type-one ghost imaging in terms of 
constructive-destructive interference between the biphoton amplitudes
of an entangled photon-pair; (4) analyze type-two ghost imaging in terms of two-photon
interference between chaotic sub-fields; and (5) discuss the physics of 
the phenomenon: whether it is a quantum interference or a classical 
intensity fluctuation correlation.

\section{Classical Imaging}

\begin{figure}[htb]
\centering
    \includegraphics[width=87mm]{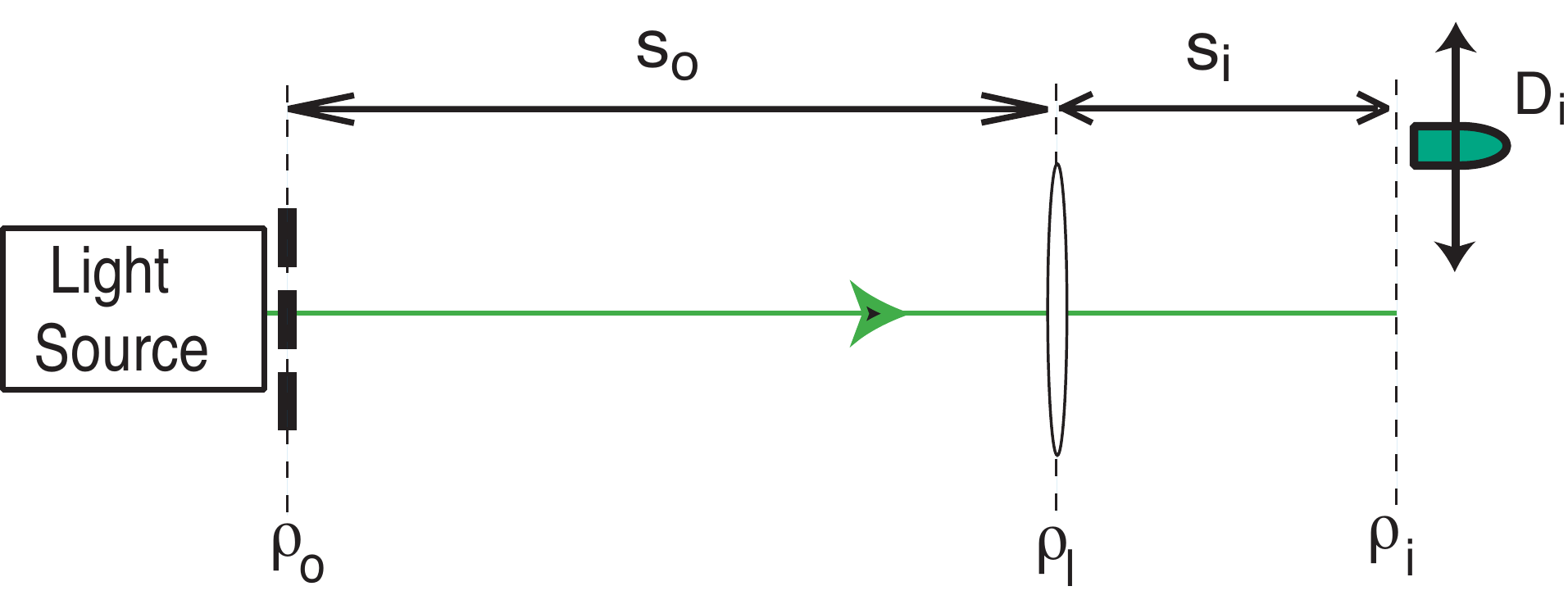}
     \parbox{14.25cm}{\caption{Typical imaging setup. A lens of finite size is used to produce 
     a magnified or demagnified image of an object with limited spatial resolution. 
}\label{fig:lithography-1}}
\end{figure}

\hspace{6.5mm}To understand the multi-photon interference nature of 
ghost imaging, it might be helpful to see the constructive-destructive
interference nature of classical imaging first.  We start from a typical classical 
imaging setup of Fig.~\ref{fig:lithography-1} and ask a simple question: 
how does the radiation field propagate 
from the object plane to the image plane?  In classical optics such
propagation is usually described by an optical transfer function
$h(\mathbf{r}-\mathbf{r}_0, t-t_0)$.  We prefer to work with 
the single-mode propagator, namely the Green's function, 
$g(\mathbf{k}, \mathbf{r}-\mathbf{r}_0, t-t_0)$ \cite{Rubin}\cite{goodman}, 
which propagates each mode of the radiation from space-time point 
($\mathbf{r}_0, t_0$) to space-time point ($\mathbf{r}, t$).  We treat the 
field $E(\mathbf{r}, t)$ as a superposition of these modes. 
A detailed discussion about $g(\mathbf{k}, \mathbf{r}-\mathbf{r}_0, t-t_0)$
is given in the Appendix.  
It is convenient to write the field $E(\mathbf{r}, t)$ as a superposition
of its longitudinal and transverse modes under the Fresnel paraxial approximation, 
\begin{eqnarray}\label{e-g}
E(\vec{\rho}, z, t) =  \int d\vec{\kappa} \, d\omega \,\tilde{E}(\vec{\kappa}, \omega) \,
g(\vec{\kappa}, \omega; \vec{\rho}, z) \, e^{-i\omega t},
\end{eqnarray}
where $\tilde{E}(\vec{\kappa}, \omega)$ is the complex amplitude for the mode 
of frequency $\omega$ and
transverse wave-vector $\vec{\kappa}$.  In Eq.~(\ref{e-g}) we have
taken $z_0=0$ and $t_0 = 0$ at the object plane as usual.
To simplify the notation, we have assumed one polarization.

Based on the experimental setup of Fig.~\ref{fig:lithography-1} and
following the Appendix, $g(\vec{\kappa}, \omega; \vec{\rho}, z)$ is found to be
\begin{align}\label{g-1}
& \ \ \ \  g(\vec{\kappa}, \omega; \vec{\rho}_i, s_o+s_i) \nonumber \\
&=  \int_{obj} d\vec{\rho}_{o} \int_{lens} d\vec{\rho}_l \, \Big\{ A(\vec{\rho}_{o}) \,
e^{i \vec{\kappa} \cdot \vec{\rho}_{o}} \Big\} \,
\Big\{\frac{-i \omega}{2 \pi c} \, 
\frac{e^{i \frac{\omega}{c} s_o} }{s_o}\, e^{i \frac{\omega}{2 c s_o} 
|\vec{\rho}_{l}-\vec{\rho}_{o}|^2} \Big\}
\,\, \Big\{ e^{-i \frac{\omega}{2 c f} \, | \vec{\rho}_l |^2} \Big\}\,
\nonumber \\ &  \ \ \ \  \times \,  \Big\{ \frac{-i \omega}{2\pi c } \, 
\frac{e^{i \frac{\omega}{c} s_i}}{s_i}\, e^{i \frac{\omega}{2 c s_i} |\vec{\rho}_i-\vec{\rho}_l |^2 } \Big\},
\end{align}
where $\vec{\rho}_{o}$, $\vec{\rho}_l$, and $\vec{\rho}_i$ are
two-dimensional vectors defined, respectively, on the object, lens, and image planes.
The first curly bracket includes the aperture function $A(\vec{\rho}_{o})$ of the
object and the phase factor $e^{i \vec{\kappa} \cdot \vec{\rho}_{o}}$ contributed 
at the object plane by each transverse mode 
$\vec{\kappa}$.  The terms in the second and fourth curly brackets describe 
free-space Fresnel propagation-diffraction from the source/object plane to the 
imaging lens, and from the imaging lens to the detection plane, respectively.
The Fresnel propagator includes a spherical wave
function $e^{i \frac{\omega}{c} (z_j -z_k)} / (z_j -z_k)$ and a Fresnel phase factor 
$e^{i \omega |\vec{\rho}_{j}-\vec{\rho}_{k}|^2 / {2 c (z_j -z_k)} } $.  
The third curly bracket adds the phase factor introduced by the imaging lens.

We now rewrite Eq.~(\ref{g-1}) into the following form
\begin{align} \label{g-2}
g(\vec{\kappa}, \omega; \vec{\rho}_i, z =s_o+s_i) 
&=  \frac{- \omega^2}{(2 \pi c)^2 s_o s_i} \, e^{i \frac{\omega}{c} (s_o+s_i)}
\, e^{i \frac{\omega}{2 c s_i} |\vec{\rho}_i|^2} 
 \int_{obj} d\vec{\rho}_{o} \, A(\vec{\rho}_{o}) \,
e^{i \frac{\omega}{2 c s_o} |\vec{\rho}_{o}|^2} \, e^{i \vec{\kappa} \cdot \vec{\rho}_{o}}
\nonumber \\ & \ \ \ \  \times  \int_{lens} d\vec{\rho}_l \, e^{i \frac{\omega}{2c}
[\frac{1}{s_o} + \frac{1}{s_i} - \frac{1}{f}] |\vec{\rho}_l|^2} \, e^{-i \frac{\omega}{c}
(\frac{\vec{\rho}_{o}}{s_o} + \frac{\vec{\rho}_i}{s_i}) \cdot \vec{\rho}_l}.
\end{align}
The image plane is defined by the Gaussian thin-lens equation
of Eq.~(\ref{Lens-Eq}). Hence, the second integral in
Eq.~(\ref{g-2}) reduces to, for a finite sized lens of
radius $R$, the so-called point-spread
function, or the Airy disk, of the imaging system: 
\begin{equation}\label{g-2-2}
\int_{lens} d\vec{\rho}_l \, e^{-i \frac{\omega}{c}
(\frac{\vec{\rho}_{o}}{s_o} + \frac{\vec{\rho}_i}{s_i}) \cdot \vec{\rho}_l}
=  \frac{2J_1(x)}{x} = somb(x), 
\end{equation}
where the sombrero-like function $somb(x) = 2J_1(x) / x$ with argument 
$x=[\frac{R}{s_o}\,\frac{\omega}{c} |\vec{\rho}_{o} + \rho_{i} / m|]$ has been 
defined in Eq.~(\ref{Image-Eq-2}).  Eq.~(\ref{g-2-2}) indicates a 
constructive interference. 

Substituting Eqs.~(\ref{g-2}) and (\ref{g-2-2}) into Eq.~(\ref{e-g}) enables one to 
obtain the classical self-correlation function of the field, or, equivalently, the intensity
on the image plane
\begin{equation}\label{i-0}
I(\vec{\rho}_i, z_i, t_i) = \langle \, E^*(\vec{\rho}_i, z_i, t_i) \, E(\vec{\rho}_i, z_i, t_i) \, \rangle,
\end{equation}
where $\langle ... \rangle$ denotes an ensemble average.  
To simplify the mathematics, monochromatic light is assumed as usual.     

\vspace{2mm}
Case (I): \textit{Incoherent imaging.}
The ensemble average 
yields zeros except when $\vec{\kappa}=\vec{\kappa'}$. The image is thus
\begin{eqnarray}\label{i-2}
I(\vec{\rho}_i) &\propto&
 \int d\vec{\rho}_{o} \, \big{|}A(\vec{\rho}_{o})\big{|}^2 \,
somb^2 [\frac{R}{s_o}\, \frac{\omega}{c} |\vec{\rho}_{o} + \frac{\vec{\rho}_{i}}{m}|].
\end{eqnarray}
An incoherent image, magnified by a factor of $m$, is
thus given by the convolution between the modulus square of the
object aperture function and the point-spread function. The
spatial resolution of the image is determined by the finite
width of the $|somb|^2$-function.

\vspace{2mm}
Case (II): \textit{Coherent imaging.} The coherent superposition of
the $\vec{\kappa}$ modes in both $E^*(\vec{\rho}_i, \tau)$ and
$E(\vec{\rho}_i, \tau)$ results in a wavepacket. The image, or the
intensity distribution on the image plane, is  
\begin{eqnarray}\label{i-1}
I(\vec{\rho}_i) \propto
\Big{|} \int_{obj} d\vec{\rho}_{o} \, A(\vec{\rho}_{o}) \,
e^{i \frac{\omega}{2 c s_o} |\vec{\rho}_{o}|^2}
somb[\frac{R}{s_o} \frac{\omega}{c} |\vec{\rho}_{o} + \frac{\vec{\rho}_{i}}{m}|] \Big{|}^2. 
\end{eqnarray}
A coherent image, magnified by a factor of $m$, is thus
given by the modulus square of the convolution between the object
aperture function (multiplied by a Fresnel phase  factor) and
the point-spread function. 

For $s_i<s_o$ and $s_o>f$, both Eqs.~(\ref{i-2}) and (\ref{i-1})
describe a real demagnified inverted image. In both cases, a
narrower $somb$-function yields a higher spatial resolution.
Therefore the use of a larger imaging lens and shorter wavelengths 
will improve the spatial resolution of an imaging system.

\section{Biphoton and type-one ghost imaging}

\hspace{6.5mm}In this section we analyze type-one ghost imaging.  Type-one ghost
imaging uses entangled photon pairs such 
as the signal-idler biphoton pairs of SPDC \cite{Klyshkobook}\cite{IEEE-03}.  
The nearly collinear signal-idler 
system generated by SPDC can be described, in the ideal case, by the following 
entangled biphoton state \cite{IEEE-03}:
\begin{eqnarray}\label{State}
|\, \Psi \, \rangle
= \Psi_0 \int d \vec{\kappa}_s \,d \vec{\kappa}_i \, 
\delta(\vec{\kappa}_s + \vec{\kappa}_i )  
\int d \omega_s \, d \omega_i \, \delta(\omega_s+\omega_i-\omega_p) 
\, a^{\dag}(\vec{\kappa}_s,\omega_s) \,
a^{\dag}(\vec{\kappa}_i,\omega_i) \, |\, 0 \, \rangle,
\end{eqnarray}
where $\omega_j$, $\vec{\kappa}_j$ ($j=s,i,p$), are the frequency and 
transverse wavevector of the signal, idler, and pump, respectively.
For simplicity a CW single mode pump with $\vec{\kappa}_p =0$ is assumed.  
Eq.~(\ref{State}) indicates that the biphoton state of the signal-idler pair is an 
entangled state.   The single-photon state of the signal and the idler can be
evaluated by taking a partial trace of its twin,
\begin{align}
\hat{\rho}_s & = tr_i \, |\, \Psi \, \rangle \langle \, \Psi \, | =  \int d \vec{\kappa}_s 
\, d \omega_s \, a^{\dag}(\vec{\kappa}_s,\omega_s) |\, 0 \, \rangle \langle \, 0 \, |
a(\vec{\kappa}_s,\omega_s), \nonumber \\
\hat{\rho}_i & = tr_s \, |\, \Psi \, \rangle \langle \, \Psi \, |  =  \int d \vec{\kappa}_i 
\, d \omega_i \ a^{\dag}(\vec{\kappa}_i,\omega_i) |\, 0 \, \rangle \langle \, 0 \, |
a(\vec{\kappa}_i,\omega_i).
\end{align}   
Although the signal-idler system is in a pure state, the 
state of the signal photon and the idler photon, respectively, are both mixed states.  

Let us imagine a measurement 
in which two point-like photon counting detectors ($D_1$ and $D_2$) are placed
at the output plane of an SPDC source for the detection of the signal photon and 
the idler photon, respectively, and for the joint-detection of the signal-idler pair.  
The probability of observing a photo-detection event in the SPDC output plane 
$\vec{\rho}_j$ at time $t_j$, $j = s, i$, is calculated from the first-order  
photo-detection theory of Glauber \cite{Glauber}
\begin{align}\label{G1-0}
G^{(1)}(\vec{\rho}_j, t_j) 
= tr \, \hat{\rho} \, E^{(-)}(\vec{\rho}_j, t_j) E^{(+)}(\vec{\rho}_j, t_j),
\end{align}
where we have chosen $z_j=0$ for the SPDC output plane as usual.  It is easy to 
find that 
\begin{align}\label{G1-1}
G^{(1)}(\vec{\rho}_s, t_s) = \text{constant}, \ \ \ \  
G^{(1)}(\vec{\rho}_i, t_i) = \text{constant},
\end{align}
which means that the signal photon and the idler photon both have equal probability
to be observed at any position in the output plane of the SPDC at any time.  
The probability of observing a joint-detection event between $D_1$ 
and $D_2$ located at $\vec{\rho}_{s}$ and $\vec{\rho}_{i}$ in the SPDC output
plane of $z_s=z_i=0$ is calculated from the second-order photo-detection theory 
of Glauber \cite{Glauber}:
\begin{align}\label{G2}
& \hspace{5mm} G^{(2)}(\vec{\rho}_s, t_s; \vec{\rho}_i, t_i) \nonumber \\
&= \langle \, \Psi \, |\, E^{(-)}_s (\vec{\rho}_s, t_s) 
E^{(-)}_i (\vec{\rho}_i, t_i)
E^{(+)}_i (\vec{\rho}_i, t_i)
E^{(+)}_s (\vec{\rho}_s, t_s) \, |\, \Psi \, \rangle \nonumber \\
&= | \langle \, 0 \, |\,
E^{(+)}_i (\vec{\rho}_i, t_i)
E^{(+)}_s (\vec{\rho}_s, t_s) \, |\, \Psi \, \rangle |^2 \nonumber \\
 &\equiv  |\, \Psi(\vec{\rho}_{s}, t_s; \vec{\rho}_{i}, t_i) \, |^2,
\end{align}
where $\Psi(\vec{\rho}_{s}, t_s; \vec{\rho}_{i}, t_i)$ is defined as the
effective biphoton wavefunction.  The transverse spatial part of the effective biphoton 
wavefunction is easily calculated to be:
\begin{eqnarray}\label{Wavefunction}
\Psi(\vec{\rho}_{s}, \vec{\rho}_{i})
 \simeq \delta(\vec{\rho}_{s}-\vec{\rho}_{i}),
\end{eqnarray}
under the condition $t_s \simeq t_i$.  Equations~(\ref{State}), (\ref{G1-1}), 
and (\ref{Wavefunction}) 
suggest that the entangled signal-idler photon pair is characterized by the EPR
correlation \cite{EPR} in transverse momentum and transverse position;
hence, similar to the original EPR state, we have \cite{disug}:
\begin{eqnarray}\label{eq_uncert}
& & \Delta(\vec{\kappa}_s+\vec{\kappa}_i)=0 \,\, \,\, \& \,\,\,
\Delta (\vec{\rho}_s- \vec{\rho}_i)=0 \,\, \\ \nonumber
&\text{with}&  \Delta \vec{\kappa}_s \sim \infty, \hspace{2mm}
\Delta \vec{\kappa}_i \sim \infty, \hspace{2mm}
\Delta \vec{\rho}_s \sim \infty, \hspace{2mm} \Delta \vec{\rho}_i \sim \infty.
\end{eqnarray}
In EPR's language, the signal photon and the idler photon may come from any
point in the output plane of the SPDC.  However, if the
signal (idler) is found in a certain position, the idler (signal)
must be observed in the same position, with certainty (100\%).  Simultaneously,
the signal photon and the idler photon may have any transverse momentum.
However, if a certain value and direction of the transverse
momentum of the signal (idler) is observed, the transverse
momentum of the idler (signal) will be uniquely determined with
equal value and opposite direction. 

\begin{figure}[hbt]
    \centering
    \includegraphics[width=90mm]{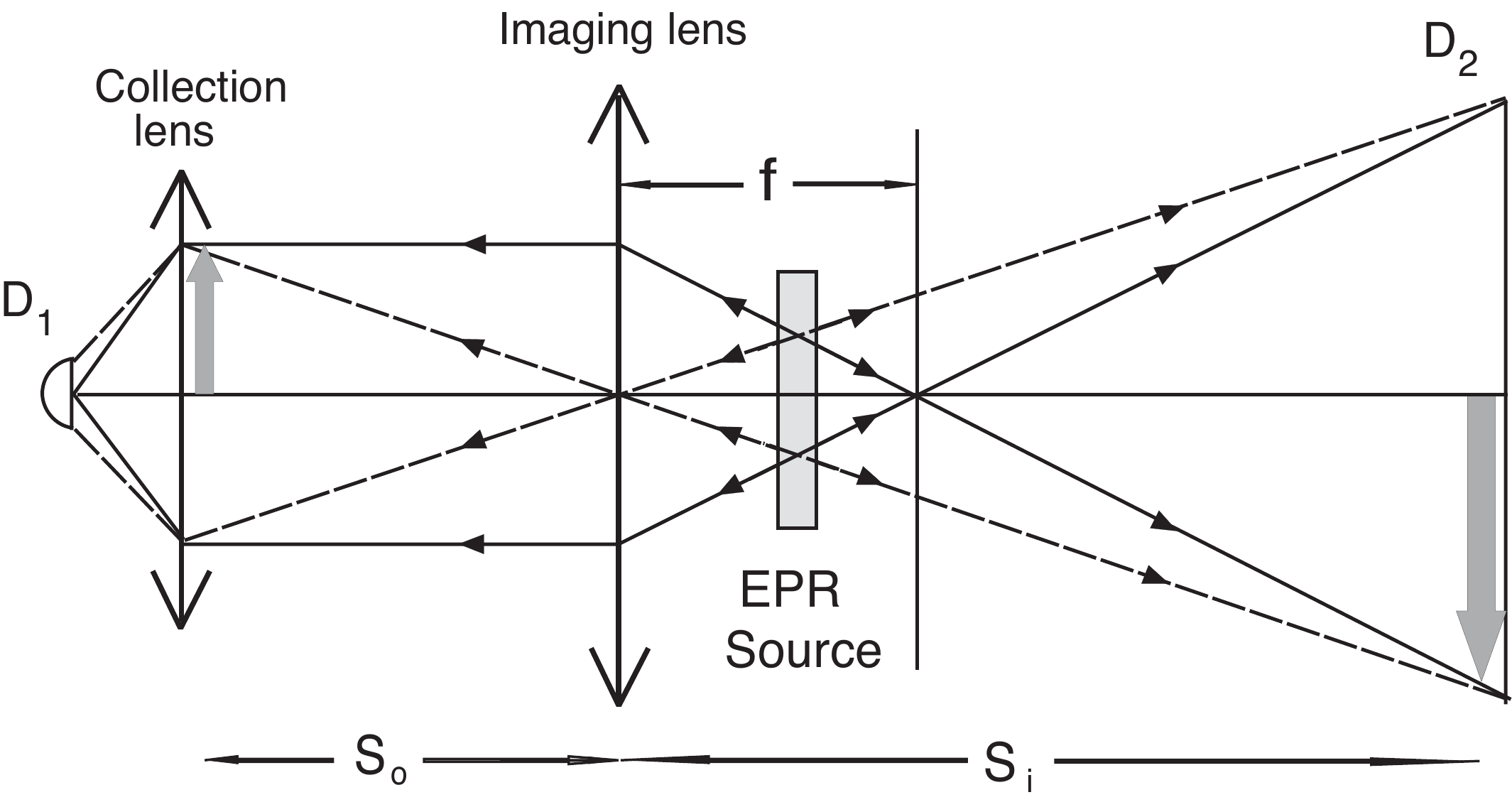}
     \parbox{14.25cm}{\caption{An unfolded schematic of the 1995 ghost imaging experiment, 
    which is helpful for understanding the physics.  Since the biphoton ``light" propagates 
    along ``straight lines", it is obvious that any point on the object plane 
    corresponds to a unique point on the image plane.  Although the placement of the lens, the 
    object, and detector $D_{2}$ obeys the Gaussian thin lens equation, it is important to 
    notice that the geometric rays in the figure actually represent the biphoton amplitudes 
    of an entangled photon pair.   The point-to-point correspondence is the result of a 
    constructive-destructive interference of these biphoton amplitudes.} 
    \label{fig:imaging-unfold}}
\end{figure}

The EPR $\delta$-functions, $\delta(\vec{\rho}_s - \vec{\rho}_i)$ and
$\delta(\vec{\kappa}_s + \vec{\kappa}_i)$ in transverse position and momentum, are 
the key to understanding the ghost imaging experiment of Pittman \emph{et al}. of 1995.  
$\delta(\vec{\rho}_s - \vec{\rho}_i)$ indicates that the signal-idler 
pair is always emitted from the same point on the output plane of the biphoton 
source.  Simultaneously, $\delta(\vec{\kappa}_s + \vec{\kappa}_i)$
defines the angular correlation of the pair:
the signal-idler pair always exists at roughly equal but opposite angles 
relative to the pump for degenerate SPDC.  This then allows for a simple
explanation of the experiment in terms of ``usual'' geometrical optics in
the following manner: we envision the nonlinear crystal as a ``hinge point'' and
``unfold'' the schematic of Fig. \ref{Imageset} into the Klyshko picture \cite{KlyshkoImg}
of Fig. \ref{fig:imaging-unfold}.  The signal-idler biphoton amplitudes can then be 
represented by straight lines (but keep in mind the different propagation directions) 
and therefore the image is reproduced in coincidences when the aperture, lens, 
and fiber tip are located according to the Gaussian thin lens equation of 
Eq.~(\ref{Lens-Eq}).  The image is exactly the same as that one
would observe on a screen placed at the fiber tip if detector $D_{1}$ were
replaced by a point-like light source and the nonlinear crystal by a reflecting
mirror. 

Comparing the ``unfolded'' schematic of the ghost imaging experiment with that of
the classical imaging setup of Fig.~\ref{fig:Projection-1}, it is not difficult to find that any 
``light point" on the object plane has a unique corresponding ``light point" on the image 
plane.  This point-to-point correspondence is the result of the constructive-destructive 
interference among these biphoton amplitudes that are illustrated as the geometrical rays
in Fig.~\ref{fig:imaging-unfold}.  Similar to the situation in classical imaging, 
these biphoton amplitudes which experience equal optical path propagation will superpose 
constructively at each pair of one-to-one points of the object plane and the image plane
for a joint-detection event, while these that 
experience unequal distance propagation will superpose
destructively at all other points on the object and image planes.   
The use of the imaging lens makes this constructive-destructive 
interference possible.  It is this unique point-to-point EPR correlation that makes the 
``ghost" image of the object-aperture function possible.  
Despite the completely different physics from classical optics, the remarkable 
feature is that the relationship between the focal length $f$ of the lens, the aperture's 
optical distance  $s_{o}$, and the image's optical distance $s_{i}$, 
satisfies the Gaussian thin lens equation of Eq.~(\ref{Lens-Eq}).  It is worth 
emphasizing again that the geometric rays in Fig.~\ref{fig:imaging-unfold}  
represent the biphoton amplitudes of a signal-idler photon pair, and the point-to-point 
correspondence is the result of the constructive-destructive interference of these 
biphoton amplitudes.  

We now calculate $G^{(2)}(\vec{\rho}_1, \vec{\rho}_2)$ for the ``ghost" imaging 
experiment in detail, where $\vec{\rho}_1$ and $ \vec{\rho}_2$ are the transverse 
coordinates of the point-like photodetector $D_1$ and $D_2$, on the object and 
image planes, respectively.  We will show that there exists a $\delta$-function-like 
point-to-point correlation between the object and image planes, 
$\delta(\vec{\rho}_1 - \vec{\rho}_2 / m)$.   We will then show 
how the object function of $A(\vec{\rho}_o)$ is transferred to the image plane as a 
magnified image $A(\vec{\rho}_2/m)$.   

\begin{figure}[hbt]
    \centering
    \includegraphics[width=95mm]{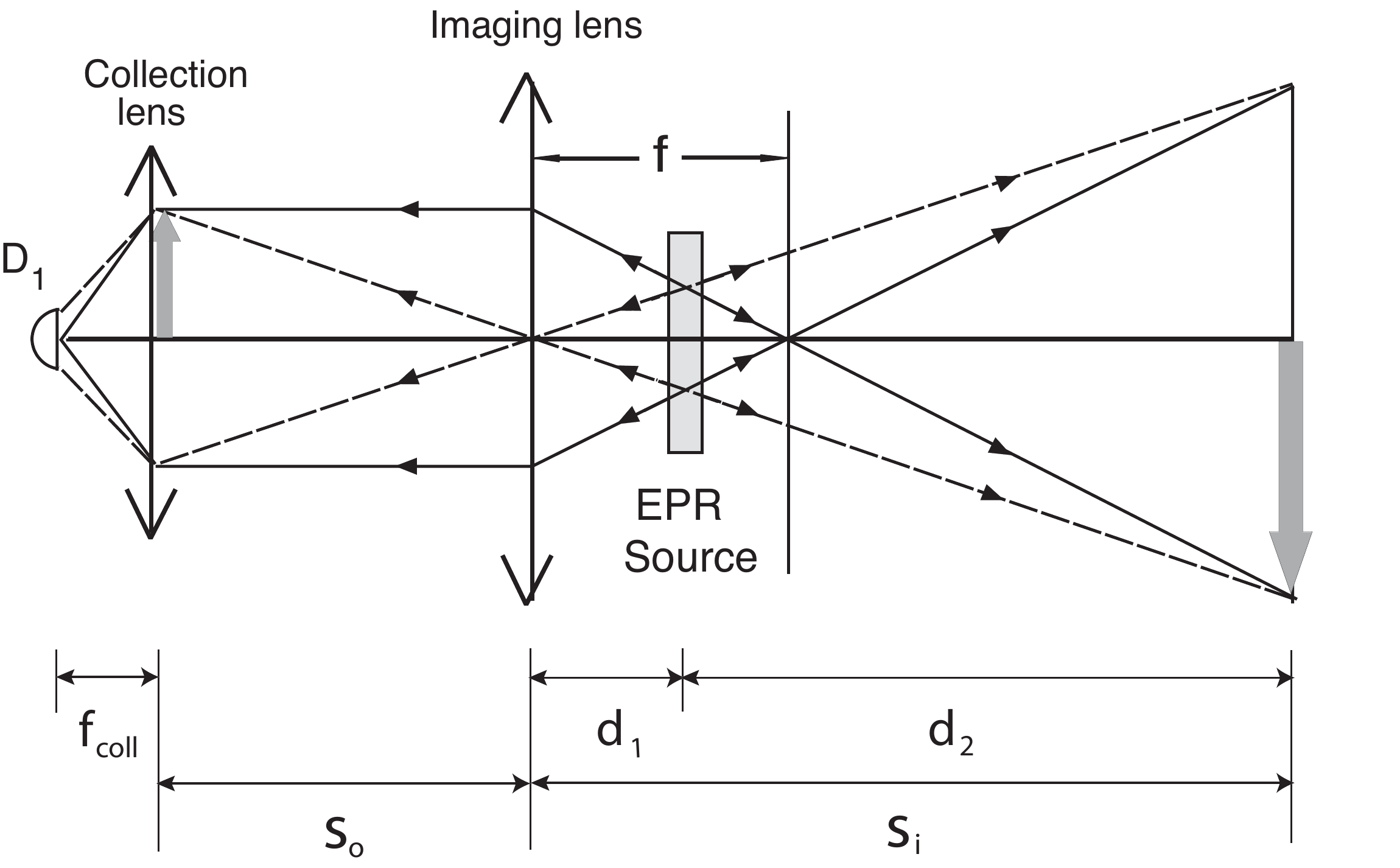}
     \parbox{14.25cm}{\caption{In arm-$1$ the signal propagates freely over a distance $d_1$ 
    from the output plane of the source to the imaging lens, passes an object aperture 
    at distance $s_o$, and then is focused onto photon-counting detector $D_1$ by a 
    collection lens.  In arm-$2$ the idler propagates freely over a distance $d_2$ from the 
    output plane of the source to a point-like photon counting detector $D_2$.} 
    \label{fig:imaging-unfold-2}}
\end{figure}

We first calculate the effective biphoton wavefunction 
$\Psi(\vec{\rho}_1, z_1, t_1; \vec{\rho}_2, z_2, t_2)$, as defined in Eq.~(\ref{G2}).    
By inserting the field operators into $\Psi(\vec{\rho}_1, z_1, t_1; \vec{\rho}_2, z_2, t_2)$, 
and considering the commutation relations of the field operators, 
the effective biphoton wavefunction is calculated to be
\begin{align}\label{psi-2}
\Psi(\vec{\rho}_1,z_1, t_1;\vec{\rho}_2, z_2, t_2) 
&= \Psi_0 \int d \vec{\kappa}_s \,d \vec{\kappa}_i \,
\delta(\vec{\kappa}_s + \vec{\kappa}_i ) 
\int d \omega_s \, d \omega_i \, \delta(\omega_s+\omega_i-\omega_p)
\\ \nonumber 
&  \hspace{5mm}\times \, 
g(\vec{\kappa}_s, \omega_s; \vec{\rho}_1, z_1) \, e^{-i\omega_s t_1}  \,
g(\vec{\kappa}_i, \omega_i; \vec{\rho}_2, z_2) \, e^{-i\omega_i t_2}.
\end{align}
Equation~(\ref{psi-2}) indicates a coherent superposition of all the
biphoton amplitudes shown in Fig.~\ref{fig:imaging-unfold}.  
Next, we follow the unfolded experimental setup of Fig.~\ref{fig:imaging-unfold-2} to 
establish the Green's functions $g(\vec{\kappa}_s, \omega_s, \vec{\rho}_1, z_1)$ and 
$g(\vec{\kappa}_i, \omega_i, \vec{\rho}_2, z_2)$.  In arm-$1$ the signal propagates freely 
over a distance $d_1$ from the output plane of the source to the imaging lens, 
passes an object aperture at distance $s_o$, and then is focused onto photon-counting 
detector $D_1$ by a collection lens.  We will evaluate 
$g(\vec{\kappa}_s, \omega_s, \vec{\rho}_1, z_1)$ by propagating the field from 
the output plane of the biphoton source to the object plane.
In arm-$2$ the idler propagates freely over a distance $d_2$ 
from the output plane of the biphoton source to a point-like detector $D_2$.     
$g(\vec{\kappa}_i, \omega_i, \vec{\rho}_2, z_2)$ is thus a free propagator.

\vspace{3mm}
\hspace*{-4mm}(I) Arm-$1$ (source to object):
\vspace{1mm}

The optical transfer function or Green's function in arm-$1$, which propagates the field 
from the source plane to the object plane, is given by:
\begin{align}\label{Arm-1}
g(\vec{\kappa}_s, \omega_s; \vec{\rho}_1, z_1 = d_{1}+s_{o}) 
&= e^{i \frac{\omega_s}{c} z_1} 
\int_{lens} d\vec{\rho}_l \, \int_{source} d\vec{\rho}_s 
\, \Big{\{} \, \frac{-i \omega_s}{2 \pi c d_1} e^{i \vec{\kappa_s} \cdot \vec{\rho}_s}  
\, e^{i \frac{\omega_s}{2 c d_1} |\, \vec{\rho}_s-\vec{\rho}_l\, |^2 } \Big{\}} 
\nonumber \\
& \ \ \ \ \times \, e^{-i \frac{\omega}{2 c f} \, | \vec{\rho}_l |^2} \, 
\Big{\{} \, \frac{-i \omega_s}{2 \pi c s_{o}} \,
e^{i \frac{\omega_s}{2 c s_o} |\,\vec{\rho}_l-\vec{\rho}_1\,|^2} \Big{\}},
\end{align}
where $\vec{\rho}_s$ and $\vec{\rho}_l$ are the transverse vectors defined, respectively, 
on the output plane of the source and on the plane of the imaging lens.  
The terms in the first and second curly brackets in Eq.~(\ref{Arm-1}) describe free space 
propagation from the output plane of the source to the imaging lens and from the imaging 
lens to the object plane, respectively.  Again, 
$ e^{i \frac{\omega_s}{2 c d_1} |\, \vec{\rho}_s-\vec{\rho}_l\, |^2 } $ and 
$e^{i \frac{\omega_s}{2 c s_o} |\,\vec{\rho}_l-\vec{\rho}_1\,|^2} $  are
the Fresnel phases as defined in the Appendix.  Here the imaging lens is treated
as a thin-lens, and the transformation function of the imaging lens is approximated as 
a Gaussian, $l(| \vec{\rho}_l |, f) \cong e^{-i \frac{\omega}{2 c f} \, | \vec{\rho}_l |^2}$.

\vspace{3mm}
\hspace{-4mm}(II) Arm-$2$ (from source to image):
\vspace{2mm}

In arm-$2$, the idler propagates freely from the source to the plane of $D_2$, which
is also the plane of the image. The Green's function is 
\begin{eqnarray}\label{Arm-2} 
g(\vec{\kappa}_i, \omega_i; \vec{\rho}_2, z_2=d_2) 
= \frac{-i \omega_i}{2 \pi c d_2} \, e^{i \frac{\omega_i}{c} d_2} 
\int_{source} d\vec{\rho'_s}  \,
e^{i \frac{\omega_i}{2 c d_2} |\, \vec{\rho'_s}-\vec{\rho}_2\, |^2} \, 
e^{i \vec{\kappa}_i \cdot \vec{\rho'_s} }
\end{eqnarray}
where $\vec{\rho'_s}$ and $\vec{\rho}_2$ are the transverse vectors defined, respectively, 
on the output plane of the source and the plane of photodetector $D_2$.  

\vspace{3mm}
\hspace{-4mm}(III) $\Psi(\vec{\rho}_1, \vec{\rho}_2)$ and 
$G^{(2)}(\vec{\rho}_1, \vec{\rho}_2)$ (object plane - image plane):
\vspace{2mm}

For simplicity, in the following calculation we consider degenerate 
($\omega_s = \omega_i = \omega$) and collinear SPDC.  The effective 
transverse biphoton wavefunction $\Psi(\vec{\rho}_1, \vec{\rho}_2)$ 
is then evaluated by substituting the Green's functions 
$g(\vec{\kappa}_s, \omega; \vec{\rho}_1, z_1) $ and 
$g(\vec{\kappa}_i, \omega; \vec{\rho}_2, z_2)$
into Eq.~(\ref{psi-2}),
\begin{align}\label{biphoton_x}
\Psi(\vec{\rho}_1,\vec{\rho}_2)
&\propto  \int d\vec{\kappa}_s \, d\vec{\kappa}_i \, 
\delta(\vec{\kappa}_s +  \vec{\kappa}_i) \, g(\vec{\kappa}_s, \omega; \vec{\rho}_1, z_1)  \, 
g(\vec{\kappa}_i, \omega; \vec{\rho}_2, z_2) \nonumber \\
&\propto e^{i \frac{\omega}{c} (s_o+s_i)}  \int d\vec{\kappa}_s \, d\vec{\kappa}_i \, 
\delta(\vec{\kappa}_s +  \vec{\kappa}_i) 
\int_{lens} d\vec{\rho}_l \, \int_{source} d\vec{\rho}_s \,
 \, e^{i \vec{\kappa_s} \cdot \vec{\rho}_s}  
e^{i \frac{\omega}{2 c d_1} |\, \vec{\rho}_s-\vec{\rho}_l\, |^2 }  \nonumber \\ 
& \ \ \ \ \times \, e^{-i \frac{\omega}{2 c f} \, | \vec{\rho}_l |^2} \,  \, 
e^{i \frac{\omega_s}{2 c s_o} |\,\vec{\rho}_l-\vec{\rho}_1\,|^2}  
 \int_{source} d\vec{\rho'_s} \,\,
e^{i \vec{\kappa}_i \cdot \vec{\rho'_s} } \,
e^{i \frac{\omega_i}{2 c d_2} |\, \vec{\rho'_s}-\vec{\rho}_2\, |^2}
\end{align}
where all the proportionality constants have been ignored.
After completing the double integral of $d\vec{\kappa}_s$ and $d\vec{\kappa}_i$
\begin{eqnarray*}\label{delta-source}
\int d\vec{\kappa}_s \, d\vec{\kappa}_i \, \delta(\vec{\kappa}_s +  \vec{\kappa}_i)\,
e^{i \vec{\kappa_s} \cdot \vec{\rho}_s} \, e^{i \vec{\kappa}_i \cdot \vec{\rho'_s} } 
\sim \, \delta(\vec{\rho}_s - \vec{\rho'_s}),
\end{eqnarray*}  
Eq.~(\ref{biphoton_x}) becomes
\begin{align*}\label{biphoton_y}
\Psi(\vec{\rho}_1,\vec{\rho}_2) 
\propto e^{i \frac{\omega}{c} (s_0+s_i)}
 \int_{lens} d\vec{\rho}_l \int_{source} d\vec{\rho}_s \,
e^{i \frac{\omega}{2 c d_2} |\, \vec{\rho}_2-\vec{\rho}_s\, |^2 } \,
e^{i \frac{\omega}{2 c d_1} |\, \vec{\rho}_s-\vec{\rho}_l\, |^2 } 
\, e^{-i \frac{\omega}{2 c f} \, | \vec{\rho}_l |^2}  \,
e^{i  \frac{\omega}{2 c s_o} |\,\vec{\rho}_l-\vec{\rho}_o\,|^2}.
\end{align*}
Next, we complete the integral for $d\vec{\rho}_s$,
\begin{equation}\label{biphoton_z}
\Psi(\vec{\rho}_1,\vec{\rho}_2) 
\propto e^{i \frac{\omega}{c} (s_0+s_i)} 
\int_{lens} d\vec{\rho}_l \, 
e^{i \frac{\omega}{2 c s_i} |\, \vec{\rho}_2-\vec{\rho}_l\, |^2} \,
 e^{-i \frac{\omega}{2 c f} \, | \vec{\rho}_l |^2}
e^{i \frac{\omega}{2 c s_o} |\,\vec{\rho}_l-\vec{\rho}_1\,|^2},
\end{equation}
where we have replaced $d_1+d_2$ with $s_i$ (as depicted in Fig.~\ref{fig:imaging-unfold-2}).
Although the signal and idler propagate in different directions along two optical arms, 
interestingly, the Green function in Eq.~(\ref{biphoton_z}) is equivalent to that of a 
classical imaging setup, as if the field is originated from a point $\vec{\rho}_1$ 
on the object plane and propagated the lens and then arrived at point $\vec{\rho}_2$ 
on the imaging plane.  The mathematics is consistent with 
our previous qualitative analysis of the experiment.  

The finite integral on $d\vec{\rho}_l$ yields a point-to-``spot" relationship between
the object plane and the image plane that is defined by the Gaussian thin-lens 
equation
\begin{align}\label{biphoton_zz}
\Psi(\vec{\rho}_1,\vec{\rho}_2) \propto \int_{lens} d\vec{\rho}_l \, e^{i \frac{\omega}{2 c}
[\frac{1}{s_o} + \frac{1}{s_i} - \frac{1}{f}] |\, \vec{\rho}_l|^2}  \,
e^{-i\frac{\omega}{c} (\frac{\vec{\rho}_1}{s_o} + \frac{\vec{\rho}_2}{s_i})\cdot \vec{\rho}_l} 
= somb\Big{(}\frac{R}{s_o}\, \frac{\omega}{c} 
|\vec{\rho}_{1} + \frac{\vec{\rho}_{2}}{m}| \Big{)}.
\end{align}
If the integral is taken to infinity, 
by imposing the condition of the Gaussian thin-lens equation the effective 
transverse biphoton wavefunction can be approximated as a $\delta$ function 
\begin{eqnarray}\label{biphoton_x_fin}
\Psi(\vec{\rho}_1,\vec{\rho}_2) \sim
\delta(\vec{\rho}_1 + \vec{\rho}_2 / m) \sim
\delta(\vec{\rho}_o + \vec{\rho}_I / m),
\end{eqnarray}
where we have replaced $\vec{\rho}_1$ and $ \vec{\rho}_2$ with $\vec{\rho}_o$ 
and $\vec{\rho}_I$, respectively,  to emphasize the point-to-point EPR 
correlation between the object and image planes.
To avoid confusion with the ``idler" we have used $\vec{\rho}_I$ to label 
the image plane.

We now include an object-aperture function, a collection lens and a photon counting
detector $D_1$ into the optical transfer function of arm-$1$ as shown in Fig.~\ref{Imageset}.   
The collection-lens$-D_1$ package can be simply treated as a 
``bucket" detector.  The ``bucket" detector integrates the biphoton amplitudes
$\Psi(\vec{\rho}_o, \vec{\rho}_2)$, which are modulated by the object aperture function 
$A(\vec{\rho}_o)$ into a joint photodetection event.  This process 
is equivalent to the following convolution
\begin{eqnarray}\label{biphoton_final-2}
R_{1, 2} \propto  \int_{object} d\vec{\rho}_o \, \big{|}  A(\vec{\rho}_o)  \big{|}^2 \,
\big{|}  \Psi(\vec{\rho}_o, \vec{\rho}_2)  \big{|}^2  
\simeq \big{|} A(\vec{\rho}_2 / m) \big{|}^{2} =
 \big{|} A(\vec{\rho}_I / m) \big{|}^{2}.
\end{eqnarray}
Again, $D_2$ is scanned in the image plane ($\vec{\rho}_2 = \vec{\rho}_I$). 
A ghost image of the object is thus reproduced on the image plane by means of
the joint-detection between the point-like-detector $D_2$ and the bucket detector
$D_1$. 

The physical process corresponding to the above convolution is rather simple.
Suppose the point detector $D_2$ is triggered by an idler photon at a transverse position of 
$\vec{\rho}_\textrm{I}$ in a joint-detection event with the bucket detector $D_1$ 
which is triggered by the signal twin that is either transmitted or reflected from 
a unique point $\vec{\rho}_\textrm{o}$ on the object plane.  This unique point-to-point 
determination comes from the non-factorizable correlation function
$\delta(\vec{\rho}_\textrm{o} + \vec{\rho}_{I} / m)$.   
Now, we move $D_2$ to another transverse 
position $\vec{\rho'}_{I}$ and register a joint-detection event.
The signal photon that triggers $D_1$ must be either transmitted or reflected 
from another unique point $\vec{\rho'}_{o}$ on
the object plane which is determined by
$\delta(\vec{\rho'}_\textrm{o} + \vec{\rho'}_\textrm{I} / m)$.  
The chances of receiving 
a joint detection event at $\vec{\rho}_\textrm{I}$ and at 
$\vec{\rho'}_\textrm{I}$ would be modulated by the values of the aperture function
$A(\vec{\rho}_\textrm{o})$ and $A(\vec{\rho'}_\textrm{o})$, respectively.  
Accumulating a large number of joint-detection events at each transverse coordinates
on the image plane, the aperture function $A(\vec{\rho}_\textrm{o})$ is thus 
reproduced in the joint-detection as a function of $\vec{\rho}_\textrm{I}$.

\vspace{5mm}

The observation of type-one ghost imaging has demonstrated a non-factorizable
point-to-point EPR correlation between the object and image planes.  This point-to-point
correlation is the result of a constructive-destructive interference 
between biphoton amplitudes,
\begin{equation}\label{Coherent-Img}
G^{(2)}(\vec{\rho}_o,\vec{\rho}_I) 
= \Big{|}  \int d\vec{\kappa}_s \, d\vec{\kappa}_i \, 
\delta(\vec{\kappa}_s +  \vec{\kappa}_i) \, g(\vec{\kappa}_s, \vec{\rho}_o)  \, 
g(\vec{\kappa}_i, \vec{\rho}_I) \Big{|}^2
= somb^2\Big{(}\frac{R}{s_o}\, \frac{\omega}{c} 
|\vec{\rho}_{O} + \frac{\vec{\rho}_{I}}{m}| \Big{)}.
\end{equation}
In this view we consider the ghost imaging experiment of Pittman \emph{et al}. 
a realization of the 1935 EPR {\em gedankenexperiment} \cite{disug} \cite{Howell}.

Classical theory has difficulties when facing type-one ghost imaging phenomenon.  
In the classical theory of light, a joint measurement between 
two photodetectors $D_1$ and $D_2$ measures the statistical 
correlation of intensity fluctuations,
\begin{equation}\label{Classical-Int}
G^{(2)}(\vec{\rho}_1,\vec{\rho}_2) 
= \langle \, I(\vec{\rho}_1) \, I(\vec{\rho}_2) \, \rangle 
= \bar{I}(\vec{\rho}_1) \bar{I}(\vec{\rho}_2) + 
\langle \, \Delta I(\vec{\rho}_1) \Delta I(\vec{\rho}_2) \rangle.
\end{equation}
Therefore, the point-to-point image-forming correlation is considered as a result of the
statistical correlation of intensity fluctuations between the object and the image planes.
Comparing Eq.~(\ref{Classical-Int}), which has a constant background 
$\bar{I}(\vec{\rho}_1) \bar{I}(\vec{\rho}_2)$, with Eq.~(\ref{Coherent-Img}), which has a
zero background, the mean intensities 
$\bar{I}(\vec{\rho}_1)$ and $\bar{I}(\vec{\rho}_2)$ must be zero, otherwise the result 
would lead to non-physical conclusions. The measurements, 
however, never yield zero mean values of $\bar{I}(\vec{\rho}_1)$ 
and $\bar{I}(\vec{\rho}_2)$ under any circumstances.  In fact, the individual-detector 
counting rates of $D_1$ and $D_2$ were monitored in the experiment of 
Pittman \emph{et al}. with much greater value than that of the coincidences.
It is clear that the classical theory of statistical correlation of intensity fluctuations 
does not reflect the correct physics behind type-one ghost imaging.   

\section{Type-two ghost imaging with chaotic radiation}\label{TypeTwo}

\hspace{6.5mm}In this section we discuss the physics of type-two 
ghost imaging.  The near-field lensless ghost imaging with chaotic radiation 
was first  demonstrated by Scarcelli \emph{et al}. in the years from 2005 to 2006 
\cite{prl2}\cite{prl1} following their experimental demonstration of two-photon interference 
of chaotic light \cite{Europhys}.  
\begin{figure}[htb]
    \centering
    \includegraphics[width=90mm]{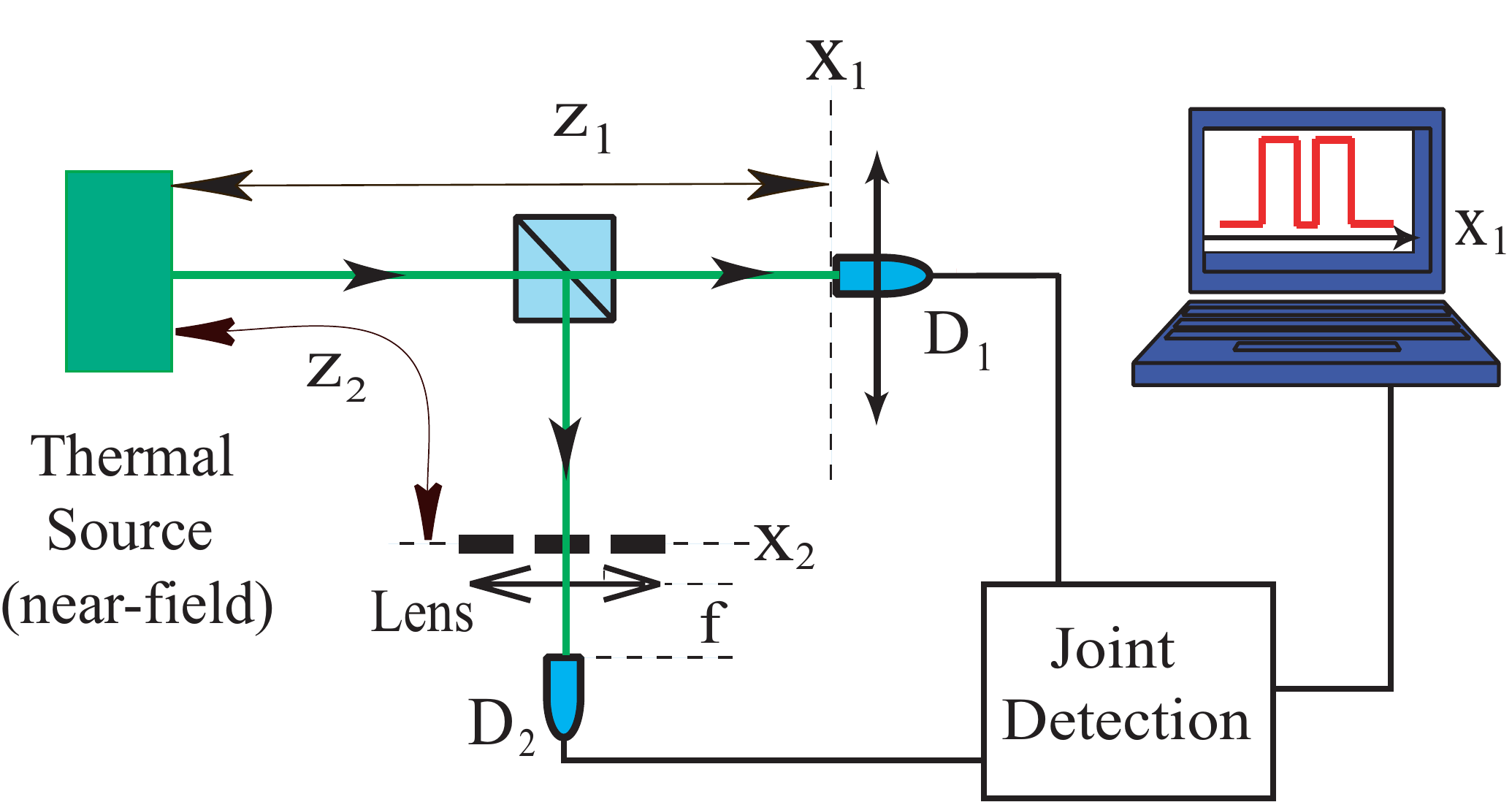} 
    \parbox{14.25cm}{\vspace{3mm}\caption{Near-field lensless ghost imaging with 
    chaotic light demonstrated in 2006 by Scarcelli \emph{et al}. $D_1$ is a point-like 
    photodetector that is scannable along the $x_1$-axis. The joint-detection between 
    $D_1$ and the bucket detector $D_2$ is realized either by a photon-counting 
    coincidence counter or by a standard HBT linear multiplier (RF mixer).  In this 
    measurement $D_2$ is fixed in the focal point of a convex lens, playing the role of a 
    bucket detector.  The counting rate or the photocurrent of $D_1$ and $D_2$,
    respectively, are measured to be constants.  Surprisingly, an image of the 1-D
    object is observed in the joint-detection between $D_1$ and $D_2$ by 
    scanning $D_1$ in the plane of $z_1 = z_2$ along the $x_1$-axis. The image,
    is blurred out when $z_1 \neq z_2$.  There is no doubt that thermal radiations 
    propagate to any transverse plane in a random and chaotic manner.  There is 
    no lens applied to force the thermal radiation ``collapsing" to a point or speckle 
    either.  What is the physical cause of the point-to-point image-forming correlation 
    in coincidences? }
    \label{fig:HBT-Fig1}}
\end{figure}
\begin{figure}
 \centering
    \includegraphics[width=70mm]{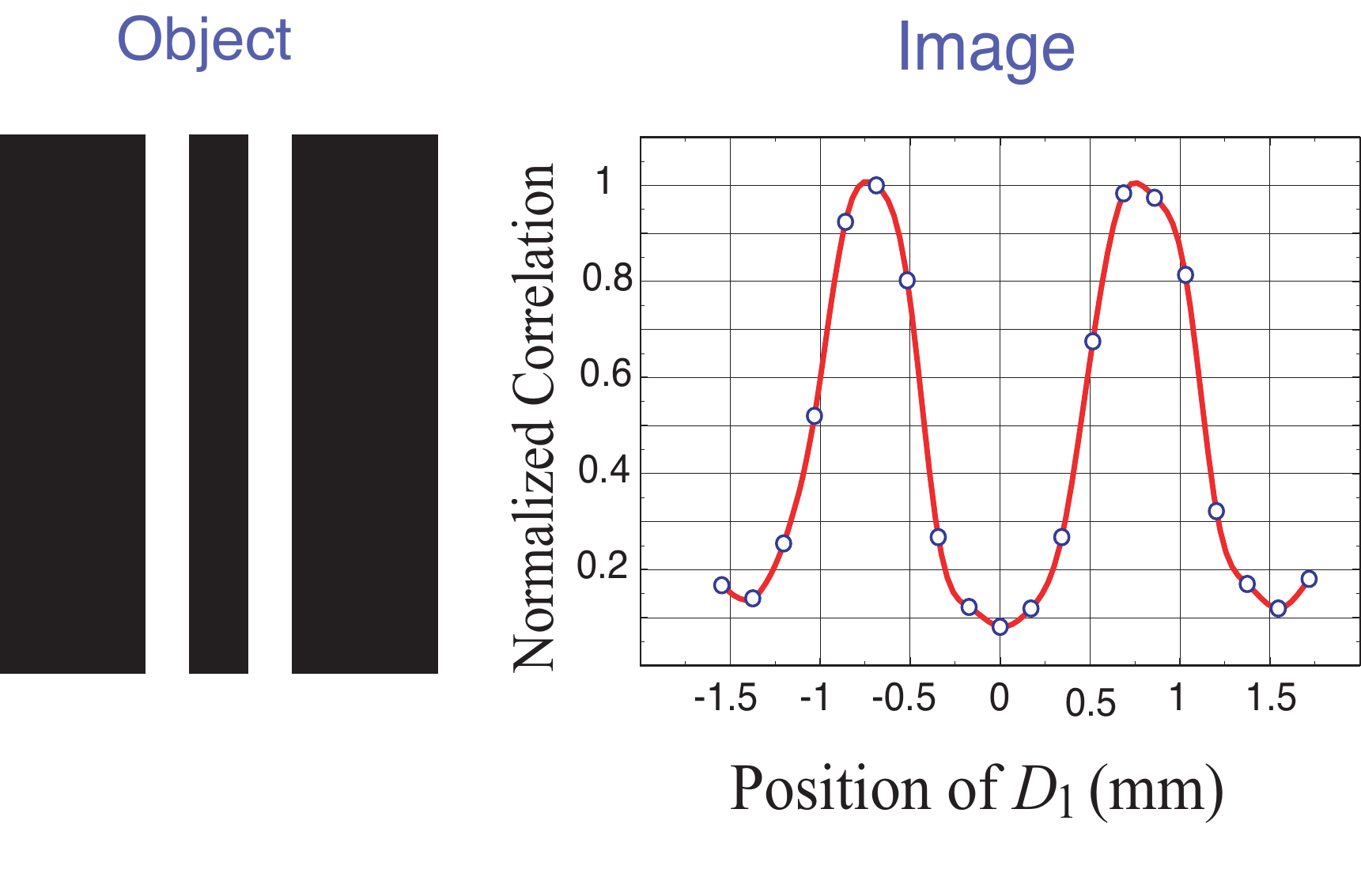}
    \parbox{14.25cm}{\caption{The double-slit and its ghost image.  Notice, the constant 
    background has been removed from the correlation.}
    \label{fig:HBT-Fig2-1}}
\end{figure}
The schematic experimental setup of their 2006 demonstration is shown in Fig.~\ref{fig:HBT-Fig1}.  
Radiation with a narrow spectral bandwidth $\Delta \omega$ of a few millimeters diameter 
from a chaotic pseudothermal source 
\cite{martienssen} was equally divided into two by a 
$50\% - 50\%$ non-polarizing beam-splitter. In the reflected arm, a
double-slit with slit separation $b=1.5$~mm and slit width
$a=0.2$~mm, was placed at a distance $z_{2}=139$~mm from the 
source and a bucket detector
$D_{2}$ was placed just behind the object.  In the transmitted arm a point detector
$D_{1}$ was scanned in the transverse plane of $z_{1}=z_{2}$.  
Scarcelli \emph{et al} tested two different joint detection schemes, namely the photon
counting coincidence circuit and the standard HBT correlator.  
In the photon counting regime two Geiger mode avalanche 
photodiodes were employed for single-photon detection.  In the bright light
condition, two silicon PIN diodes were used with a standard analog HBT linear multiplier.
The bucket detector $D_{2}$ was simulated by using a short focal length lens 
($f=25mm$) to focus the light coming from the object onto the active area of
the detector while the point detector $D_{2}$ was simulated by a pinhole like aperture. 
After a large number of reaped measurements for different experimental schemes and
conditions, Scarcelli \emph{et al} reported the following observations.

\hspace{-3mm}Observation (1):  A typical measured ghost image of the double-slit is 
shown in Fig.~\ref{fig:HBT-Fig2-1}.  The measured curve reports the joint-detection
counting rate between $D_1$ and $D_2$, or the output current of a HBT linear 
multiplier, as a function of the transverse position of the point detector $D_1$ along
$x_1$ axis.   Notice, in Fig.~\ref{fig:HBT-Fig2-1} the constant background has been 
removed from the correlation.  

\hspace{-3mm}Observation (2): The measured contrasts vary significantly under different 
experimental schemes and conditions.  It was found that the image contrast can achieve 
$\sim50\%$ in photon counting measurement if no more than one joint-detection event 
occurring within the time window of the coincidence circuit. $50\%$ is the 
maximum image contrast we expect for thermal light ghost imaging.  

\hspace{-3mm}Observation (3): To achieve less than one joint-detection event per 
coincidence time window, weak light source is not a necessary condition. 
It can be easily achieved under bright light condition by using adjustable ND-filters 
with $D_1$ and $D_2$.    

\begin{figure}[hbt]
    \centering
    \includegraphics[width=95mm]{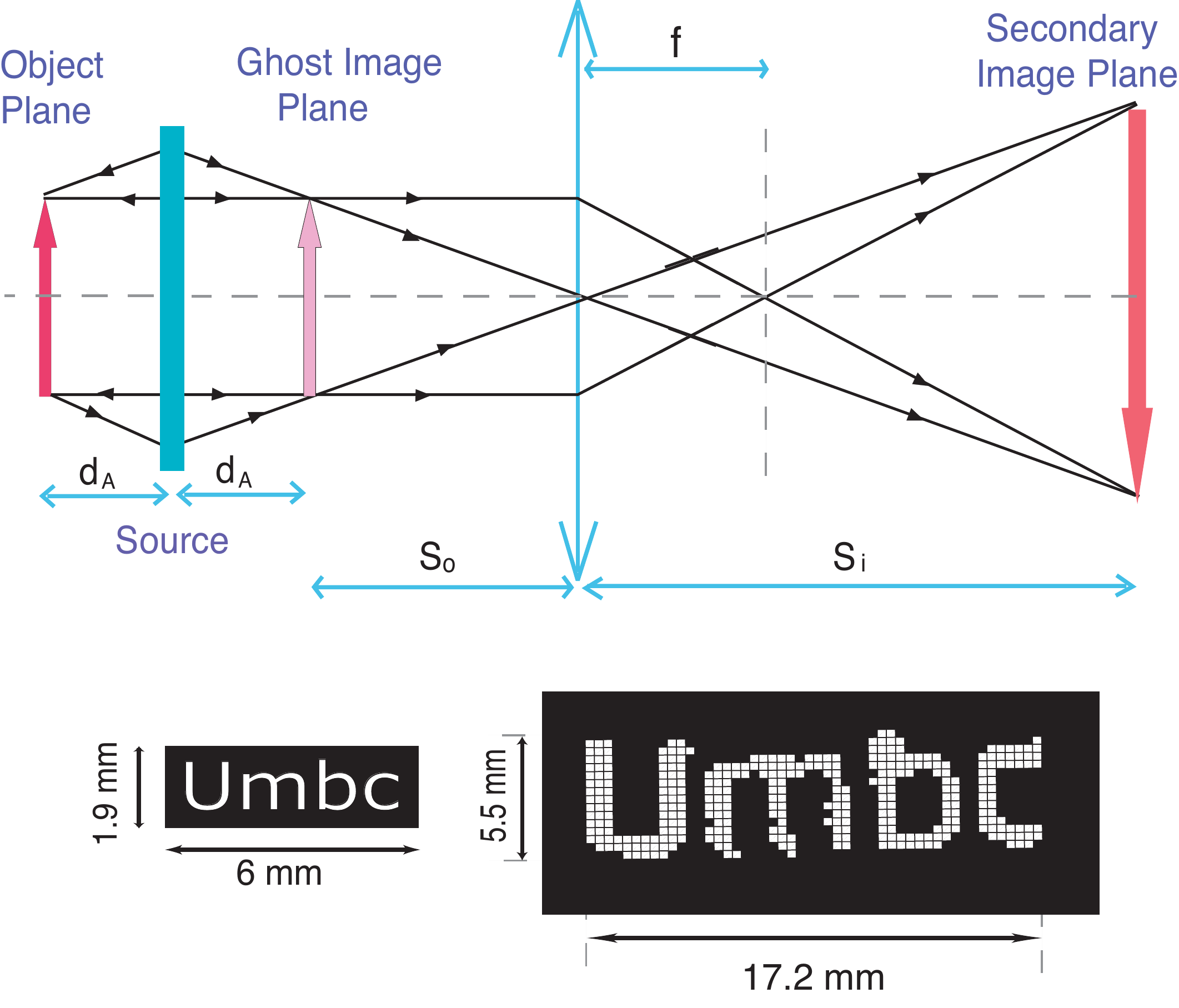} 
    \parbox{14.25cm}{\caption{Unfolded schematic experimental setup of 
    a secondary image measurement of the ghost image and the measured secondary 
    images. By using a convex lens of focal length $f$, the ghost image is imaged onto 
    a secondary image plane, which is
    defined by the Gaussian thin-lens equation, $1/s_o + 1/s_i = 1/f$, with 
    magnification $m = -s_i/s_o$.  This setup is useful for distant large scale ghost
    imaging applications.}
    \label{fig:Thermal}}
\end{figure}
To confirm the observations are imaged images, and not ``projection shadows", 
Scarcelli \emph{et al}. made two additional measurements. In the first 
measurement, photodetector
$D_1$ was moved away from the ghost image plane of $z_1 = z_2$.  Whether 
moved in the direction of $z_1 > z_2$ or  $z_1 < z_2$, 
the ghost image became ``blurred".  
The measurement also showed that the depth of the image is a function
of the angular size of the thermal source: a larger angular sized source (opening
angle $\Delta \theta$ relative to the photodetectors) produces sharper image with 
shorter image depth.   In the second measurement, Scarcelli \emph{et al}.
constructed a secondary imaging system, illustrated schematically in 
Fig.~\ref{fig:Thermal}.  By using a convex lens of focus length $f$ 
the ghost image is imaged onto a secondary image plane,
which is defined by the Gaussian thin-lens equation, $1/s_o + 1/s_i = 1/f$, 
with magnification $m = -s_i/s_o$.   In this measurement
the scanning photodetector $D_1$ is placed on the secondary imaging
plane.  The secondary image of the ghost image is observed in the 
joint-detection between $D_1$ and $D_2$ by means of either a
photon-counting coincidence counter or a HBT linear multiplier.

\subsection{What is the physical cause of chaotic light ghost imaging?}

\hspace{6.5mm}It is the partial point-to-point correlation between the object plane 
and the image plane that makes ghost imaging with thermal light possible.
Similar but different from classical imaging and type-one ghost imaging, mathematically, 
type-two ghost imaging is the result of a convolution between the aperture 
function $|A(\vec{\rho}_2)|^2$ and a $\delta$-function like partial point-to-``spot" 
correlation function 
\begin{align}\label{Ghost-3-D}
R_{12} \propto  \int_{object} d\vec{\rho}_2 \, \big{|} A(\vec{\rho}_2)\big{|}^2  
\, \big{[} 1 + 
somb^2 \big{(} \, \frac{\pi \Delta \theta \, |\vec{\rho}_{1} 
- \vec{\rho}_{2}|}{\lambda} \big{)} \big{]}  
\end{align} 
in 2-D, where $\Delta \theta$ is the angular diameter of the radiation source viewed 
from the photodetector, $\vec{\rho}_{1}$ and $\vec{\rho}_{2}$ are the transverse 
coordinates on the object plane and the image plane, respectively, or 
\begin{align}\label{Ghost-1-D}
R_{12} \propto  
\int_{object} dx_2 \, \big{|} A(x_2)\big{|}^2  
\, \big{[} 1 + sinc^2 \big{(} \, \frac{\pi \Delta \theta \, (x_1 - x_2)}{\lambda} \big{)}  
\big{]}. 
\end{align} 
in 1-D.  For a chosen wavelength, the spatial resolution of the ghost
image is determined by the angular diameter of the light source: the larger the size 
of the source in transverse dimensions, the higher the spatial resolution of the lensless
ghost image.  The point-to-``spot" image-forming functions in Eqs.~(\ref{Ghost-3-D}) 
and (\ref{Ghost-1-D}) have been verified experimentally by Scarcelli \emph{et al}.

The physical process corresponding to the convolution of Eq.~(\ref{Ghost-3-D}) 
and (\ref{Ghost-1-D}) is similar to that of  the type-one ghost imaging.   
Suppose the point detector $D_1$ or a CCD element is triggered by a 
photon at a transverse position of $\vec{\rho}_1$ in a joint-detection event with the bucket 
detector $D_2$ which is triggered by another photon that is either transmitted or reflected 
from the object.  According to Eq.~(\ref{Ghost-3-D}) and (\ref{Ghost-1-D}), under condition 
of $z_1 = z_2$, the photon from the object would have twice greater chance to be found 
at $\vec{\rho}_\textrm{obj} = \vec{\rho}_1$.  Now, we move $D_1$ to another transverse 
position $\vec{\rho'}_1$, or locate another CCD element at $\vec{\rho'}_1$ for joint-detection.  
The photon that triggers $D_2$ would have twice greater chance of been located 
at $\vec{\rho'}_\textrm{obj} = \vec{\rho'}_1$.  The probabilities of receiving 
a joint detection event at $\vec{\rho}_1 = \vec{\rho}_\textrm{obj}$ and at 
$\vec{\rho'}_1 = \vec{\rho'}_\textrm{obj}$ would be modulated by the values of the 
aperture function $A(\vec{\rho}_\textrm{obj})$ and $A(\vec{\rho'}_\textrm{obj})$, 
respectively.  Accumulating a large number of joint-detection events for each transverse 
coordinates $\vec{\rho}_1$, or for each CCD element in the image plane,
a 50\% contrast aperture function $A(\vec{\rho}_1) = A(\vec{\rho}_\textrm{obj})$ is thus 
reproduced in the joint-detection as a function of $\vec{\rho}_1$.\footnote{
To observe thermal light ghost image with maximum 50\% contrast requires achieving
a necessary experimental condition: no more than one joint detection event within 
the coincidence time window. }

\begin{figure}[hbt]
    \centering
    \includegraphics[width=66mm]{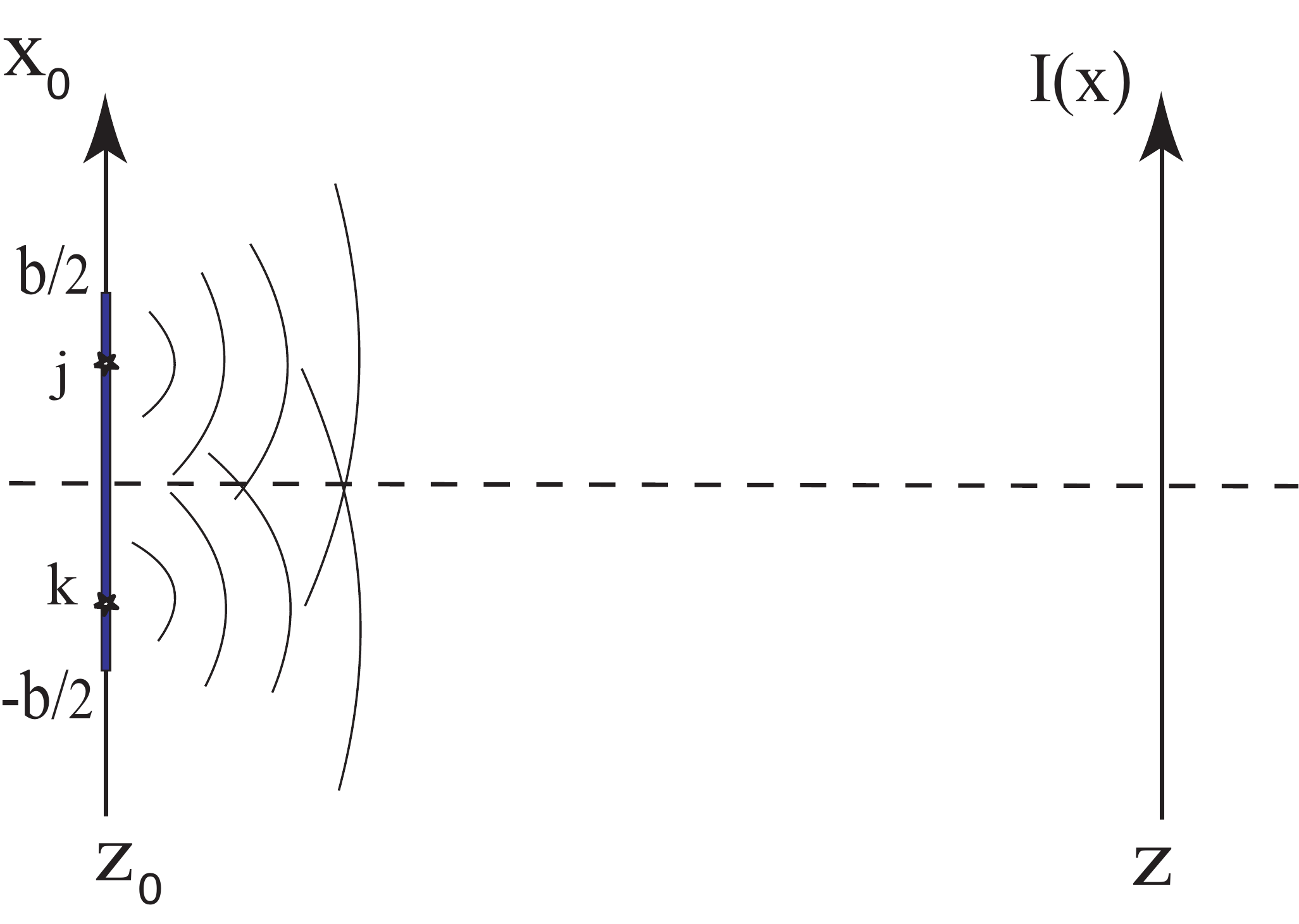} 
    \parbox{14.25cm}{\caption{A large number of independent point sub-sources, 
    such as the $j$th and $k$th, are randomly distributed on the plane of a
    thermal source.  These point sub-sources randomly radiate independent spherical 
    waves to the object and image planes, respectively.  Due to the chaotic nature 
    of the source, these independent 
    sub-intensities simply added together yielding a constant total intensity in space and 
    in time on any transverse planes.   
    }\label{fig:Near-Field-2}}
\end{figure}

To achieve thermal light ghost image with $50\%$ contrast, we need 
(1) randomly distributed radiations on the object plane and on the image plane, respectively; 
and (2) for any photoelectron event at $\vec{\rho}_{1}$ there exists a unique corresponding
point $\vec{\rho}_\textrm{obj} = \vec{\rho}_{1}$ on the object plane which has twice 
chance of observing another photoelectron event jointly and simultaneously.  There is 
no doubt that random and chaotic radiation would propagate to any transverse plane 
in a random and chaotic manner.  Therefore, condition (1) is satisfied automatically for 
chaotic thermal radiation.  However, it is not easy to understand condition (2).  
We have been asking ourself a question since the first observation of lensless
thermal light ghost image: what is the physical cause of the 
non-factorizable partial point-to-point image-forming function of
$1 + \delta \big{(} \, \vec{\rho}_{1} - \vec{\rho}_\textrm{obj} \,\big{)}$?  There seems no reason 
to have such a statistical correlation for thermal light.  
Figure~\ref{fig:Near-Field-2} schematically illustrates this situation.  
To simplify the picture we assume the source in 1-D with a large number of 
independent point sub-sources randomly distributed from $-b/2$ to $b/2$. 
Each point sub-source,
such as the $j$th and the $k$th sub-source, randomly radiates independent spherical 
waves to the object and image planes, respectively.  Due to the chaotic nature 
of the source, these independent and incoherent
sub-intensities simply add together yielding a constant total intensity spatially and 
temporally on any transverse plane.  The more chaotic sub-fields that contribute to the 
intensity sum, the less value of $\Delta I / I$ is expected.  
For any two transverse planes, such as the object and the image planes in 
Fig.~\ref{fig:HBT-Fig1}, each with independent and randomly distributed intensities,
statistically, there is no reason to expect any spatial or temporal correlations.  
What is the physical cause that forces a twice large probability for the thermal 
radiation to jointly appear at $\vec{\rho}_{1} = \vec{\rho}_\textrm{obj}$?  

In fact, we have been facing this question since 1956, after the discovery of  
Hanbury Brown and Twiss (HBT).  
The lensless ghost imaging setup looks similar to that of the historical HBT 
spatial interferometer which was used for measuring the angular 
size of distant stars.   A significant difference is that the lensless ghost imaging 
measurement is in near-field\footnote{The concept of
``near-field" was defined by Fresnel to be distinct from the Fraunhofer
far-field.  The Fresnel near-field is different from the ``near-surface-field" which
considers a  distance of a few wavelengths from a surface.} 
for imaging purposes \cite{prl2}.  
The HBT experiment created quite a surprise in the physics community with an 
enduring debate about the classical or quantum nature of the phenomenon 
\cite{hbt-book}\cite{Scully-book}.  
\begin{figure}[htb]
 \centering
    \includegraphics[width=100mm]{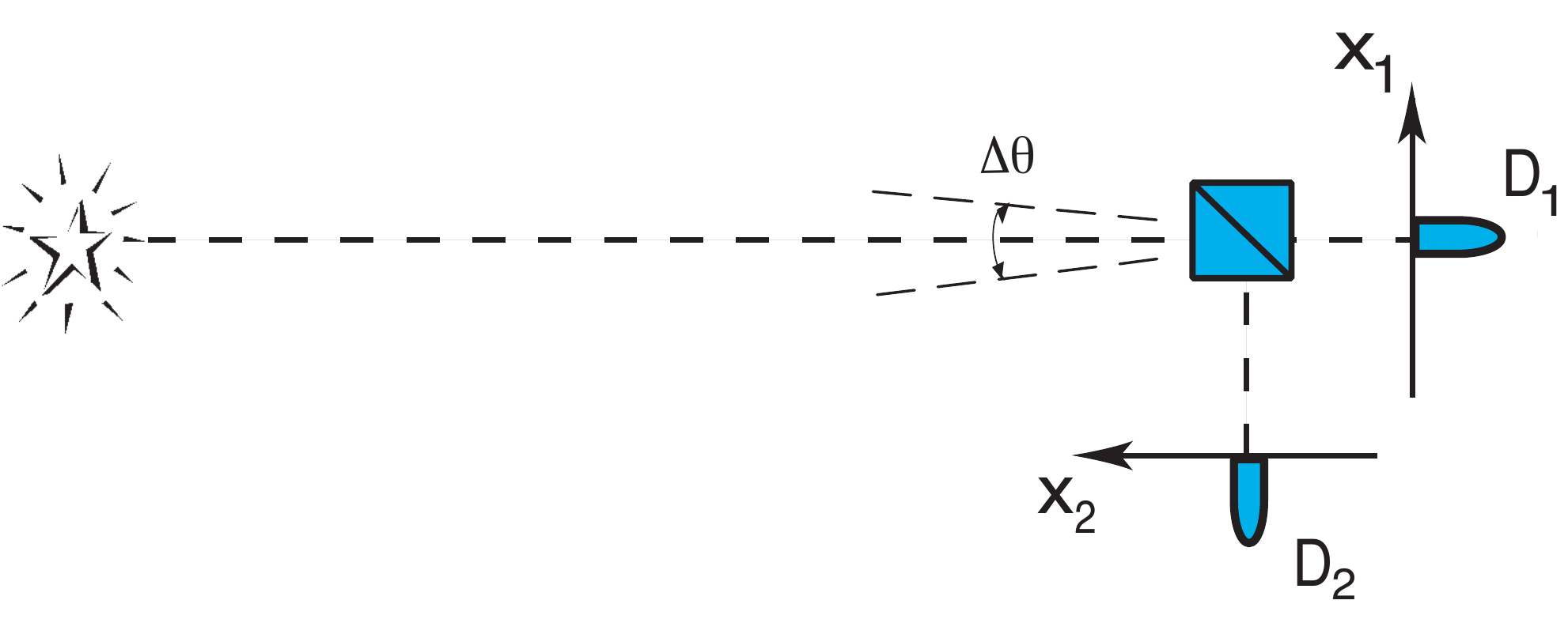}
    \parbox{14.25cm}{\caption{Schematic of the historical Hanbury Brown and 
    Twiss experiment which measures the transverse spatial correlation 
    of far-field thermal radiation. }
    \label{fig:HBT-Fig1-2}}
\end{figure}
\begin{figure}
    \centering
    \includegraphics[width=85mm]{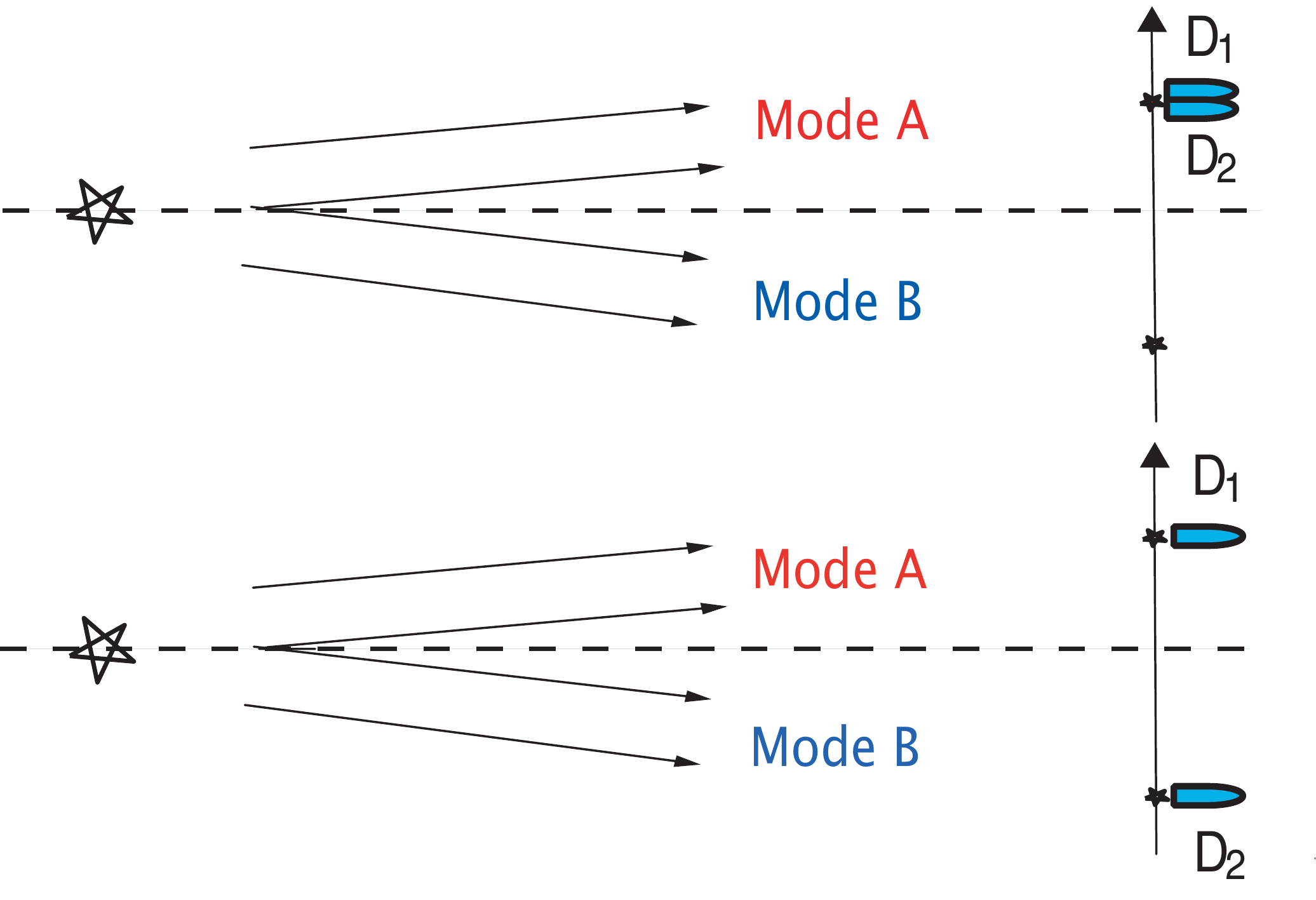} 
    \parbox{14.25cm}{\caption{A phenomenological interpretation of the historical 
    HBT experiment. 
    Upper: the two photodetectors receive identical modes of the far-field radiation 
    and thus experience identical intensity fluctuations. 
    The joint measurement of $D_1$ and $D_2$ gives a
    maximum value of $\langle\Delta I_{1}\Delta I_{2}\rangle$.  Lower: the two
    photodetectors receive different modes of the far-field radiation.  
    In this case the joint measurement gives $\langle\Delta I_{1}\Delta I_{2}\rangle=0$.
    Unfortunately, this hand-waving interpretation does not reflect the correct physics 
    in the case of $\Delta \theta \neq 0$.  For a finite angular sized source, 
    there is no chance, at least realistically, for $D_1$ and $D_2$ to 
    receive radiation from a single radiation mode only.  Nevertheless, the above 
    theory has convinced us to believe that the observation of the intensity fluctuation 
    correlation only takes place in the far-field zone of the thermal source.  
     }\label{fig:HBT-Physics-2}}
\end{figure}
\begin{figure}
    \centering
    \includegraphics[width=85mm]{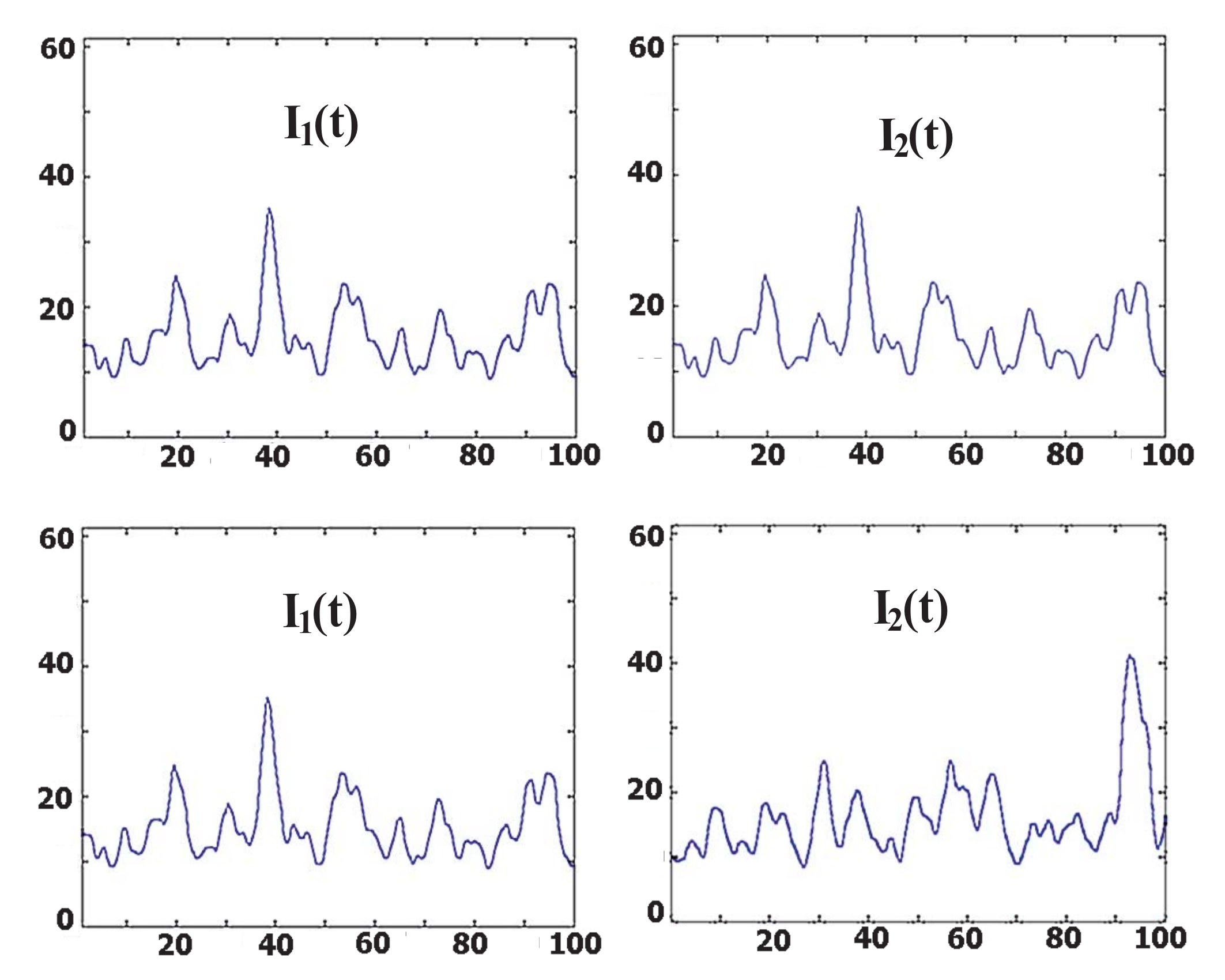} 
    \parbox{14.25cm}{\caption{Schematic illustration of the light intensities $I_1(t)$ at 
    $D_1$ and $I_2(t)$ at $D_2$.  The two upper (lower) curves of $I(t)$ corresponds 
    to the upper (lower) configuration in Fig~\ref{fig:HBT-Physics-2}. } 
    \label{fig:HBT-Physics-3}}
\end{figure}
Figure~\ref{fig:HBT-Fig1-2} is a schematic of the historical HBT
experiment which measures the transverse spatial 
correlation of far-field thermal radiation. 
Performing the measurement in 1-D by scanning 
photodetectors $D_1$ and/or $D_2$ along
the axes $x_1$ and $x_2$, the second-order transverse spatial correlation 
function $G^{(2)}(x_1, x_2)$ was found to be
\begin{eqnarray}\label{HBT-1}
G^{(2)}(x_1, x_2) \sim I_{0}^{2} \,
\Big\{1+sinc^{2}\big{[}\,\frac{\pi\Delta\theta(x_1- x_2)}{\lambda}\,\big{]}\Big\},
\end{eqnarray}
where $\Delta\theta$ is the angular size of the star, $\lambda$ the wavelength
of the radiation.
The far-field HBT correlation of Eq.~(\ref{HBT-1}) has been interpreted 
as the result of classical statistical correlation of the intensity fluctuations
\begin{equation*}\label{Intensity-1}
 \langle \, I_{1} I_{2} \, \rangle 
=  \langle (\bar{I}_{1} + \Delta I_{1}) (\bar{I}_{2} + \Delta I_{2}) \rangle
= \bar{I}_{1} \bar{I}_{2} + \langle \, \Delta I_{1} \Delta I_{2} \, \rangle,
\end{equation*}
where $\bar{I}_1$ and $\bar{I}_2$ are the mean intensities of the radiation 
measured by photodetectors $D_1$ and $D_2$, respectively.
The second term in Eq.~(\ref{HBT-1}), 
$I_{0}^{2} \, sinc^{2}[\pi\Delta\theta(x_1-x_2)/\lambda]$, 
is phenomenologically interpreted as the intensity fluctuation correlation
$\langle \Delta I_{1} \Delta I_{2} \rangle$ in classical theory.  For visible 
wavelengths and large values of $\Delta\theta$ this function quickly drops 
from its maximum to minimum when $x_1-x_2$ moves from zero to a value 
such that $\Delta\theta(x_1-x_2)/\lambda=1$. In this situation we effectively 
have a ``point-to-point" relationship between the $x_1$ and $x_2$ axes:  
for each point on the $x_1$ there exists only one point on the $x_2$ that 
may have a nonzero intensity fluctuation correlation. 

The well-accepted interpretation of the HBT phenomenon is the 
following: in HBT the measurement is taken in the far-field zone of the 
radiation source, which is equivalent to the Fourier transform plane.  
When $D_1$  ($D_2$) is scanned in the neighborhood of $x_1 = x_2$, 
the two detectors measure the same mode of the radiation
field.  The measured intensities have the same fluctuations and  
yield a maximum value of 
$\langle\Delta I_{1}\Delta I_{2}\rangle$.   The two upper curves of $I(t)$ in 
Fig.~\ref{fig:HBT-Physics-3} schematically illustrate this situation. 
When the two photodetectors move apart from $x_1 = x_2$, $D_1$ and $D_2$ 
measure different modes of the radiation field.  In this case, the measured 
two modes may have completely different fluctuations.
The measurement yields $\langle\Delta I_{1}\Delta I_{2}\rangle=0$ and gives 
$\langle I_{1}I_{2}\rangle=\bar{I}_{1}\bar{I}_{2}$.  This situation is 
illustrated in the two lower curves of $I(t)$ in Fig.~\ref{fig:HBT-Physics-3}.
Unfortunately, this hand-waving interpretation does not reflect the correct physics 
in the case of $\Delta \theta \neq 0$.  For a finite angular sized source, 
there is no chance, at least realistically, for $D_1$ and $D_2$ to 
receive radiation from a single radiation mode only.
Nevertheless, the above theory has convinced us to believe that the 
observation of the intensity fluctuation correlation 
only takes place in the far-field zone of the thermal source.  

\begin{figure}[hbt]
    \centering
    \includegraphics[width=85mm]{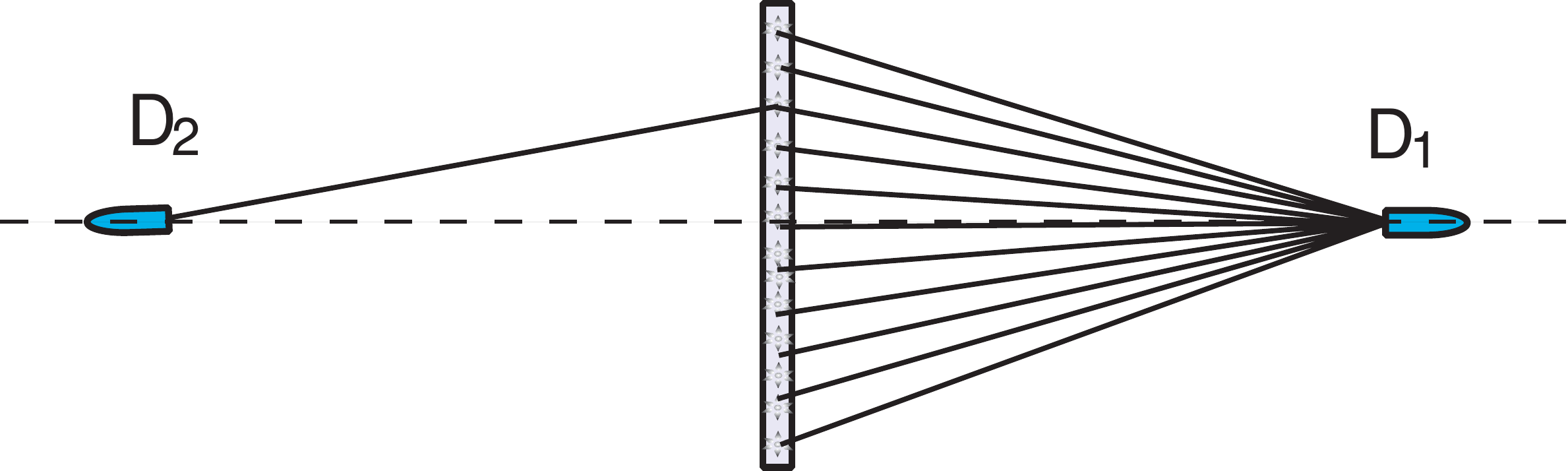} 
    \parbox{14.25cm}{\caption{Modified near-field HBT measurement -  
    an unfolded Klyshko picture of the setup.  Assuming a large sized disk-like 
    near-field chaotic source, 
    each point on the disk can be considered as an independent sub-source. 
    It is easy to see that (1) $D_1$ and $D_2$ are capable of receiving radiation 
    from a large number of sub-sources; and (2) $D_1$ and $D_2$ have more chances 
    to be triggered jointly by radiations from different sub-sources;
    (3) The ratio between the joint-detections triggered by radiation from a single 
    sub-source and from different sub-sources is roughly $N/N^{2} = 1/N$ in any 
    transverse position of $D_1$ and $D_2$.  }\label{fig:Near-Field}}
\end{figure}
What will happen if we move the photodetectors $D_1$ and $D_2$
to the ``near-field" as shown in the unfolded schematic of Fig.~\ref{fig:Near-Field}? 
Does this hand-waving argument still predict the point-to-point 
correlation in this situation?   We consider a disk-like thermal source with a large
number of independent and randomly radiating point sub-sources and assume the 
radiations coming from the same sub-source have the same intensity fluctuation,
and the radiations coming from different sub-sources have different intensity 
fluctuations.   It is easy to see that in the near-field, 
(1) each photodetector, $D_1$ and $D_2$, is capable of receiving 
radiations from a large number of sub-sources; and (2)  
$D_1$ and $D_2$, have more chances to be triggered jointly by radiation 
from different sub-sources;  (3) The ratio between the joint-detections triggered by
radiation from a single sub-source and from different sub-sources is roughly 
$N/N^{2} = 1/N$ in any transverse position of $D_1$ and $D_2$.  For a large 
value of $N$ the contribution of joint-detections triggered by radiation 
from a single sub-source in any transverse position of $D_1$ and $D_2$ has the  
same negligible value
$\langle\Delta I_{1}\Delta I_{2}\rangle/\bar{I}_1 \bar{I}_2 \sim 0$.  Following the
above philosophy, the near-field $G^{(2)}(\vec{\rho}_1, \vec{\rho}_2)$ should be a 
constant for any chosen transverse coordinates $\vec{\rho}_1$ and $\vec{\rho}_2$.
The experimental observations, however, have shown a different story.

\begin{figure}[htb]
    \centering
    \includegraphics[width=80mm]{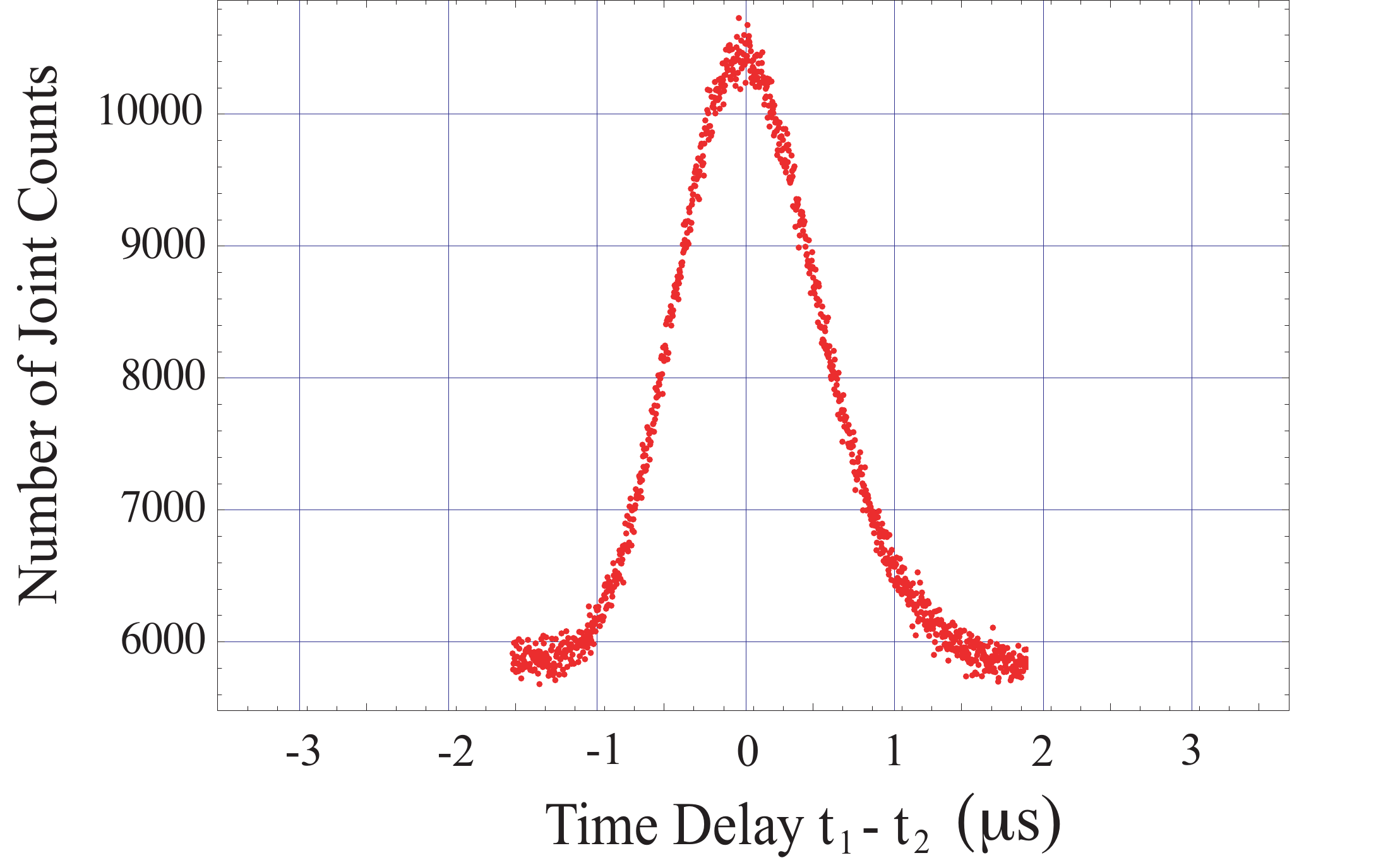} 
    \parbox{14.25cm}{\caption{$G^{(2)}(t_1 - t_2)$ of a chaotic source.  
    The temporal correlation width is measured $\sim 0.5 \mu$s, which means that 
    unless $t_1 - t_2 > 0.5 \mu$s, the value of $G^{(2)}(t_1 - t_2)$ will stay at the 
    neighborhood of its maximum.} \label{fig:Width}}
\end{figure}

The nontrivial near-field point-to-point correlation was experimentally observed in a 
modified HBT experiment by Scarcelli \emph{et al}. in 2005 before the near-field 
lensless ghost imaging demonstration. 
The modified HBT has a similar experimental setup as that of the historical 
HBT of Fig.~\ref{fig:HBT-Fig1-2}, except replacing the distant star
with a near-field disk-like chaotic source.  This light source has a considerably 
large angular diameter from the view of the photodetectors $D_1$ and $D_2$. 
The point photodetectors $D_1$ and $D_2$ are scannable along the axes of $x_1$ and 
$x_2$, respectively.   The frequency bandwidth $\Delta \omega$ of this thermal source 
is chosen to be narrow enough to achieve $\sim$$\mu$s correlation width of  
$G^{(2)}(t_1 - t_2)$ which is shown in Fig.~\ref{fig:Width}.  
\vspace{3mm}
This means to change $G^{(2)}$ from its maximum 
(minimum) value to minimum (maximum) value requires a few hundred meters 
optical delay in the arm of either $D_1$ or $D_2$.  
The transverse intensity distributions were examined before the measurement of 
transverse correlation.  The counting rate (weak light condition) or the 
output photocurrent (bright light condition) of each individual photodetector 
was found to be constant, i.e., $I(\vec{\rho}_1) \sim$~constant 
and $I(\vec{\rho}_2) \sim$~constant by scanning $D_1$ and $D_2$ in the transverse
planes of $z_1 = z_0$ and $z_2 = z_0$, respectively.  There is no surprise to
have constant $I(\vec{\rho}_1)$ and $I(\vec{\rho}_2)$.  The physics has been  
clearly illustrated in Fig.~\ref{fig:Near-Field-2}.   By using this kind of chaotic 
source, Scarcelli \emph{et al}. measured the 1-D near-field normalized transverse 
spatial correlation of $g^{(2)}(x_1 - x_2)$ by scanning $D_1$ in the
neighborhood of $x_1 = x_2$.  The measurements confirmed the
point-to-``spot" correlation of 
\begin{equation}
g^{(2)}(x_1 - x_2) \sim 1+ 
sinc^{2}\big{[}\,\frac{\pi\Delta\theta(x_1 - x_2)}{\lambda}\,\big{]},
\end{equation}
where, again, $\Delta\theta$ is the angular diameter of the near-field disk-like 
chaotic source.   It is worth emphasizing that $g^{(2)}(x_1 - x_2)$
dependents on $x_1 - x_2$ only.  Taking $x_1 - x_2 =$~constant,
$g^{(2)}(x_1 - x_2)$ is invariant under the displacements of transverse coordinates.   

\begin{figure}[hbt]
    \centering
    \includegraphics[width=80mm]{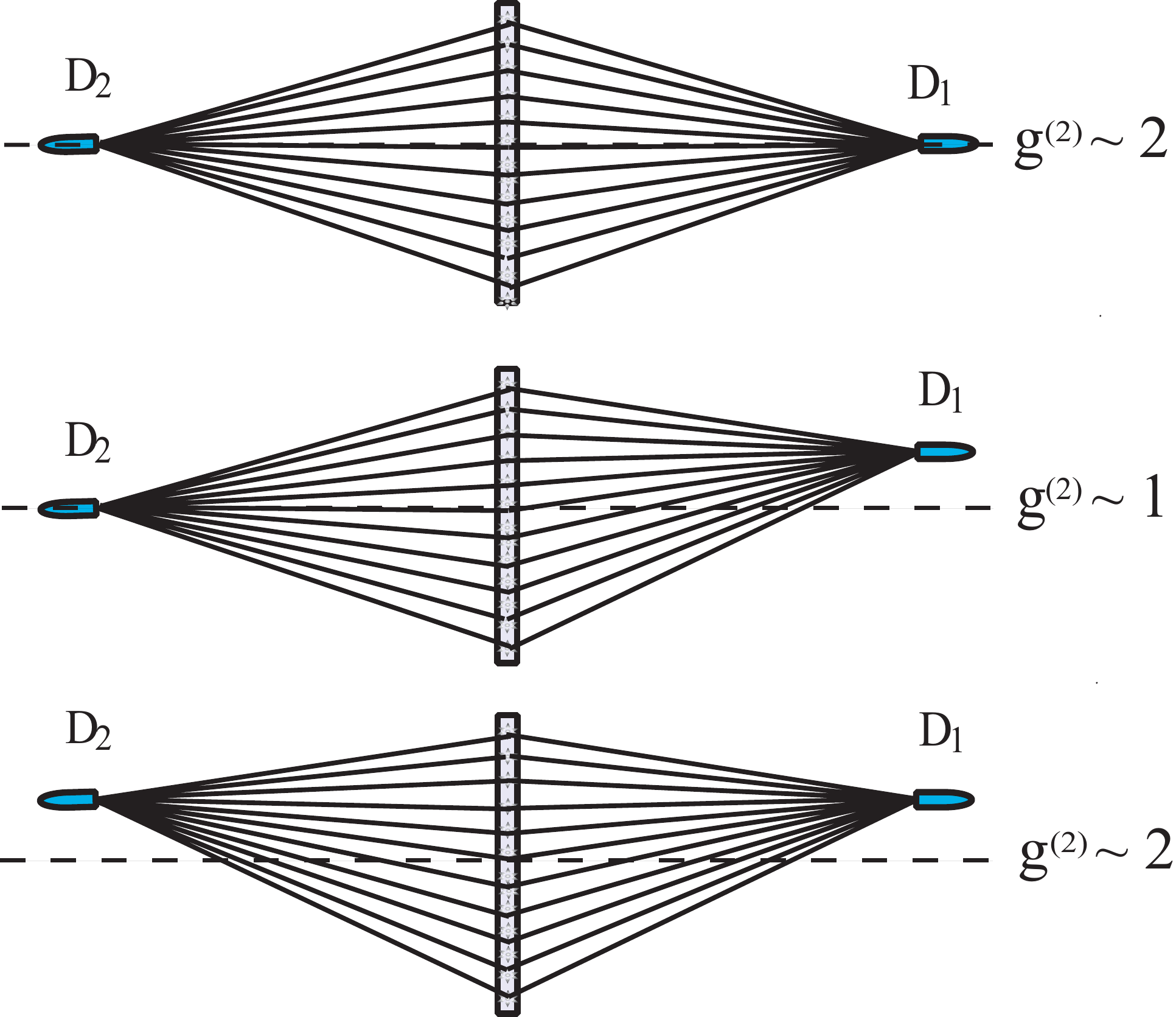} 
    \parbox{14.25cm}{\caption{Schematic of the near-field spatial correlation measurement 
    of Scarcelli \emph{et al}.  
    Upper: $g^{(2)} \sim 2$, where $D_1$ and $D_2$ are placed at equal distances 
    from the source and aligned symmetrically on the optical axis. In the spirit of the 
    traditional interpretation of HBT the intensities measured by $D_1$ and $D_2$
    must have same fluctuations as shown in the figure.  Middle: $g^{(2)} \sim 1$, 
    where $D_1$ is moved up a few millimeters to an asymmetrical position. In the spirit 
    of the traditional interpretation of HBT the intensities measured by $D_1$ and $D_2$
    must have different fluctuations.  Lower: $g^{(2)}\sim 2$, where $D_2$ is moved up 
    to a symmetrical position with respect to $D_1$, again. In the spirit of the 
    traditional interpretation of HBT the intensities measured by $D_1$ and $D_2$
    must have same fluctuations again. 
    What is the physical cause of the changes of the intensity fluctuations then?  
    Remember the $G^{(2)}(t_1 - t_2)$ function has a width of $ \sim 0.5 \mu$, see 
    Fig.~\ref{fig:Width}.}  \label{fig:Interference}}
\end{figure}

A simplified summary of the experimental observation is shown in Fig.~\ref{fig:Interference}: 
(1) In the upper figure, $D_1$ and $D_2$ are placed at equal distances from the source 
and aligned symmetrically on the optical axis.  The normalized joint-detection, or the 
value of $g^{(2)}$ achieved its maximum of $\sim 2$.  
(2) In the middle figure, $D_1$ is moved up a few millimeters to a non-symmetrical position, 
the normalized joint-detection, or the value of $g^{(2)}$ is measured to be $\sim 1$. 
(3) In the lower figure, $D_2$ is moved a few millimeters up to a symmetrical position with 
respect to $D_1$.  The normalized joint-detection, or the value of $g^{(2)}$ turned back 
to its maximum of $\sim 2$ again.

It is easy to see that the classical theory of statistical correlation of intensity 
fluctuations is facing difficulties in explaining the experimental results.  
In near-field $D_1$ and $D_2$ receive the same large number of modes 
at any $\vec{\rho}_1$ and $\vec{\rho}_2$.
In the spirit of the traditional interpretation of HBT,
there seems no reason to have a different intensity fluctuation correlation 
between $\vec{\rho}_1 = \vec{\rho}_2$ and $\vec{\rho}_1 \neq \vec{\rho}_2$
for the $G^{(2)}$ function shown in Fig.~\ref{fig:Width}.  In the upper measurement, 
we have obtained the maximum value of $g^{(2)} \sim 2$ at $z_1 = z_2$
and $x_1 = x_2$, which indicates the achievement of a maximum 
intensity fluctuation correlation as shown in Fig.~\ref{fig:Width} with $|t_1 - t_2| \sim 0$.  
In the middle measurement, $g^{(2)} \sim 1$ indicates a minimum intensity 
fluctuation correlation by moving $D_1$ a few millimeters up, which means the 
intensities measured by $D_1$ and $D_2$ must have different fluctuations.
In the lower measurement $D_2$ is moved up a few millimeters
to a new symmetrical position with respect to $D_1$, the measurements obtain 
$g^{(2)}\sim 2$ again.   The 
intensities measured by $D_1$ and $D_2$ must have same fluctuations again. 
What is the physical cause of the changes of the intensity fluctuations then?  
Remember the $G^{(2)}(t_1 - t_2)$ function has a width of $ \sim 0.5 \mu$.

For half a century since 1956, it has been believed that the HBT correlation 
is observable in the far-field only.  It was quite a surprise 
that in 2005 Scarcelli \emph{et al}. successfully demonstrated 
a near-field point-to-point transverse correlation of chaotic light, 
indicating that the nontrivial HBT spatial correlation is 
observable in the near-field and is useful for reproducing ghost images
in a nonlocal manner.\footnote{We cannot help but stop to ask: What has been 
preventing this simple move from far-field to near-field for half a century?}
The experiment of Scarcelli \emph{et al}. raised a 
question: ``Can two-photon correlation of chaotic 
light be considered as correlation of intensity fluctuations?" \cite{prl2}  
At least, this experiment suggested we reexamine the relationship between 
the quantum mechanical concept of joint-detection probability with 
the classical concept of intensity fluctuation correlation.  It seems that jointly observing
a pair of photons at space-time point $(\mathbf{r}_1, t_1)$ and $(\mathbf{r}_2, t_2)$ 
is perhaps only phenomenologically connected but not physically caused by 
the classical statistical correlation of intensity fluctuations.  The point-to-point
image-forming correlation is more likely the result of an interference.  In the 
view of two-photon interference,
far-field is not a necessary condition for observing the partial point-to-point 
correlation of thermal light.  Furthermore, 
it is quite common in two-photon interference type experiments to observe constant 
counting rates or intensities in individual photodetectors $D_1$ and $D_2$, respectively,
and simultaneously observe nontrivial space-time correlation in the joint-detection 
between $D_1$ and $D_2$.   These observations are consistent with the quantum
theory of two-photon interferometry \cite{IEEE-03}.

\subsection{Quantum theory of thermal light ghost imaging}

\hspace{6.5mm}According to the quantum theory of light, the observed partial 
point-to-point image-forming correlation is the result of multi-photon interference.   
In Glauber's theory of photo-detection \cite{Glauber}, an
idealized point photodetector measures the probability of observing
a photo-detection event at space-time point $(\mathbf{r}, t)$
\begin{equation}\label{G1-000}
G^{(1)}(\mathbf{r},t) = tr{\big{\{}\hat{\rho} \, E^{(-)}(\mathbf{r},t)E^{(+)}(\mathbf{r},t)\big{\}}},
\end{equation}
where $\hat{\rho}$ is the density
operator which characterizes the state of the quantized
electromagnetic field, $E^{(-)}(\mathbf{r},t)$ and
$E^{(+)}(\mathbf{r},t)$ the negative and positive field operators at
space-time coordinate $(\mathbf{r},t)$. The counting rate of a point
photon counting detector, or the output current of a point analog
photodetector, is proportional to $G^{(1)}(\mathbf{r},t)$. A
joint-detection of two independent point photodetectors measures the
probability of observing a joint-detection event of two photons at
space-time points $(\mathbf{r}_1, t_1)$ and $(\mathbf{r}_2, t_2)$
\begin{equation}\label{G2-000}
G^{(2)}(\mathbf{r}_1,t_1; \mathbf{r}_2,t_2)
= tr{\big{\{}\hat{\rho} \, E^{(-)}(\mathbf{r}_1,t_1)E^{(-)}(\mathbf{r}_2,t_2)
E^{(+)}(\mathbf{r}_2,t_2)E^{(+)}(\mathbf{r}_1,t_1)\big{\}}},
\end{equation}
where $(\mathbf{r}_j,t_j)$, $j = 1,2$, is
the space-time coordinate of the $j$th photo-detection event.  The coincidence 
counting rate of two photon counting detectors, or the output reading of a 
linear multiplier (RF mixer) between two photodetectors, is proportional to 
$G^{(2)}(\mathbf{r}_1,t_1; \mathbf{r}_2,t_2)$.
To calculate the partial point-to-point correlation between the object plane 
and the image plane, we need (1) to estimate the state, 
or the density matrix, of the thermal radiation; and (2) to propagate the field 
operators from the radiation source to the object and the image planes.  
We will first calculate the state of thermal radiation at the single-photon level for 
photon counting measurements to explore the physics behind ghost imaging 
as two-photon interference and then generalize the result to any intensity of 
thermal radiation.  

We assume a large transverse sized chaotic source consisting of 
a large number of independent and randomly radiating point sub-sources.  
Each point sub-source may also consist of a large number of
independent atoms that are ready for two-level atomic transitions in a 
random manner. 
Most of the time, the atoms are in their ground state. There is, however, a 
small chance for each atom to be excited to a higher energy level $E_{2}$ 
($\Delta E_{2} \neq 0$) and later return
back to its ground state $E_{1}$. It is reasonable to assume that each
atomic transition generates a field in the following single-photon state
\begin{eqnarray}
|\, \Psi\, \rangle  \simeq|\, 0 \, \rangle+ \epsilon\,
\sum_{\mathbf{k},s} \, f(\mathbf{k},s) \, \hat{a}^{\dag}_{\mathbf{k},s}\,|\, 0
\, \rangle,
\end{eqnarray}
where $|\epsilon| \ll1$ is the probability amplitude for the atomic
transition, $f(\mathbf{k},s) = \langle\,
\Psi_{\mathbf{k},s} \, |\, \Psi\, \rangle$ is the probability amplitude for
the radiation field to be in the single-photon state of wave number
${\mathbf{k}}$ and polarization $s$: $|\, \Psi_{\mathbf{k},s} \, \rangle=
|\,1_{\mathbf{k},s} \, \rangle= \hat{a}^{\dag}_{\mathbf{k},s}\,|\, 0 \,
\rangle$.
For this simplified two-level system, the density matrix that characterizes
the state of the radiation field excited by a large number of possible atomic
transitions is thus
\begin{align}\label{B-2}
\hat{\rho}  &  =\prod_{t_{0j}}\,\Big\{|\,0\,\rangle+\epsilon\sum
_{\mathbf{k},s}\,f(\mathbf{k},s)\,e^{-i\omega t_{0j}}\,\hat{a}_{\mathbf{k}%
,s}^{\dag}\,|\,0\,\rangle\,\Big\}\\ \nonumber
&  \times\,\prod_{t_{0k}}\,\Big\{\langle\,0\,|\,+\epsilon^{\ast}%
\sum_{\mathbf{k}^{\prime},s^{\prime}}\,f(\mathbf{k}^{\prime},s^{\prime
})\,e^{i\omega^{\prime}t_{0k}}\,\langle\,0\,|\hat{a}_{\mathbf{k}^{\prime
},s^{\prime}}\,\Big\}\\ \nonumber
&  \simeq\Big\{|\,0\,\rangle+\epsilon[\sum_{t_{oj}}\sum_{\mathbf{k}%
,s}\,f(\mathbf{k},s)\,e^{-i\omega t_{0j}}\,\hat{a}_{\mathbf{k},s}^{\dag
}\,|\,0\,\rangle] + \epsilon^2[...] \Big\} \\ \nonumber
&  \times\,\,\Big\{\langle\,0\,|\,+\epsilon^{\ast}[\sum_{t_{ok}}\sum_{\mathbf{k}^{\prime
},s^{\prime}}\,f(\mathbf{k}^{\prime},s^{\prime})\,e^{i\omega^{\prime}t_{0k}%
}\,\langle\,0\,|\hat{a}_{\mathbf{k}^{\prime},s^{\prime}}] + \epsilon^{\ast 2}[...] \Big\},
\end{align}
where $e^{-i\omega t_{0j}}$ is a random phase factor associated with the 
jth atomic transition. Since $\left\vert\epsilon\right\vert \ll1$, it is a good approximation 
to keep the necessary lower-order terms of $\epsilon$ in Eq.~(\ref{B-2}).  
After summing over $t_{0j}$ ($t_{0k}$) by taking into account all its
possible values we obtain
\begin{align}\label{Density-2}
\hat{\rho} &\simeq|\,0\,\rangle\langle\,0\,|+ \left\vert \epsilon\right\vert
^{2}\sum_{\mathbf{k,s}}
\left\vert f(\mathbf{k,s})\right\vert ^{2}
|\,1_{\mathbf{k,s}}\,\rangle
\langle\,1_{\mathbf{k,s}}\,|  \nonumber \\
& + \left\vert \epsilon\right\vert
^{4}\sum_{\mathbf{k,s}}\sum_{\mathbf{k}^{\prime}, s^{\prime}}
\left\vert f(\mathbf{k,s})\right\vert ^{2} \left\vert f(\mathbf{k}^{\prime},s^{\prime})\right\vert ^{2}
|\,1_{\mathbf{k,s}}1_{\mathbf{k}^{\prime},s^{\prime}}\,\rangle
\langle\,1_{\mathbf{k},s}1_{\mathbf{k}^{\prime},s^{\prime}}\,| 
+ ... 
\end{align}

Similar to our earlier discussion we will focus our calculation on the transverse 
correlation by assuming a narrow enough frequency 
bandwidth in Eq.~(\ref{Density-2}).  In the experiments of Scarcelli \emph{et al}.
the coherence time of the radiation was chosen $\sim$$\mu$s, the maximum 
achievable optical path differences $\sim$ps by the scanning of $D_1$ and $D_2$,
and the response time of the photodetectors is much less than the coherence time. 
The transverse spatial correlation measurement is under the condition of 
achieving a maximum temporal coherence of $G^{(2)}(t_1 - t_2) \sim 2$ 
during the  scanning of $D_1$ and $D_2$ at any $\vec{\rho}_1$ and 
$\vec{\rho}_2$.  In the \emph{{photon counting}} regime, under the above condition, 
it is reasonable to model the thermal light in the following mixed state 
\begin{eqnarray}\label{densitythermal}
\hat{\rho} \simeq |\,0\,\rangle\langle\,0\,| + 
\left\vert \epsilon\right\vert ^{2}
\sum_{\vec{\kappa}} 
\hat{a}^{\dagger}(\vec{\kappa})  | \,0 \, \rangle
\langle \, 0 \, |  \hat{a}(\vec{\kappa})
+\left\vert \epsilon\right\vert ^{4}
\sum_{\vec{\kappa}}\sum_{\vec{\kappa}'} 
\hat{a}^{\dagger}(\vec{\kappa}) \hat{a}^{\dagger}(\vec{\kappa}') \, | \,0 \, \rangle
\langle \, 0 \, | \, \hat{a}(\vec{\kappa}') \hat{a}(\vec{\kappa}).
\end{eqnarray}
Basically we are modeling the light source as an incoherent statistical mixture 
of single-photon states and two-photon states with equal probability of
having any transverse momentum.  The spatial part of the 
second-order coherence function is thus calculated as 
\begin{align}\label{G2-11}
G^{(2)}(\vec{\rho}_{1}, z_1;\vec{\rho}_{2}, z_2) &=
tr{[\, \hat{\rho} \, E^{(-)}(\vec{\rho}_1, z_1)E^{(-)}(\vec{\rho}_2, z_2)
E^{(+)}(\vec{\rho}_2, z_2)E^{(+)}(\vec{\rho}_1, z_1) \,]} \nonumber \\
&= \sum_{\vec{\kappa},\vec{\kappa}'} \langle 1_{\vec{\kappa}}1_{\vec{\kappa}'} |
E^{(-)}(\vec{\rho}_{1}, z_1)E^{(-)}(\vec{\rho}_{2}, z_2)
E^{(+)}(\vec{\rho}_{2}, z_2)E^{(+)}(\vec{\rho}_{1}, z_1)
|1_{\vec{\kappa}}1_{\vec{\kappa}'}\rangle  \nonumber \\
&\equiv \sum_{\vec{\kappa},\vec{\kappa}'} 
\big{|} \Psi_{\vec{\kappa},\vec{\kappa}'}(\vec{\rho}_{1}, z_1;\vec{\rho}_{2}, z_2) \big{|}^2,
\end{align}
where we have defined an effective two-photon wavefunction in transverse spatial
coordinates
\begin{equation}\label{G2-2-222}
\Psi_{\vec{\kappa},\vec{\kappa}'}(\vec{\rho}_{1}, z_1;\vec{\rho}_{2}, z_2)
= \langle 0 | E^{(+)}(\vec{\rho}_{2}, z_2)E^{(+)}(\vec{\rho}_{1}, z_1)
|1_{\vec{\kappa}}1_{\vec{\kappa}'}\rangle.
\end{equation}
The transverse part of the electric field operator can be written as 
\begin{eqnarray}\label{E12}
E^{(+)}(\vec{\rho}_{j}, z_j)\propto\sum_{\vec{\kappa}} \,
g_{j}(\vec{\rho}_{j}, z_j;\vec{\kappa}) \, \hat{a}(\vec{\kappa}),
\end{eqnarray}
again, $g_{j}(\vec{\rho}_{j}, z_j;\vec{\kappa})$ is the Green's function.
Substituting the field operators into Eq.~(\ref{G2-2-222}) we have
\begin{align}\label{G2-2-22220}
\Psi_{\vec{\kappa},\vec{\kappa}'}(\vec{\rho}_{1}, z_1;\vec{\rho}_{2}, z_2)
= \frac{1}{\sqrt{2}} \big{[} 
g_{2}(\vec{\rho}_{2},z_2;\vec{\kappa})g_{1}(\vec{\rho}_{1},z_1;\vec{\kappa}')+
g_{2}(\vec{\rho}_{2},z_2;\vec{\kappa}')g_{1}(\vec{\rho}_{1},z_1;\vec{\kappa}) \big{]}
\end{align}
and
\begin{eqnarray}\label{G2-2-2222}
G^{(2)}(\vec{\rho}_{1}, z_1;\vec{\rho}_{2}, z_2) 
=\sum_{\vec{\kappa},\vec{\kappa}'}
\Big{|}  \frac{1}{\sqrt{2}} \big{[} 
g_{2}(\vec{\rho}_{2},z_2;\vec{\kappa})g_{1}(\vec{\rho}_{1},z_1;\vec{\kappa}')+
g_{2}(\vec{\rho}_{2},z_2;\vec{\kappa}')g_{1}(\vec{\rho}_{1},z_1;\vec{\kappa}) 
\big{]} \Big{|}^{2},
\end{eqnarray}
\begin{figure}[htb]
    \centering
    \includegraphics[width=92mm]{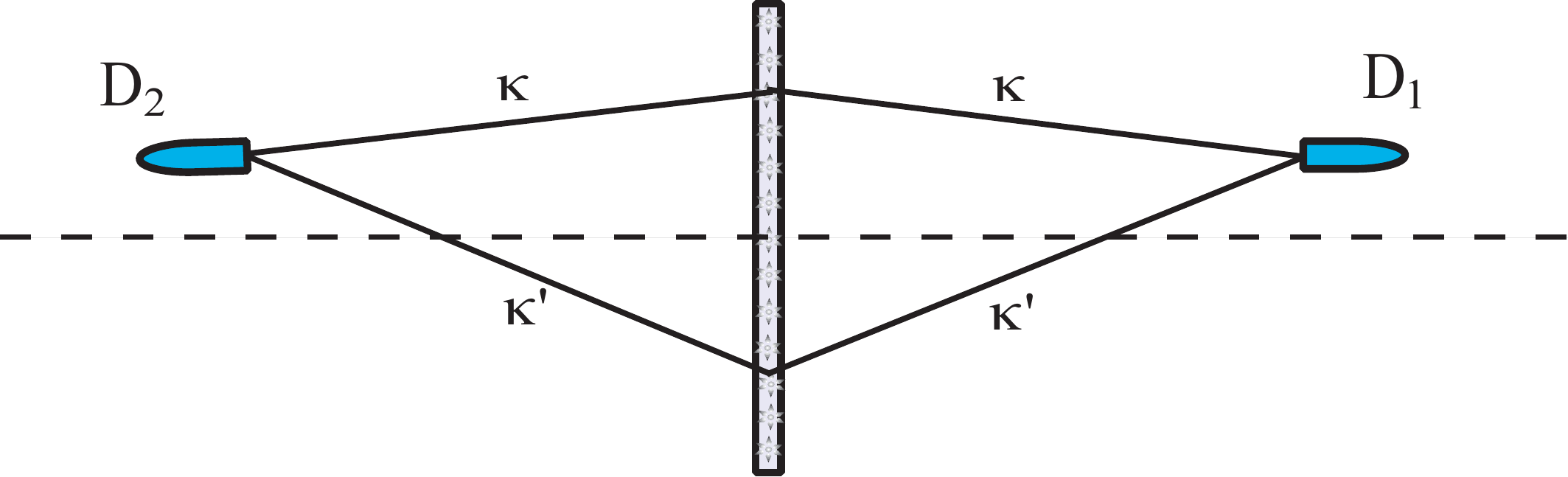} 
    \parbox{14.25cm}{\caption{Schematic illustration of  two-photon interference: 
    a superposition between two-photon amplitudes 
    $g_{2}(\vec{\rho}_{2},z_2\vec{\kappa})g_{1}(\vec{\rho}_{1},z_1;\vec{\kappa}')$ and 
    $g_{2}(\vec{\rho}_{2},z_2;\vec{\kappa}')g_{1}(\vec{\rho}_{1},z_1;\vec{\kappa})$.  
    It is clear that the amplitudes 
    $g_{2}(\vec{\rho}_{2},z_2;\vec{\kappa})g_{1}(\vec{\rho}_{1},z_1;\vec{\kappa}')$ and 
    $g_{2}(\vec{\rho}_{2},z_2;\vec{\kappa}')g_{1}(\vec{\rho}_{1},z_1;\vec{\kappa})$ 
    will experience equal optical path propagation and superpose constructively when 
    $D_1$ and $D_2$ are located at $\vec{\rho}_1 \simeq \vec{\rho}_2$ and $z_1 \simeq z_2$.
    This nonlocal superposition has no classical correspondence.
    }\label{fig:Field-Field-2}}
\end{figure}
representing the key result for our understanding of the phenomenon.  
Eqs.~(\ref{G2-2-22220}) and (\ref{G2-2-2222}) indicates an interference between
two alternatives, different yet indistinguishable, which leads to a joint
photo-detection event. This interference phenomenon is not, as in
classical optics, due to the superposition of electromagnetic
fields at a local point of space-time. This interference is the result of the superposition
between $g_{2}(\vec{\rho}_{2},z_2;\vec{\kappa})g_{1}(\vec{\rho}_{1},z_1;\vec{\kappa}')$ and
$g_{2}(\vec{\rho}_{2},z_2;\vec{\kappa}')g_{1}(\vec{\rho}_{1},z_1;\vec{\kappa})$, the
so-called two-photon amplitudes, non-classical entities that
involve both arms of the optical setup as well as two distant photo-detection
events at ($\vec{\rho}_{1},z_1$) and ($\vec{\rho}_{2},z_2$), respectively.  
Examining the effective wavefunction of Eq.~(\ref{G2-2-22220}), we find this 
symmitrized effective wavefunction plays the same role as that of the 
symmitrized wavefunction of identical particles in quantum mechanics. 
This peculiar nonlocal superposition has no classical correspondence, and makes 
the type-two ghost image turbulence-free,  i.e., any phase 
disturbance in the optical path has no influence on the ghost image \cite{ARL-2}.
Fig.~\ref{fig:Field-Field-2} schematically illustrates the two alternatives 
for a pair of mode $\vec{\kappa}$ and $\vec{\kappa}'$ to produce a joint 
photo-detection event: $\vec{\kappa}1 \times \vec{\kappa}'2$ and 
$\vec{\kappa}2 \times \vec{\kappa}'1$.  
The superposition of each pair of these amplitudes produces
an individual sub-interference-pattern in the joint-detection space of 
$(\vec{\rho}_{1},z_1,t_1; \vec{\rho}_{2},z_2,t_2)$.  A large number of these 
sub-interference-patterns simply add together resulting in a nontrivial 
$G^{(2)}(\vec{\rho}_{1}, z_1;\vec{\rho}_{2}, z_2)$ function.  
It is easy to see that each pair of 
the two-photon amplitudes, illustrated in Fig.~\ref{fig:Field-Field-2}, will 
superpose constructively whenever $D_1$ and $D_2$ are placed in the positions  
satisfying $\vec{\rho}_1 \simeq \vec{\rho}_2$ and $z_1 \simeq z_2$; and
consequently, $G^{(2)}(\vec{\rho}_{1}, z_1;\vec{\rho}_{2}, z_2)$ achieves 
its maximum value as the result of the sum of these individual constructive 
interferences.  In other coordinates, however, the superposition of each individual 
pair of the two-photon amplitudes may yield different values between constructive 
maximum and destructive minimum due to unequal optical path 
propagation, resulting in an averaged sum. 

Before calculating $G^{(2)}(\vec{\rho}_{1}, z_1;\vec{\rho}_{2}, z_2)$ we examine
the single counting rate of the point photodetectors $D_1$ and $D_2$ which are
placed at ($\vec{\rho}_{1}, z_1)$ and ($\vec{\rho}_{2}, z_2$), respectively. 
With reference to the experimental setup of
Fig.~\ref{fig:HBT-Fig1}, the Green's function of free-propagation is derived in 
the Appendix
\begin{eqnarray*}
g_{1}(\vec{\rho}_{1},z_1;\vec{\kappa}) = \int d \vec{\rho}_s \, 
\Big\{\frac{-i \omega}{2 \pi c} \, 
\frac{e^{i \frac{\omega}{c} z_1} }{z_1}\, e^{i \frac{\omega}{2 c z_1} 
|\vec{\rho}_{1}-\vec{\rho}_{s}|^2} \Big\} \, 
e^{-i \vec{\kappa} \cdot \vec{\rho}_s}, \\
g_{2}(\vec{\rho}_{2},z_2;\vec{\kappa}) =  \int d \vec{\rho'_s} \, 
\Big\{\frac{-i \omega}{2 \pi c} \, 
\frac{e^{i \frac{\omega}{c} z_2} }{z_2}\, e^{i \frac{\omega}{2 c z_2} 
|\vec{\rho}_{2}-\vec{\rho'_s}|^2} \Big\} \, 
e^{-i \vec{\kappa} \cdot \vec{\rho'_s}}, 
\end{eqnarray*}
where $\vec{\rho}_s$ is the transverse vector in the source plane,
and the field has propagated from the source to the $\vec{\rho}_{1}$
plane and $\vec{\rho}_2$ plane in arms 1 and 2, respectively. 
The single detector counting rate or the output photocurrent
is proportional to $G^{(1)}(\mathbf{r},t)$ as shown in Eq.~(\ref{G1-000}),
\begin{align}\label{G1-111}
G^{(1)}(\vec{\rho}_{j},z_j) &= tr{\big{\{}\hat{\rho} \, 
E^{(-)}(\vec{\rho}_{j},z_j)E^{(+)}(\vec{\rho}_{j},z_j)\big{\}}} \nonumber \\
&\propto \sum_{\vec{\kappa}} \langle\,0 \,| \, \hat{a}(\vec{\kappa})\,
E^{(-)}(\vec{\rho}_{j},z_j)E^{(+)}(\vec{\rho}_{j},z_j) \, \hat{a}^{\dagger}(\vec{\kappa})
| \,0\, \rangle \nonumber \\
& \sim \text{constant},
\end{align}
where $j = 1,2$ indicating the $j$th photodetector.  

Although $G^{(1)}(\vec{\rho}_{1},z_1) $ and $G^{(1)}(\vec{\rho}_{2},z_2) $ 
are both constants, $G^{(2)}(\vec{\rho}_{1}, z_1;\vec{\rho}_{2}, z_2)$ turns
to be a nontrivial function of $(\vec{\rho}_{1}, z_1)$ and ($\vec{\rho}_{2}, z_2$), 
\begin{align}
G^{(2)}(\vec{\rho}_{1}, z_1;\vec{\rho}_{2}, z_2) 
&=\sum_{\vec{\kappa},\vec{\kappa}'}
\Big{|}  \frac{1}{\sqrt{2}} \big{[} 
g_{2}(\vec{\rho}_{2},z_2;\vec{\kappa})g_{1}(\vec{\rho}_{1},z_1;\vec{\kappa}')+
g_{2}(\vec{\rho}_{2},z_2;\vec{\kappa}')g_{1}(\vec{\rho}_{1},z_1;\vec{\kappa}) 
\big{]} \Big{|}^{2} \nonumber \\
&\equiv G^{(1)}(\vec{\rho}_{1}, z_1)G^{(1)}(\vec{\rho}_{2}, z_2)+
\big{|}G^{(1)}(\vec{\rho}_{1}, z_1;\vec{\rho}_{2}, z_2)\big{|}^{2},
\end{align} 
where 
\begin{align*}
\big{|} G^{(1)}(\vec{\rho}_{1}, z_1;\vec{\rho}_{2}, z_2) \big{|}^2
&=\big{|} \int d\vec{\kappa} \,\, 
g^{*}_{1}(\vec{\rho}_{1},z_1;\vec{\kappa}) \, 
g_{2}(\vec{\rho}_{2},z_2;\vec{\kappa}) \big{|}^2 \nonumber \\
&\propto \big{|} \int d \vec{\rho}_s \, e^{-i \frac{\omega}{c} z_1} \, 
e^{-i \frac{\omega}{2 c z_1} |\vec{\rho}_{1}-\vec{\rho}_{s}|^2} \, 
e^{i \frac{\omega}{c} z_2} \, 
e^{i \frac{\omega}{2 c z_2} |\vec{\rho}_{2}-\vec{\rho}_{s}|^2} \big{|}^2.
\end{align*}
If we choose the distances from the source to the two detectors to be equal
($z_{1}=z_{2} = d$), the above integral of $d \vec{\rho}_s$ 
yields a point-to-point correlation between the transverse planes 
$z_1 = d$ and $z_2 =d$, 
\begin{align}\label{G12-Integral-2}
\big{|}\, G^{(1)}_{12}(\vec{\rho}_{1};\vec{\rho}_{2}) \, \big{|}^2
&\propto \Big{|} \int d \vec{\rho}_s \, 
e^{i \frac{\omega}{c d} (\vec{\rho}_{1}-\vec{\rho}_{2}) \cdot \vec{\rho}_{s}} \, \Big{|}^2
= somb^2 \big{[} \frac{R}{d}\frac{\omega}{c} | \vec{\rho}_{1} - \vec{\rho}_{2}| \big{]}
\sim \delta (\vec{\rho}_{1} - \vec{\rho}_{2}), 
\end{align}
where the $\delta$-function is an approximation by assuming a large enough
thermal source of angular size $\Delta \theta \sim R/d$ and high enough 
frequency $\omega$, such as a visible light source.  The nontrivial $G^{(2)}$ 
function is therefore,
\begin{align}
G^{(2)}(\vec{\rho}_{1};\vec{\rho}_{2}) \sim 1 +  \delta (\vec{\rho}_{1} - \vec{\rho}_{2}).
\end{align}
In the ghost imaging experiment, the joint-detection counting rate is thus
\begin{eqnarray}\label{imagemir}
R_{12}  \propto \int
d\vec{\rho}_{2} \, |A(\vec{\rho}_{2})|^2 \,G^{(2)}(\vec{\rho}_{1}; \vec{\rho}_{2})
\sim R_0 +  |A(\vec{\rho}_{1})|^{2},
\end{eqnarray}
where $R_0$ is a constant and $A(\vec{\rho}_{2})$ is the aperture function of the 
object.

So far, we have successfully derived an analytical solution for ghost imaging with 
thermal radiation at the single-photon level.  We have shown that the partial point-to-point 
correlation of thermal radiation is the result of a constructive-destructive interference 
caused by the superposition of two two-photon amplitudes, corresponding to two 
alternative ways for a pair of jointly measured photons to produce a joint-detection 
event.  In fact the above analysis is not restricted to single-photon states.  The partial 
point-to-point correlation of $G^{(2)}(\vec{\rho}_{1};\vec{\rho}_{2})$ is generally true 
for any order of quantized thermal radiation \cite{Liu}.  Now we generalize the 
calculation to an arbitrary quantized thermal field with occupation number from 
$n_{\mathbf{k,s}} = 0$ to $n_{\mathbf{k,s}} \gg 1$ by keeping all higher order 
terms in Eq.~(\ref{B-2}).  After summing over $t_{0j}$ 
and $t_{0k}$ the density matrix can be written as 
\begin{equation}\label{ThermalState}
\hat{\rho}=\sum_{\{n\}} \, p_{\{n\}} \, |\{n\} \rangle \langle \{n\}|, 
\end{equation}
where $p_{\{n\}}$ is the probability for the thermal field in the state 
$$
|\{n\} \rangle \equiv \prod_{\mathbf{k}, s} |n_{\mathbf{k},s} \rangle
= |n_{\mathbf{k}, s} \rangle 
|n_{\mathbf{k}',s'} \rangle ... 
|n_{\mathbf{k}^{\prime \prime ... \prime},s^{\prime \prime ... \prime}} \rangle.
$$   
The summation of Eq.~(\ref{ThermalState}) includes all possible modes 
$\mathbf{k}$, polarizations $s$, occupation numbers $n_{\mathbf{k},s}$ 
for the mode $(\mathbf{k}, s)$ and all possible combinations of occupation
numbers for different modes in a set of $\{n\}$.
Substituting the field operators and the density operator of Eq.~(\ref{ThermalState}) 
into Eq.~(\ref{G1-000}) we obtain the constant $G^{(1)}(\vec{\rho}_j, z_j, t_j)$,
$j =1,2$, which corresponds to the intensities $I(\vec{\rho}_1, z_1, t_1)$ and
$I(\vec{\rho}_2, z_2, t_2)$,
\begin{align}\label{Intensity-constant}
& \ \ \ \ G^{(1)}(\vec{\rho}_j, z_j, t_j) \nonumber \\
&= \sum_{\{n\}}  \int d\vec{\kappa}\int d\vec{\kappa}' 
g_{j}^*(\vec{\rho}_j, z_1, t_j; \vec\kappa)
g_{j}(\vec{\rho}_j, z_1,t_j; \vec{\kappa}') 
\, p_{\{n\}} \, \langle \{n\}|
\, a(\vec{\kappa})a^{\dagger}(\vec{\kappa}') \, |\{n\}\rangle \nonumber \\
& \propto \sum_{\{n\}} n_{\vec{\kappa}} \, n_{\vec{\kappa}'} \int d\vec{\kappa} \
\big{|} \, g_{j}^*(\vec{\rho}_j, z_j, t_j; \vec{\kappa}) \, \big{|}^2 \nonumber \\
& \simeq \text{constant}. 
\end{align}
Although $G^{(1)}(\vec{\rho}_1, z_1, t_1)$ and $G^{(1)}(\vec{\rho}_2, z_2, t_2)$ are
both constants, substituting the field operators and the density operator of 
Eq.~(\ref{ThermalState}) into Eq.~(\ref{G2-000}), 
we obtain a nontrivial point-to-point correlation function of
$G^{(2)}(\vec{\rho _1}; \vec{\rho_2})$ at the two 
transverse planes $z_1 = d$ and $z_2 = d$,
\begin{align}\label{ThermalCorrelation}
& \ \ \ \ G^{(2)}(\vec{\rho _1}; \vec{\rho_2}) \nonumber \\
&= \sum_{\{n\}}  \int d\vec{\kappa}\int d\vec{\kappa}' 
\int d\vec{\kappa}'' \int d\vec{\kappa}''' g_{1}^*(\vec{\rho}_1, z_1; \vec\kappa)
g_{2}^*(\vec{\rho}_2, z_2; \vec{\kappa}')
g_{2}(\vec{\rho}_2, z_2; \vec{\kappa}'')
g_{1}(\vec{\rho}_1, z_1; \vec{\kappa}''') \nonumber \\ 
& \ \ \ \  \times \, p_{\{n\}} \, \langle \{n\}|
\, a(\vec{\kappa})a(\vec{\kappa}')a^{\dagger}(\vec{\kappa}'')
a^{\dagger}(\vec{\kappa}''') \, |\{n\}\rangle \nonumber \\
& \propto \sum_{\{n\}} n_{\vec{\kappa}} \, n_{\vec{\kappa}'} \int d\vec{\kappa}\int
d\vec{\kappa}' \int d\vec{\kappa}'' \int
d\vec{\kappa}''' g_{1}^*(\vec{\rho}_1, z_1; \vec{\kappa})
g_{2}^*(\vec{\rho}_2, z_2; \vec{\kappa}')g_{2}(\vec{\rho}_2, z_2; \vec{\kappa}'')
g_{1}(\vec{\rho}_1, z_1; \vec{\kappa}''') \nonumber \\
& \ \ \ \  \times \,  p_{\{n\}}
(\delta_{\vec{\kappa}\vec{\kappa}'''}\delta_{\vec{\kappa}'  \vec{\kappa}'' }
+\delta_{\vec{\kappa}\vec{\kappa}''}\delta_{\vec{\kappa}' \vec{\kappa}'''})
\nonumber \\
& = \sum_{n_{\vec{\kappa}} n_{\vec{\kappa}'}}  
p_{\{... n_{\vec{\kappa}} ... n_{\vec{\kappa}'} ...\}} \, 
n_{\vec{\kappa}} \, n_{\vec{\kappa}'} \nonumber \\
& \ \ \ \ \times \Big{\{} \int d\vec{\kappa} 
\int d\vec{\kappa}' \, \Big{|}  \frac{1}{\sqrt{2}} \, \big{[}
g_{1}(\vec{\rho}_1, z_1; \vec{\kappa})g_{2}(\vec{\rho}_2, z_2; \vec{\kappa}')
+ g_{2}(\vec{\rho}_2, z_2; \vec{\kappa})
g_{1}(\vec{\rho}_1, z_1; \vec{\kappa}') \big{]} \Big{|}^2 \Big{\}}
\nonumber \\
& \propto \Big\{1+somb^2\big{[} \frac{R}{d}\frac{\omega}{c} 
(\vec{\rho}_1 -\vec{\rho}_2)\big{]} \Big\}. 
\end{align}
It is clear that in Eq.~(\ref{ThermalCorrelation}),
the partial point-to-point correlation of thermal light
is the result of a constructive-destructive interference between 
two quantum-mechanical amplitudes.  We also note
from Eq.~(\ref{ThermalCorrelation}) that the partial point-to-point correlation 
is independent of the occupation numbers, $\{n\}$, and the probability 
distribution, $p_{\{n\}}$, of the quantized thermal radiation.  

It is interesting but not surprising to see that the effective two-photon wavefunction in 
bright light condition
$$
\Psi_{\vec{\kappa},\vec{\kappa}'}(\vec{\rho}_{1}, z_1;\vec{\rho}_{2}, z_2)
= \frac{1}{\sqrt{2}} \big{[} 
g_{2}(\vec{\rho}_{2},z_2;\vec{\kappa})g_{1}(\vec{\rho}_{1},z_1;\vec{\kappa}')+
g_{2}(\vec{\rho}_{2},z_2;\vec{\kappa}')g_{1}(\vec{\rho}_{1},z_1;\vec{\kappa}) \big{]}
$$
is the same as that of weak light at single-photon level.   In fact, the above effective
wavefunction does play the same role in specifying two different yet indistinguishable 
alternatives for the two annihilated photons contributing to a joint-detection event 
of $D_1$ and $D_2$, which implies that the partial point-to-point correlation is the 
result of two-photon interference in bright light condition.  This nonlocal partial 
correlation indicates that a $50\%$ contrast ghost image is observable at bright 
light condition provided registering no more than one coincidence event within the 
joint-detection time window.  This requirement can be easily 
achieved by using adjustable ND-filters with $D_1$ and $D_2$.

Quantum theory predicts and calculates the probability of observing a certain 
physical event.   The output photocurrent of an 
idealized point photodetector is proportional to the probability of observing a 
photo-detection event at space-time point ($\mathbf{r},t$).   The joint-detection 
between two idealized point photodetectors is proportional to
the probability of observing a joint photo-detection event at space-time points 
($\mathbf{r}_1,t_1$) and ($\mathbf{r}_2,t_2$).  In most of the experimental situations, 
there exists more than one possible alternative ways to produce a photo-detection
event, or a joint photo-detection event.  These probability amplitudes, 
which are defined as the single-photon amplitudes and the two-photon amplitudes, 
respectively, are superposed to contribute to the final measured probability, and 
consequently determine the probability of observing  a photo-detection
event or a joint photo-detection event.   In the view of quantum 
theory, whenever the state of the quantum system and the alternative ways to 
produce a photo-detection event or a joint photo-detection event are determined, 
the result of a measurement is determined.   We may consider this as a basic 
criterion of quantum measurement theory.

\subsection{A semiclassical model of nonlocal interference}
\label{ClassicalModel}
  
\hspace{6.5mm}The multi--photon interference nature of type-two ghost imaging can be 
seen intuitively from the superposition of paired-sub-fields of chaotic radiation.  
Let us consider a similar experimental setup as that of the modified HBT experiment of 
Scarcelli \emph{et al}..  We assume a large angular sized disk-like chaotic 
source that contains a large number of randomly radiating independent point ``sub-sources", 
such as trillions of independent atomic transitions randomly distributed spatially and 
temporally.   It should be emphasized that a large number of independent or incoherent 
sub-sources is the only requirement for type-two ghost imaging.   What we need is an 
ensemble of point-sub-sources with random relative phases so that the sub-fields coming 
from these sub-sources are able to take all possible values of relative phases in their 
superposition.  It is \emph{unnecessary} to require the radiation source to have either 
nature or artificial intensity fluctuations at all.   In this model,
each point sub-source contributes to the measurement an independent spherical wave 
as a sub-field of complex amplitude $E_j = a_j e^{i\varphi_j}$, where $a_j$ is the real 
and positive amplitude of the $j$th sub-field and $\varphi_j$ is a \emph{random} phase 
associated with the $j$th sub-field.  We have the following picture for the source: 
(1) a large number of independent point-sources distribute randomly on the transverse
plane of the source (counted spatially); (2) each point-source  
contains a large number of independently and randomly radiating atoms 
(counted temporally); (3) a large number of sub-sources, either counted spatially or 
temporally, may contribute to each of the independent radiation mode 
($\vec{\kappa}, \omega$) at $D_1$ and $D_2$ (counted by mode).   
The instantaneous intensity at space-time $(\mathbf{r}_j, t_j)$, measured by the $j$th 
idealized point photodetector $D_j$, $j =1,2$, is calculated as   
\begin{align}\label{Model-11}
I(\mathbf{r}_{j},t_{j}) &= E^{*}(\mathbf{r}_{j},t_{j}) E(\mathbf{r}_{j},t_{j}) 
=\sum_{l} E^{*}_{l}(\mathbf{r}_{j},t_{j}) \sum_{m} E_{m}(\mathbf{r}_{j},t_{j}) \nonumber \\
&= \sum_{l=m} E^{*}_{l}(\mathbf{r}_{j},t_{j}) E_{l}(\mathbf{r}_{j},t_{j}) 
+  \sum_{l\neq m} E^{*}_{l}(\mathbf{r}_{j},t_{j}) E_{m}(\mathbf{r}_{j},t_{j}),
\end{align}
where the sub-fields are identified by the index $l$ and $m$ originated from the 
$l$ and $m$ sub-sources.  
The first term is a constant representing the sum of the sub-intensities, where 
the $l$th sub-intensity is originated from the $l$th sub-source.  The second term 
adds the ``cross" terms corresponding to different sub-sources.   When 
\emph{taking into account all possible realizations of the fields}, it is easy to find that 
the only surviving terms in the sum are these terms in which the field and its conjugate 
come from the same sub-source, i.e., the first term in Eq.~(\ref{Model-11}).  
The second term in Eq.~(\ref{Model-11}) vanishes 
if $\varphi_l - \varphi_m$ \emph{takes all possible values}.   We may write  
Eq.~(\ref{Model-11}) into the following form
\begin{align}\label{Model-11-1}
I(\mathbf{r}_{j},t_{j}) = \langle I(\mathbf{r},t) \rangle + \Delta I(\mathbf{r},t),
\end{align}
where 
\begin{equation}\label{Model-12}
\langle I(\mathbf{r},t) \rangle \equiv \langle 
\sum_{l} E^{*}_{l}(\mathbf{r}_{j},t_{j}) \sum_{m} E_{m}(\mathbf{r}_{j},t_{j}) \, \rangle
=  \sum_{l} E^{*}_{l}(\mathbf{r}_{j},t_{j}) E_{l}(\mathbf{r}_{j},t_{j}).
\end{equation}
The notation $\langle ... \rangle$ denotes
the mathematical expectation, when \emph{taking into 
account all possible realizations of the fields}, i.e., taking into account all possible 
complex amplitudes for the large number of sub-fields in the superposition.  
In the probability theory, the expectation value of a measurement equals the 
mean value of an ensemble.  In a real measurement, the superposition 
may not take all possible realizations of the fields and consequently the measured 
instantaneous intensity $I(\mathbf{r}, t)$ may differ from its expectation value 
$\langle I(\mathbf{r}, t) \rangle$ from time to time.  The variation $\delta I(\mathbf{r}, t)$ 
turns to be a random function of time. The measured $I(\mathbf{r}, t)$ fluctuate 
randomly in the neighborhood of $\langle I(\mathbf{r}, t) \rangle$ non-deterministically.

In the classical limit, a large number of independent and randomly
radiated sub-sources contribute to the instantaneous intensity $I(\mathbf{r}_{j},t_{j})$.  
These large number of independent randomly distributed sub-fields may have 
\emph{taken all possible realizations of their complex amplitudes} in the superposition.  
In this case the sum of the cross terms vanishes, 
\begin{equation}
\Delta I(\mathbf{r},t) = \sum_{l\neq m} E^{*}_{l}(\mathbf{r}_{j},t_{j}) E_{m}(\mathbf{r}_{j},t_{j}) 
\simeq 0,
\end{equation} 
therefore,
$$
I(\mathbf{r}_{j},t_{j}) \simeq \sum_{l} E^{*}_{l}(\mathbf{r}_{j},t_{j}) E_{l}(\mathbf{r}_{j},t_{j})
= \langle I(\mathbf{r},t) \rangle.
$$ 

Now we calculate the second-order correlation function 
$G^{(2)}(\mathbf{r}_{1},t_{1}; \mathbf{r}_{2},t_{2})$, which is defined as
\begin{align}\label{G2-1}
G^{(2)}(\mathbf{r}_{1},t_{1};\mathbf{r}_{2},t_{2}) 
&\equiv \langle \sum_{j,k,l,m}
E^{*}_{j}(\mathbf{r}_{1},t_{1}) \,E_{k}(\mathbf{r}_{1},t_{1})
E^{*}_{l}(\mathbf{r}_{2},t_{2})\, E_{m}(\mathbf{r}_{2},t_{2}) \, \rangle,
\end{align}
where the notation $\langle \, ... \, \rangle$, again, 
denotes an expectation operation by \emph{taking into account all possible realizations 
of the fields}, i.e., averaging all possible complex amplitudes for the sub-fields in the 
superposition.  In the following calculation we only take into account the random phases 
of the sub-fields without considering the amplitude variations.  
 Due to the chaotic nature of the independent sub-sources, after taking into 
account all possible realizations of the phases associated with the sub-fields, 
the only surviving terms in the summation
are those with: (1) $j=k, l=m$, (2)  $ j=m, k=l$.  Therefore,
$G^{(2)}(\mathbf{r}_{1},t_{1};\mathbf{r}_{2},t_{2})$ reduces
to the sum of the following two groups:
\begin{align}\label{G2-2}
G^{(2)}(\mathbf{r}_{1},t_{1};\mathbf{r}_{2},t_{2}) 
&=  \langle \, \sum_{j} \,
E^{*}_{j}(\mathbf{r}_{1},t_{1}) \,E_{j}(\mathbf{r}_{1},t_{1}) \,
 \sum_{l}\,  E^{*}_{l}(\mathbf{r}_{2},t_{2})\, E_{l}(\mathbf{r}_{2},t_{2}) 
 \nonumber \\ & \hspace{5mm} +  \sum_{j}\, 
E^{*}_{j}(\mathbf{r}_{1},t_{1}) \,E_{j}(\mathbf{r}_{2},t_{2}) \,
 \sum_{ l} \, E^{*}_{l}(\mathbf{r}_{2},t_{2})\, 
E_{l}(\mathbf{r}_{1},t_{1}) \, \rangle \nonumber \\
& =\langle \, \sum_j \sum_l \, \Big{|} \frac{1}{\sqrt{2}} \big{[} 
E_{j}(\mathbf{r}_{1},t_{1}) E_{l}(\mathbf{r}_{2},t_{2})
+  E_{l}(\mathbf{r}_{1},t_{1}) E_{j}(\mathbf{r}_{2},t_{2}) \big{]} \Big{|}^2 \, \rangle.
\end{align}

It is not difficult to see the nonlocal nature of the superposition shown 
in Eq.~(\ref{G2-2}).  In Eq.~(\ref{G2-2}), 
$G^{(2)}(\mathbf{r}_{1},t_{1};\mathbf{r}_{2},t_{2})$ is
written as a superposition between the paired
sub-fields $E_{j}(\mathbf{r}_{1},t_{1}) E_{l}(\mathbf{r}_{2},t_{2})$ and
$E_{l}(\mathbf{r}_{1},t_{1}) E_{j}(\mathbf{r}_{2},t_{2})$.  The first term in the 
superposition corresponds to the situation in which the field 
at $D_1$ was generated by the $jth$ sub-source, and the field at $D_2$ was generated 
by the $lth$ sub-source.   The second term in the superposition corresponds to 
a different yet indistinguishable situation in which the field at $D_1$ was generated 
by the $lth$ sub-source, and the field at $D_2$ was generated by the 
$jth$ sub-source.  Therefore, an interference is concealed in 
the joint measurement of $D_1$ and $D_2$, which physically occurs at two 
space-time points $(\mathbf{r}_{1},t_{1})$ and $(\mathbf{r}_{2},t_{2})$.
The interference corresponds to $|E_{j1}E_{l2} + E_{l1}E_{j2}|^2$.
\begin{figure}[htb]
    \centering
    \vspace{5mm}
    \includegraphics[width=95mm]{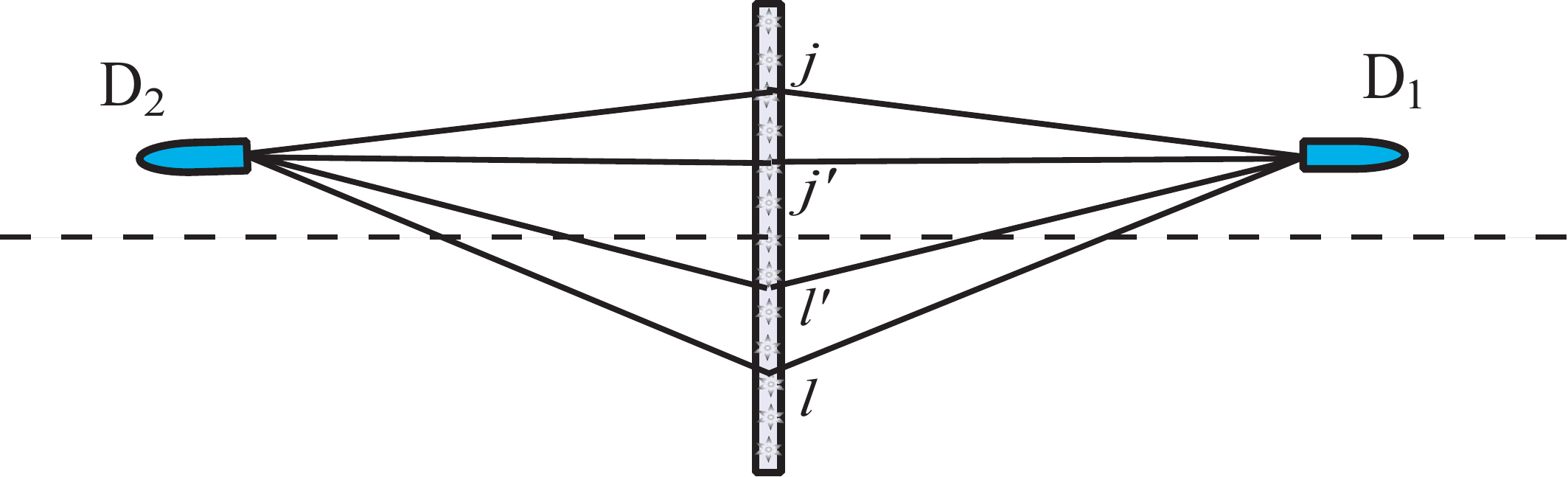} 
    \parbox{14.25cm}{\vspace{3mm}\caption{Schematic illustration of  
    $\sum_{j,l}|E_{j1}E_{l2} + E_{l1}E_{j2}|^2$.  
    It is clear that the amplitude pairs $j1 \times l2$ with
    $l1 \times j2$, where $j$ and $l$ represent all point sub-sources, pair by pair,
    will experience equal optical path propagation and superpose constructively 
    when $D_1$ and $D_2$ are located at $\vec{\rho}_1 \simeq \vec{\rho}_2$,
    $z_1 \simeq z_2$.  This interference is similar to symmetrizing the wavefunction 
    of identical particles in quantum mechanics.
    }\label{fig:Field-Field}}
\end{figure}
It is easy to see from Fig.~\ref{fig:Field-Field}, the amplitude pairs  
$j1 \times l2$ with $l1 \times j2$, 
$j'1 \times l'2$ with $l'1 \times j'2$,  $j1 \times l'2$ with $l'1 \times j2$,
and $j'1 \times l2$ with $l1 \times j'2$, etc., pair by pair, experience 
equal total optical path propagation, which involves two arms of 
$D_1$ and $D_2$, 
and thus superpose constructively when $D_1$ and $D_2$ are placed in the 
neighborhood of  
$\vec{\rho}_1 = \vec{\rho}_2$, $z_1 = z_2$.   Consequently, the summation 
of these individual constructive interference terms will yield a maximum value.  
When $\vec{\rho}_1 \neq \vec{\rho}_2$, $z_1 = z_2$, however, each pair of the 
amplitudes may achieve different relative phase and contribute a different value 
to the summation, resulting in an averaged constant value.  

It does not seem to make sense to claim a nonlocal interference between 
[($E_j$ goes to $D_1$) $\times$ ($E_l$ goes to $D_2$)] and [($E_l$ goes to $D_1$) 
$\times$ ($E_j$ goes to $D_2$)] in the framework 
of Maxwell's electromagnetic wave theory of light.  
This statement is more likely adapted from particle physics,  
similar to symmetrizing the wavefunction of identical particles,
and is more suitable to describe the interference between quantum amplitudes:  
[(particle-$j$ goes to $D_1$) $\times$ (particle-$l$ goes to $D_2$)] and 
[(particle-$l$ goes to $D_1$) $\times$ (particle-$j$ goes to $D_2$)], rather than 
waves.  Classical waves do not behave in such a way.   In fact,  in this model 
each sub-source corresponds to an independent spontaneous atomic transition 
in nature, and consequently corresponds to the creation of a photon.  Therefore,    
the above superposition corresponds to the superposition between two indistinguishable 
two-photon amplitudes, and is thus called \emph{two-photon interference} \cite{IEEE-03}. 
In Dirac's theory, this interference is the result of a measured pair of photons 
interfering with itself.

In the following we attempt a near-field calculation to derive the point-to-point
correlation of $G^{(2)}(\vec{\rho}_{1},z_1; \vec{\rho}_{2},z_2)$.  
We start from Eq.~(\ref{G2-2}) and concentrate to the transverse spatial correlation
\begin{equation}\label{HBT-near-field-0}
G^{(2)}(\vec{\rho}_{1},z_1; \vec{\rho}_{2},z_2) 
= \langle \, \sum_j \sum_l \, \Big{|} \frac{1}{\sqrt{2}} \big{[} 
E_{j}(\vec{\rho}_{1},z_1) E_{l}(\vec{\rho}_{2},z_2)
+  E_{l}(\vec{\rho}_{1},z_1) E_{j}(\vec{\rho}_{2},z_2) \big{]} \Big{|}^2 \, \rangle. 
\end{equation}
In the near-field we apply the Fresnel approximation as usual to propagate the field 
from each sub-source to the photodetectors.  
$ G^{(2)}(\vec{\rho}_{1},z_1; \vec{\rho}_{2},z_2)$ can be formally written in terms  
of the Green's function,
\begin{align}\label{HBT-near-field-1}
& \hspace{5mm} G^{(2)}(\vec{\rho}_{1},z_1; \vec{\rho}_{2},z_2) \nonumber \\
&= \langle \int d\vec{\kappa} \, d\vec{\kappa}'
\big{|}  \frac{1}{\sqrt{2}} \big{[} g(\vec{\rho}_{1},z_1,\vec{\kappa})g(\vec{\rho}_{2},z_2,\vec{\kappa}') +
g(\vec{\rho}_{2},z_2,\vec{\kappa}) g(\vec{\rho}_{1},z_1,\vec{\kappa}') \big{]} \big{|}^{2} \,
\rangle \nonumber \\
& = \langle \int d\vec{\kappa} \ \big{|}g(\vec{\rho}_{1},z_1,\vec{\kappa}) \big{|}^2 
 \int d\vec{\kappa}' \, \big{|}g(\vec{\rho}_{2},z_2,\vec{\kappa}') \big{|}^2 
+  \big{|} \int d\vec{\kappa} \ g^{*}(\vec{\rho}_{1},z_1,\vec{\kappa}) \,
g(\vec{\rho}_{2},z_2,\vec{\kappa}) \, \big{|}^2 \, \rangle \\ \nonumber 
& \equiv G^{(1)}(\vec{\rho}_{1},z_1) 
G^{(1)}(\vec{\rho}_{2},z_2) 
+ \big{|}G^{(1)}(\vec{\rho}_{1},z_1; \vec{\rho}_{2},z_2) \big{|}^2.
\end{align}
In Eq.~(\ref{HBT-near-field-1}) we have formally written 
$G^{(2)}$ in terms of the first-order correlation functions $G^{(1)}$, but keep
in mind that the first-order correlation function $G^{(1)}$ and the second-order 
correlation function $G^{(2)}$ represent 
different physics based on different measurements.    
Substituting the Green's function derived in the Appendix  for free propagation 
\begin{equation*}
g(\vec{\rho}_{j}, z_j, \vec{\kappa}) = \frac{-i \omega}{2 \pi c} \, 
\frac{e^{i \frac{\omega}{c} z_j} }{z_j} \int d \vec{\rho}_0 \, 
a(\vec{\rho}_0) \, e^{i \varphi (\vec{\rho}_0) }
\, e^{i \frac{\omega}{2 c z_j} |\vec{\rho}_{j}-\vec{\rho}_{0}|^2} 
\end{equation*}
into Eq.~(\ref{HBT-near-field-1}), we obtain 
$G^{(1)}(\vec{\rho}_{1},z_1)G^{(1)}(\vec{\rho}_{2},z_2)\sim$~constant and
\begin{align*}
\big{|}G^{(1)}(\vec{\rho}_{1}, z_1;\vec{\rho}_{2}, z_2) \big{|}^2
\propto  \, \big{|} \langle \frac{1}{z_1 z_2}
\int d \vec{\rho}_0 \, a^2(\vec{\rho}_0) \, e^{-i \frac{\omega}{c} z_1} \, 
e^{-i \frac{\omega}{2 c z_1} |\vec{\rho}_{1}-\vec{\rho}_{0}|^2} \, 
e^{i \frac{\omega}{c} z_2} \, e^{i \frac{\omega}{2 c z_2} 
|\vec{\rho}_{2}-\vec{\rho}_{0}|^2} \, \rangle \big{|}^2.
\end{align*} 
Assuming $a^2(\vec{\rho}_0)\sim$~constant, and taking $z_1 = z_2 = d$, we obtain
\begin{align}\label{HBT-near-field-11}
\big{|} G^{(1)}_{12}(\vec{\rho}_{1};\vec{\rho}_{2}) \big{|}^2
& \propto \big{|} \int d \vec{\rho}_0 \, a^2(\vec{\rho}_0) \,  
e^{-i \frac{\omega}{2 c d} |\vec{\rho}_{1}-\vec{\rho}_{0}|^2} \, 
 e^{i \frac{\omega}{2 c d} 
|\vec{\rho}_{2}-\vec{\rho}_{0}|^2} \big{|}^2 \nonumber \\
&\propto \big{|} e^{-i \frac{\omega}{2 c d} (|\vec{\rho}_{1}|^2 - |\vec{\rho}_{2}|^2)}
\int d \vec{\rho}_0 \, a^2(\vec{\rho}_0) \, 
e^{i \frac{\omega}{cd} (\vec{\rho}_{1}-\vec{\rho}_{2})\cdot \vec{\rho}_{0}} \big{|}^2
\nonumber \\
& \propto somb^2 \big{[} \, \frac{R}{d}\, \frac{\omega}{c} \, 
|\vec{\rho}_{1} - \vec{\rho}_{2}| \,\big{]},
\end{align} 
where we have assumed a disk-like light source with a finite radius of $R$.
The transverse spatial correlation function 
$G^{(2)}(\vec{\rho}_{1};\vec{\rho}_{2})$ is thus
\begin{equation}\label{HBT-near-field-22}
G^{(2)}(\big{|}\vec{\rho}_{1}-\vec{\rho}_{2}\big{|})  = I_{0}^2 \, \big{[} 1 + 
somb^2 \big{(} \, \frac{R}{d}\, \frac{\omega}{c} \, |\vec{\rho}_{1} 
- \vec{\rho}_{2}| \,\big{)} \big{]}.
\end{equation}
Consequently, the degree of the second-order spatial coherence is
\begin{equation}\label{HBT-near-field-222}
g^{(2)}(\big{|}\vec{\rho}_{1}-\vec{\rho}_{2}\big{|})  = 1 + 
somb^2 \big{(} \, \frac{R}{d}\, \frac{\omega}{c} \, |\vec{\rho}_{1} 
- \vec{\rho}_{2}| \,\big{)}.
\end{equation}
For a large value of $2R/d \sim \Delta \theta$, where $\Delta\theta$ is the angular
size of the radiation source viewed at the photodetectors, the point-spread 
$somb$-function can be approximated as a $\delta$-function of 
$|\vec{\rho}_{1} - \vec{\rho}_{2}|$.  We effectively have a 
``point-to-point" correlation between the transverse planes of $z_1 = d$ and 
$z_2 = d$.  
In 1-D Eqs.~(\ref{HBT-near-field-22}) and (\ref{HBT-near-field-222}) become
\begin{equation}\label{HBT-near-field-33}
G^{(2)}(x_{1}-x_{2})  = I_{0}^2 \, \big{[} 1 + 
sinc^2 \big{(} \, \frac{\pi \Delta \theta (x_1 - x_2)}{\lambda} \big{)} \big{]}
\end{equation}
and
\begin{equation}\label{HBT-near-field-333}
g^{(2)}(x_{1}-x_{2})  = 1 + 
sinc^2 \big{(} \, \frac{\pi \Delta \theta (x_1 - x_2)}{\lambda} \big{)},
\end{equation}
which has been experimentally demonstrated  and reported 
in Fig.~\ref{fig:Interference}.  

\vspace{6mm}
We have thus derived the same second-order correlation and coherence functions 
as that of the quantum theory.  The non-factorizable point-to-point correlation is expected 
at any intensity.  The only requirement is a large number 
of point sub-sources with random relative phases participating to the measurement, 
such as trillions of independent atomic transitions.   There is no surprise to derive 
the same result as that of the quantum theory from this simple model.  Although 
the fields are not quantized and no quantum formula was used in the above calculation, 
this model has implied the same nonlocal two-photon interference mechanism as that of the 
quantum theory.   Different from the phenomenological 
theory of intensity fluctuations, this semiclassical model explores the physical cause of 
the phenomenon.

\section{Classical simulation }
 
\hspace{6.5mm}There have been quite a few classical approaches to simulate
type-one and type-two ghost imaging.  Different from the natural non-factorizable type-one 
and type-two point-to-point imaging-forming correlations, 
classically simulated correlation functions are all factorizable.     
We briefly discuss two of these man-made factoriable classical correlations 
in the following.   

\vspace{3mm}
\hspace{-4mm} (I) Correlated laser beams.
\vspace{3mm}

\begin{figure}[hbt]
 \centering
    \includegraphics[width=85mm]{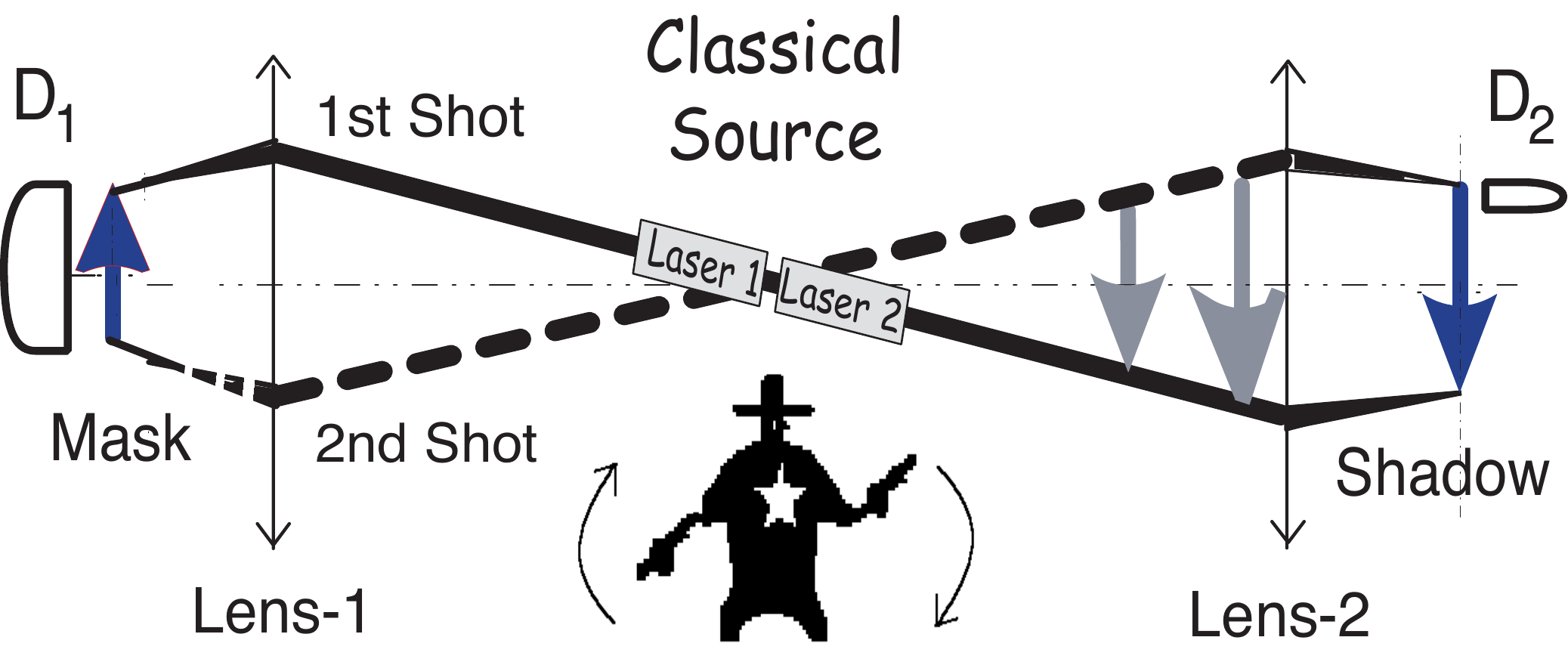}
    \parbox{14.25cm}{\caption{A ghost shadow can be made in coincidences by 
    ``blocking-unblocking" of the correlated laser beams, or simply by  
    ``blocking-unblocking" two correlated gun shots.  The man-made trivial ``correlation"
    of either laser beams or gun shots are deterministic, i.e., the laser beams or the bullets know 
    where to go in each shot, which are fundamentally different from the quantum mechanical
    nontrivial nondeterministic multi-particle correlation.} \label{fig:boyd}}
\end{figure}

In 2002, Bennink \emph{et al.} simulated ghost imaging by two correlated laser beams 
\cite{boyd}.  In this experiment, the authors intended to show that two correlated rotating 
laser beams can simulate the same physical effects as entangled states.  
Figure \ref{fig:boyd} is a schematic picture of the experiment of  Bennink 
\emph{et al}..  Different from type-one and type-two ghost imaging,  
here the point-to-point correspondence between the object plane and the 
``image plane" is made artificially by two  co-rotating laser beams ``shot by shot".  
The laser beams propagated 
in opposite directions and focused on the object and image planes, respectively.  
If laser beam-1 is blocked by the object mask there would be no joint-detection 
between $D_1$ and $D_2$ for that ``shot", while if laser beam-1 is unblocked, 
a coincidence count will be recorded against that angular position of the 
co-rotating laser beams.  A shadow of the object mask is then reconstructed in 
coincidences by the blocking$-$unblocking of laser beam-1.   

A man-made factorizable correlation of laser beam is not only different from the 
non-factorizable correlations in type-one and type-two ghost imaging, but also  
different from the standard statistical correlation of intensity fluctuations.   
Although the experiment of Bennink \emph{et al}. obtained a ghost shadow,  which 
may be useful for certain purposes, it is clear that the physics shown in their experiment 
is fundamentally different from that of ghost imaging.   
In fact, this experiment can be considered as 
a good example to distinguish a man-made trivial deterministic classical intensity-intensity 
correlation from quantum entanglement and from a natural nonlocal nondeterministic 
multi-particle correlation. 

\vspace{3mm}
\hspace{-4mm} (II) Correlated speckles.
\vspace{3mm}

Following a similar philosophy, Gatti \emph{et al}. proposed 
a factorizable ``speckle-speckle" classical correlation between 
two distant planes, $\vec{\rho}_1$ and $\vec{\rho}_2$, by imaging the speckles of the 
common light source onto the distant planes of $\vec{\rho}_1$ and $\vec{\rho}_2$, \cite{gatti}
\begin{equation}\label{eq4}
G^{(2)}(\vec{\rho}_1, \vec{\rho}_2) \propto \delta(\vec{\rho}_o - \vec{\rho}_1/m) 
\delta(\vec{\rho}_o - \vec{\rho}_2/m),
\end{equation}
where $\vec{\rho}_0$ is the transverse coordinate in the plane of the light source.\footnote{
The original publications of Gatti \emph{et al}. choose 2f-2f classical imaging systems with 
$1/2f + 1/2f = 1/f$ to image the speckles of the source onto the object plane and the ghost 
image plane.  The man-mde speckle-speckle image-forming correlation of Gatti \emph{et al}. 
shown in Eq.~(\ref{eq4}) is factorizeable, which is fundamentally different from the natural non-factorizable 
image-formimg correlations in type-one and type-two ghost imaging.  In fact, it is very easy to distinguish 
a classical simulation from ghost imaging by examining its experimental setup and operation.  
The man-made speckle-speckle correlation
needs to have two sets of identical speckles observable (by the detectors or CCDs) on the object and the 
image planes.  In thermal light ghost imaging, when using pseudo-thermal light source, the classical 
simulation requires a slow rotating ground grass in order to image the speckles of the source onto the 
object and image planes (typically, sub-Hertz to a few Hertz).  However, 
to achieve a natural HBT non-factorizable correlation of chaotic light for type-two ghost imaging, we 
need to rotate the ground grass as fast as possible (typically, a few thousand Hertz, the 
higher the batter). }  

\begin{figure}[hbt]
 \centering
    \includegraphics[width=100mm]{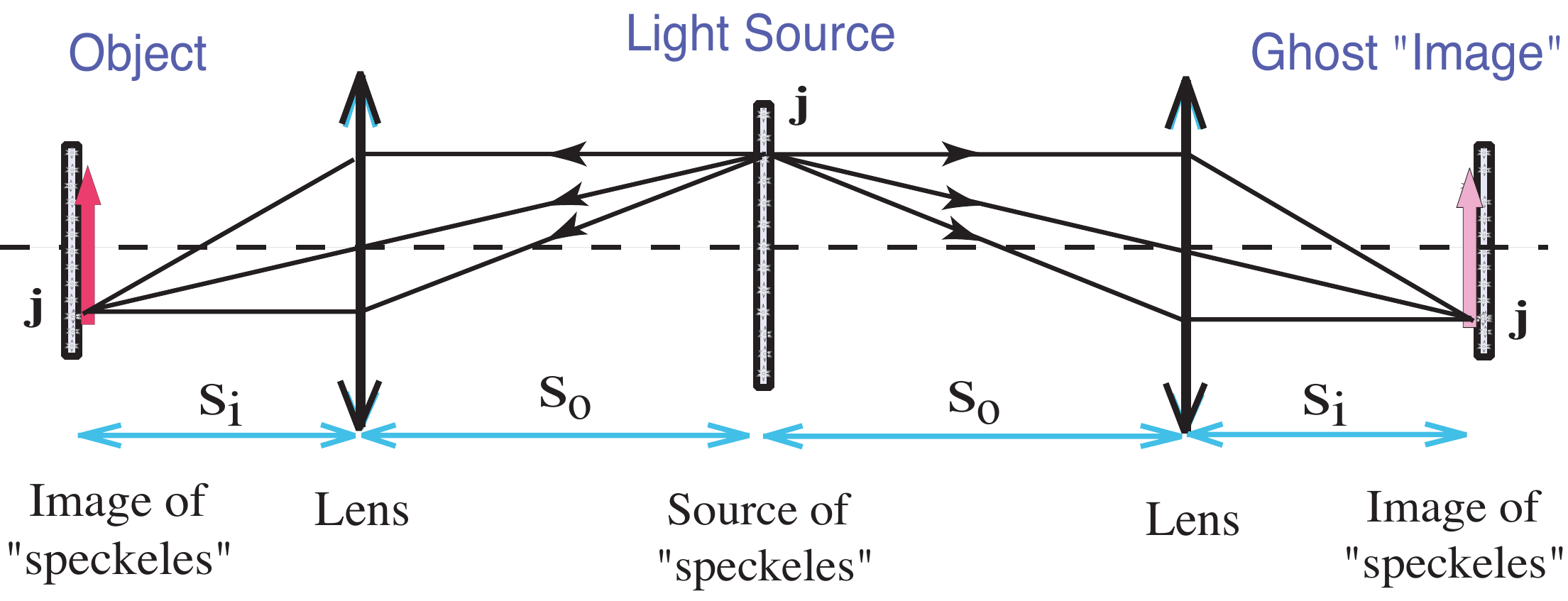}
    \parbox{14.25cm}{\caption{A ghost ``imager" is made by 
    blocking-unblocking the correlated speckles.  
    The two identical sets of speckles on the object plane and the image plane, 
    respectively, are the \emph{classical images} of the speckles 
    of the source plane.  The lens, which may be part of a CCD \emph{camera} used
    for the joint measurement, reconstructs classical images of the speckles 
    of the source onto the object plane and the image plane, respectively.  $s_o$ and 
    $s_i$ satisfy the Gaussian thin lens equation $1/s_o + 1/s_i = 1/f$.}
    \label{fig:Speckle}}
\end{figure}

The schematic setup of the classical simulation of 
Gatti \emph{et al}. is depicted in Fig.~\ref{fig:Speckle} \cite{gatti}.  Their 
experiment used either entangled photon pairs of spontaneous parametric 
down-conversion (SPDC) or chaotic light for obtaining ghost 
shadows in coincidences.   
To distinguish from ghost imaging, Gatti \emph{et al.} named their work ``ghost imager". 
The ``ghost imager" comes from a man-made classical speckle-speckle correlation.   
The speckles observed on the object and image planes are the classical images of the 
speckles of the radiation source, reconstructed by the imaging lenses shown in the figure 
(the imaging lens may be part of a CCD \emph{camera} used for the joint measurement).   
Each speckle on the source, such as the $j$th speckle near the top of the source, 
has two identical images on the object plane and on the image plane.   
Different from the non-factorizeable nonlocal image-forming correlation in type-one 
and type-two ghost imaging, mathematically, the speckle-speckle correlation is 
factorizeable into a product of two classical images of speckles.
If two point photodetectors $D_1$ and $D_2$ are scanned on the object plane and 
the image plane, respectively, $D_1$ and $D_2$ will have more ``coincidences"
when they are in the position within the two identical speckles, 
such as the two $j$th speckles near the bottom of the object plane and the 
image plane.   The blocking-unblocking of the speckles on the object plane 
by a mask will project a ghost \emph{shadow} of the mask in the coincidences of $D_1$ 
and $D_2$.  It is easy to see that the size of the identical speckles determines
the spatial resolution of the ghost shadow.  This observation has been confirmed
by quite a few experimental demonstrations.  
There is no surprise that Gatti \emph{et al.} consider ghost imaging classical \cite{comm}. 
Their speckle-speckle 
correlation is a man-made classical correlation and their ghost imager is indeed classical.
The classical simulation of  Gatti \emph{et al.} might be useful for certain applications,
however, to claim the nature of 
ghost imaging in general as classical, perhaps, is too far  \cite{comm}.   
The man-made factorizable speckle-speckle correlation of  Gatti \emph{et al.} is  
a classical simulation of the natural nonlocal point-to-point image-forming correlation 
of ghost imaging, despite the use of either entangled photon source or classical light.

\section{Local?  Nonlocal?}
\hspace{6.5mm} We have discussed the physics of both type-one 
and type-two ghost imaging.  Although different radiation sources
are used for different cases, these
two types of experiments demonstrated a similar non-factorizable
point-to-point image-forming correlation:

\vspace{2mm}
Type-one:
\begin{align}\label{Type-one}
\delta(\vec{\rho}_1 - \vec{\rho}_2) 
\sim \Big{|}  \int d\vec{\kappa}_s \, d\vec{\kappa}_i \, 
\delta(\vec{\kappa}_s +  \vec{\kappa}_i) \, g_{1}(\vec{\kappa}_s, \vec{\rho}_1)  \, 
g_{2}(\vec{\kappa}_i, \vec{\rho}_2) \Big{|}^2, 
\end{align}

Type-two:
\begin{align}\label{Type-two}
1 + \delta(\vec{\rho}_1 - \vec{\rho}_2) 
&\propto \langle \, \sum_j \sum_l \, \big{|}  \frac{1}{\sqrt{2}} \big{[} 
E_{j}(\vec{\rho}_{1}) E_{l}(\vec{\rho}_{2})
+  E_{l}(\vec{\rho}_{1}) E_{j}(\vec{\rho}_{2}) \big{]}
\big{|}^2 \, \rangle  \\ \nonumber  
& = \langle \int d\vec{\kappa} \, d\vec{\kappa}'
\big{|}  \frac{1}{\sqrt{2}} \big{[}  
g_{1}(\vec{\kappa}, \vec{\rho}_{1})g_{2}(\vec{\kappa}', \vec{\rho}_{2}) +
g_{2}(\vec{\kappa}, \vec{\rho}_{2}) g_{1}(\vec{\kappa}', \vec{\rho}_{1}) 
\big{]} \big{|}^{2} \rangle.
\end{align}

\vspace{2mm}

\hspace{-6.5mm}Equations~(\ref{Type-one}) and (\ref{Type-two}) indicate that the 
point-to-point correlation of ghost imaging, either type-one or type-two, is
the results of two-photon interference.  
Unfortunately, neither of them is in the form of 
$| \sum_j E_j |^2$ or $|E_1 + E_2|^2$, and neither is measured at 
a local space-time point.  The interference shown in Eqs.~(\ref{Type-one}) and 
(\ref{Type-two}) occurs at different space-time points through the 
measurements of two spatially separated independent photodetectors.   

In type-one ghost imaging, the $\delta$-function in Eq.~(\ref{Type-one}) 
means a typical EPR position-position correlation of an entangled photon pair.  
In EPR's language: when the pair is generated at the source the momentum and 
position of neither photon is determined, and neither photon-one nor photon-two  
``knows" where to go. However, if one of them 
is observed at a point at the object plane the other one must be found at a unique 
point in the image plane.  In type-two ghost imaging, although the position-position
determination in Eq.~(\ref{Type-two}) is only partial, it generates more surprises 
because of the chaotic nature of the radiation source.  Photon-one and photon-two, 
emitted from a thermal source, are completely random and independent, i.e.,
both propagate freely to any direction and may arrive at any position in the object 
and image planes.  Analogous to EPR's language: when the measured two photons 
were emitted from the thermal source, neither the momentum nor the position of any 
photon is determined.  However, if one of them is observed at a point on the object 
plane the other one must have twice large probability to be found at a unique 
point in the image plane.   Where 
does this partial correlation come from?  If one insists on the view point of
intensity fluctuation correlation,  then, it is reasonable to ask
why the intensities of the two light beams exhibit fluctuation correlations at 
$\vec{\rho}_1 = \vec{\rho}_2$ only?  Recall that in the experiment of 
Sarcelli \emph{et al}. the ghost image is measured in the near-field.  
Regardless of position, $D_1$ and $D_2$ receive light from all 
(a large number) point sub-sources of the thermal source, and
all sub-sources fluctuate randomly and independently.  If $\Delta I_1 \Delta I_2 = 0$ for  
$\vec{\rho}_1 \neq \vec{\rho}_2$, what is the physics to cause 
$\Delta I_1 \Delta I_2 \neq 0$ at $\vec{\rho}_1 = \vec{\rho}_2$?  

The classical superposition is considered ``local". 
The Maxwell electromagnetic field theory requires the superposition 
of the electromagnetic fields, either $| \sum_j E_j |^2$ or $|E_1 + E_2|^2$, takes place 
at a local space-time point $(\mathbf{r}, t)$.   However, the superposition shown 
in Eqs.~(\ref{Type-one}) and (\ref{Type-two}) happens at two different space-time
points $(\mathbf{r}_1, t_1)$ and $(\mathbf{r}_2, t_2)$ and is measured by 
two independent photodetectors.  Experimentally, it is not 
difficult to make the two photo-detection events space-like separated events.  
Following the definition given by EPR-Bell, we consider the 
superposition appearing in Eqs.~(\ref{Type-one}) and (\ref{Type-two}) \emph{nonlocal}.   
Although the two-photon interference of thermal light can be written and calculated
in terms of a semiclassical model, the nonlocal 
superposition appearing in Eq.~(\ref{Type-two}) has no counterpart in the classical
measurement theory of light, unless one forces a nonlocal classical 
theory by allowing the superposition to occur at a distance through the 
measurement of independent photodetectors, as we have done in 
Eq.~(\ref{G2-2}).   Perhaps, it would be
more difficult to accept a nonlocal classical measurement theory of thermal light
rather than to apply a quantum mechanical concept to ``classical" thermal radiation.    

\vspace{5mm}

\hspace{-6.5mm}\textbf{Conclusion}: In summary, we may conclude that 
ghost imaging is the result of quantum interference. 
Either type-one or type-two, ghost imaging is characterized by a
non-factorizable point-to-point image-forming correlation which is caused by  
constructive-destructive interferences involving the nonlocal superposition of 
two-photon amplitudes, a nonclassical entity corresponding to different yet 
indistinguishable alternative ways of producing a joint photo-detection event.
The interference happens within a pair of photons and at 
two spatially separated coordinates.   The multi-photon interference nature of 
ghost imaging determines its peculiar features: (1) it is nonlocal; (2) its imaging 
resolution differs from that of classical; and (3) the type-two ghost image is 
turbulence-free. Taking advantage of its quantum interference nature, 
a ghost imaging system may turn a local ``bucket" 
sensor into a nonlocal imaging camera with 
classically unachievable imaging resolution.  For instance, using the Sun as light
source for type-two ghost imaging, we may achieve an imaging spatial resolution equivalent 
to that of a classical imaging system with a lens of 92-meter diameter when taking pictures 
at 10 kilometers.\footnote{The angular size of Sun is about $0.53^\circ$.  To achieve a 
compatible image spatial resolution, a traditional camera must have a lens of 92-meter 
diameter when taking pictures at 10 kilometers.}  Furthermore, any phase disturbance 
in the optical path has no influence on the ghost image.  To achieve these features
the realization of multi-photon interference is necessary.

\vspace{6mm}

\hspace{-6.5mm}\textbf{Acknowledgment}: The author thanks M. D'Angelo,  
G. Scarcelli, J.M. Wen, T.B. Pittman, M.H. Rubin, and L.A. Wu for helpful discussions.  
This work is partially supported by AFOSR and ARO-MURI program.

\section*{Appendix: Fresnel free-propagation}
\def\theequation{$A-$\arabic{equation}}
\setcounter{equation}{0}
\def\thefigure{$A-$\arabic{figure}}
\setcounter{figure}{0}

\hspace{5.5mm} 
We are interested in knowing how a known  field  $E(\mathbf{r}_0, t_0)$
on the plane $z_0 =0$ propagates or diffracts into $E(\mathbf{r}, t)$
on another plane $z =$~constant.   We assume the field 
$E(\mathbf{r}_0, t_0)$ is excited by an arbitrary source, either point-like or spatially 
extended. The observation plane of $z =$~constant is located at an arbitrary 
distance from plane $z_0 =0$, either far-field or near-field.  Our goal is to find out a 
general solution $E(\mathbf{r}, t)$, or $I(\mathbf{r}, t)$, on the observation plane, 
based on our knowledge of $E(\mathbf{r}_0, t_0)$ and the laws of the 
Maxwell electromagnetic wave theory.   
It is not easy to find such a general solution. However, the use of the 
Green's function or the field transfer function, which describes the propagation 
of each mode from the plane of $z_0 =0$ to the observation plane 
of $z =$~constant, makes this goal formally achievable.  

Unless $E(\mathbf{r}_0, t_0)$ is a non-analytic function in the space-time region 
of interest, there must exist a Fourier integral representation for 
$E(\mathbf{r}_0, t_0)$
\begin{align}\label{Fourier-22}
E(\mathbf{r}_0, t_0) = \int d\mathbf{k} \, E(\mathbf{k}) \, 
w_{\mathbf{k}}(\mathbf{r}_0,t_0) \, e^{-i \omega t_0},
\end{align}
where $w_{\mathbf{k}}(\mathbf{r}_0,t_0)$ is a solution of the Helmholtz wave equation
under appropriate boundary conditions.   The solution of the Maxwell wave equation
$w_{\mathbf{k}}(\mathbf{r}_0,t_0) \, e^{-i \omega t_0}$, namely the Fourier mode, 
can be a set of plane-waves or spherical-waves depending on the chosen boundary 
condition.  In Eq.~(\ref{Fourier-22}), $E(\mathbf{k}) = a(\mathbf{k})e^{i \varphi (\mathbf{k})}$
is the complex amplitude of the Fourier mode $\mathbf{k}$. 
In principle we should be able to find an appropriate 
Green's function which propagates each mode under the Fourier integral 
point by point from the plane of $z_0 =0$ to the plane of observation, 
\begin{align}\label{Fourier-33}
E(\mathbf{r}, t) &= \int d\mathbf{k} \, E(\mathbf{k}) \, 
g(\mathbf{k}, \mathbf{r}-\mathbf{r}_0, t - t_0)
\, w_{\mathbf{k}}(\mathbf{r}_0,t_0) \, e^{-i \omega t_0} \nonumber \\
&= \int d\mathbf{k} \, g({\bf k}, {\bf r}-{\bf r}_0, t-t_0) \,
E({\bf k}, {\bf r}_0, t_0),
\end{align}
where $E({\bf k}, {\bf r}_0, t_0) = E(\mathbf{k}) 
\, w_{\mathbf{k}}(\mathbf{r}_0,t_0) \, e^{-i \omega t_0}$.
The secondary wavelets that originated from each point on the plane of $z_0 =0$ 
are then superposed \emph{coherently} on each point on the observation plane 
with their after-propagation amplitudes and phases.  
It is convenient to write Eq.~(\ref{Fourier-33}) in the following form
\begin{equation}\label{Green-1}
E(\vec{\rho}, z, t) 
=  \int  d\omega \, d\vec{\kappa} \,
g(\vec{\kappa}, \omega;  \vec{\rho} - \vec{\rho}_0, z - z_0, t - t_0) \,
E(\vec{\kappa}, \omega; \vec{\rho}_0, z_0, t_0),
\end{equation}
where we have used the transverse-longitudinal coordinates in space-time
($\vec{\rho}$ and $z$) and in momentum ($\vec{\kappa}$, $\omega$).  

\begin{figure}[hbt]
 \centering
    \includegraphics[width=75mm]{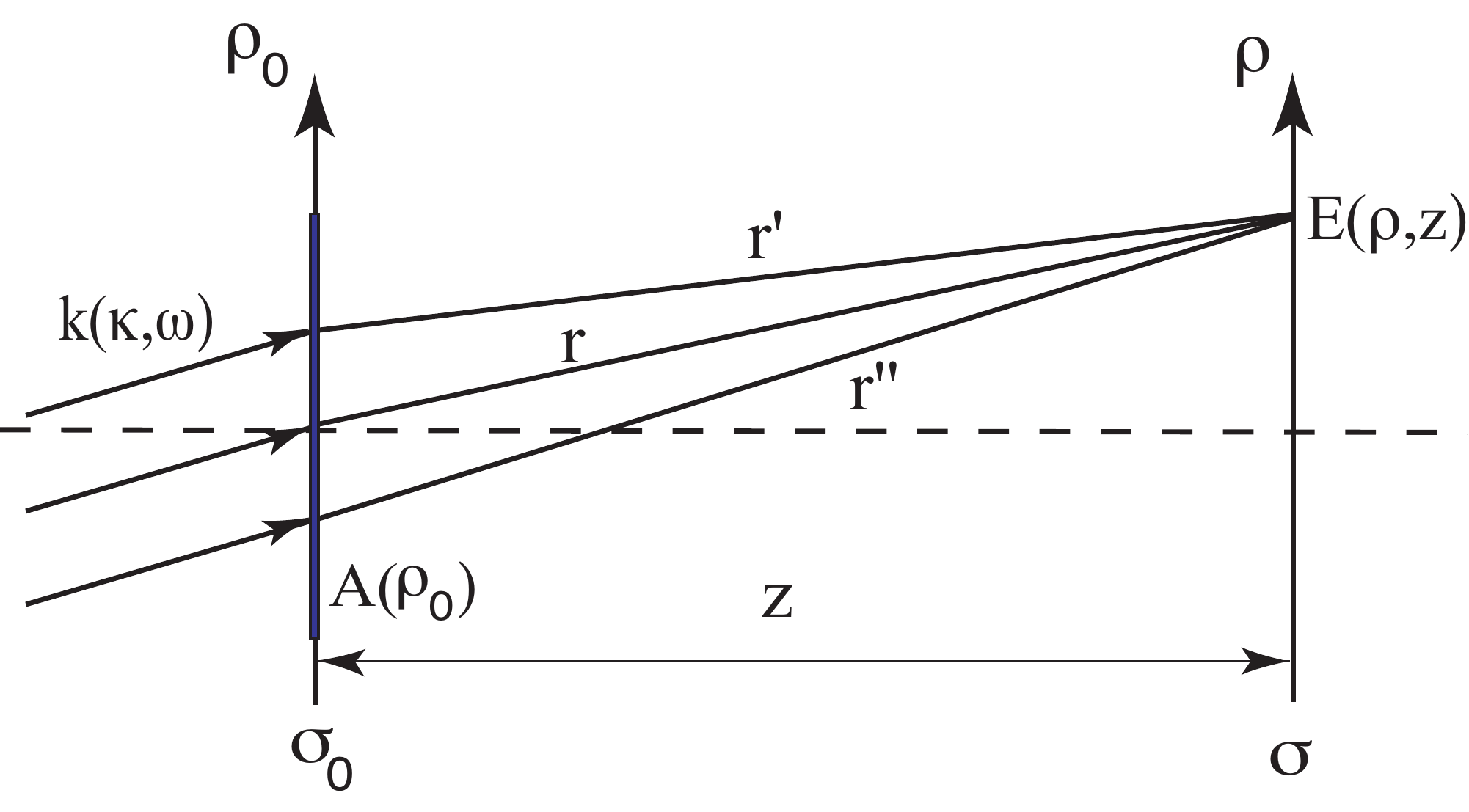}
     \parbox{14.25cm}{\caption{Schematic of free-space Fresnel propagation.  
    The complex amplitude $\tilde{A}(\vec{\rho}_0)$ 
    is composed of a real function $A(\vec{\rho}_0)$ and a phase 
    $e^{-i \vec{\kappa} \cdot \vec{\rho}_0}$ associated with each of the transverse 
    wavevectors $\vec{\kappa}$ in the plane of $\sigma_0$.  Notice: only one mode of 
    wavevector $\mathbf{k}(\vec{\kappa}, \omega)$ is shown in the figure.}
    \label{fig:Fresnel}}
\end{figure}

Fig.~\ref{fig:Fresnel} is a simple example in which
the field propagates freely from an aperture $A$ of finite size on the
plane $\sigma_0$ to the observation plane $\sigma$.   
Based on Fig.~\ref{fig:Fresnel} we evaluate  
$g(\vec{\kappa}, \omega; \vec{\rho}, z)$, namely the Green's function  
for free-space Fresnel propagation-diffraction.  
According to the Huygens-Fresnel principle the field at a given space-time point
$(\vec{\rho}, z, t)$ is the result of a superposition of the spherical 
secondary wavelets that originated from each point on the $\sigma_0$ plane
(see Fig.~\ref{fig:Fresnel}),
\begin{eqnarray}\label{gg-00}
E(\vec{\rho}, z, t) = \int  d\omega \, d\vec{\kappa} \
E(\vec{\kappa}, \omega; 0, 0) \int_{\sigma_0} \, d\vec{\rho}_0 \, 
\frac{\tilde{A}(\vec{\rho}_0)}{r} \, e^{-i (\omega t - k r)},
\end{eqnarray}
where we have set $z_0 = 0$ and $t_0 =0$ at plane $\sigma_0$, and defined 
$r = \sqrt{z^2 + |\vec{\rho} - \vec{\rho}_0|^2}$.  In Eq.~(\ref{gg-00}),
$\tilde{A}(\vec{\rho}_0)$ is the complex amplitude or relative distribution of the field 
on the plane of $\sigma_0$, which may be written as a simple aperture function
in terms of the transverse coordinate 
$\vec{\rho}_0$, as we have done in the earlier discussions.  

In the near-field Fresnel paraxial approximation, when $|\vec{\rho} - \vec{\rho}_0|^2 \ll z^2$
we take the first-order expansion of $r$ in terms of $z$ and $\vec{\rho}$,
\begin{equation}
r = \sqrt{z^2 + |\vec{\rho} - \vec{\rho}_0|^2} \simeq z(1 + 
\frac{|\vec{\rho} - \vec{\rho}_0|^2}{2 z^2}),
\end{equation}
so that $E(\vec{\rho}, z, t)$ can be approximated as
\begin{eqnarray*}
E(\vec{\rho}, z, t) \simeq  \int d\omega \,  d\vec{\kappa} \
E(\vec{\kappa}, \omega, 0, 0) \int d\vec{\rho}_0 \, \frac{\tilde{A}(\vec{\rho}_0)}{z}
\, e^{i \frac{\omega}{c} z} \, 
e^{i \frac{\omega}{2 c z} |\vec{\rho} - \vec{\rho}_0|^2} e^{-i \omega t},
\end{eqnarray*}
where $e^{i \frac{\omega}{2 c z} |\vec{\rho} - \vec{\rho}_0|^2}$ is named
the Fresnel phase factor.

Assuming that the complex amplitude $\tilde{A}(\vec{\rho}_0)$ is composed of a real 
function $A(\vec{\rho}_0)$ and a phase $e^{-i \vec{\kappa} \cdot \vec{\rho}_0}$, 
associated with the transverse
wavevector and the transverse coordinate on the plane of $\sigma_0$, 
as is reasonable for the setup of Fig.~\ref{fig:Fresnel},  we can then 
write $E(\vec{\rho}, z, t)$ in the  form
\begin{eqnarray*}
E(\vec{\rho}, z, t) 
=  \int d\omega \, d\vec{\kappa} \ E(\vec{\kappa}, \omega; 0, 0) \,
e^{-i \omega t} \, \frac{e^{i \frac{\omega}{c} z}}{z} 
\int d\vec{\rho}_0 \, A(\vec{\rho}_0) \, e^{i \vec{\kappa} \cdot \vec{\rho}_0} \,  
e^{i \frac{\omega}{2 c z} |\vec{\rho} - \vec{\rho}_0|^2}.
\end{eqnarray*}
The Green's function $g(\vec{\kappa}, \omega; \vec{\rho}, z)$ for free-space
Fresnel propagation is thus 
\begin{equation}\label{gg-10}
g(\vec{\kappa}, \omega; \vec{\rho}, z) = \frac{e^{i \frac{\omega}{c} z}}{z} 
\int_{\sigma_0} \, d\vec{\rho}_0 \, A(\vec{\rho}_0) \, 
e^{i \vec{\kappa} \cdot \vec{\rho}_0} \, 
G(|\vec{\rho} - \vec{\rho}_0|, \frac{\omega}{c z}).
\end{equation}

In Eq.~(\ref{gg-10}) we have defined a Gaussian function 
$G(|\vec{\alpha|}, \beta)= e^{i (\beta/2) |\alpha|^2}$, namely the Fresnel phase factor.  
It is straightforward to find that
the Gaussian function $G(|\vec{\alpha|}, \beta)$ has the following properties:
\begin{eqnarray}\label{Gaussian-10}
G^*(|\vec{\alpha}|, \beta) &=& G(|\vec{\alpha}|, - \beta ), \nonumber \\
G(|\vec{\alpha}|, \beta_1 + \beta_2) &=& G(|\vec{\alpha}|, \beta_1 ) \,
G(|\vec{\alpha}|, \beta_2), \nonumber \\
G(|\vec{\alpha}_1+\vec{\alpha}_2|, \beta)
 &=& G(|\vec{\alpha}_1|, \beta) \, G(|\vec{\alpha}_2|, \beta) \,
e^{i \beta \vec{\alpha}_1 \cdot \vec{\alpha}_2}, \nonumber \\
\int d\vec{\alpha} \,\,
G(|\vec{\alpha}|, \beta ) \, e^{i \vec{\gamma} \cdot \vec{\alpha}} &=& i \frac{2\pi}{\beta} \,
G(|\vec{\gamma}|, -\frac{1}{\beta} ).
\end{eqnarray}
Notice that the last equation in Eq.~(\ref{Gaussian-10}) is the Fourier transform of the 
$G(|\vec{\alpha|}, \beta)$ function. As we shall see in the following, these properties 
are very useful in simplifying the calculations of the Green's functions 
$g(\vec{\kappa}, \omega; \vec{\rho}, z)$.

\vspace{6mm}

Next, we consider inserting an imaginary plane $\sigma'$ between $\sigma_0$ and 
$\sigma$.  This is equivalent to having two consecutive Fresnel propagations with a 
diffraction-free $\sigma'$ plane of infinity.   Thus, the calculation of these consecutive
Fresnel propagations should yield the same Green's function as that of the 
above direct Fresnel propagation shown in Eq.~(\ref{gg-10}):
\begin{eqnarray}\label{gg-11}
&& g(\omega, \vec{\kappa}; \vec{\rho}, z) \nonumber \\
&=& C^2 \frac{e^{i \frac{\omega}{c}( d_1+ d_2)}}{d_1 d_2} 
\int_{\sigma'}  d\vec{\rho'}  \int_{\sigma_0} d\vec{\rho}_0 \, \tilde{A}(\vec{\rho}_0) 
G(|\vec{\rho'} - \vec{\rho}_0|, \frac{\omega}{c d_1})
G(|\vec{\rho} - \vec{\rho'}|, \frac{\omega}{c d_2}) \nonumber \\
&=& C \, \frac{e^{i \frac{\omega}{c} z}}{z} 
\int_{\sigma_0} \, d\vec{\rho}_0 \, \tilde{A}(\vec{\rho}_0) \, 
G(|\vec{\rho} - \vec{\rho}_0|, \frac{\omega}{c z})
\end{eqnarray}
where $C$ is a necessary normalization constant for a valid Eq.~(\ref{gg-11}),
and $z = d_1 + d_2$.  The double integral of $d\vec{\rho}_0 $ and 
$d\vec{\rho'}$ in Eq.~(\ref{gg-11}) can be evaluated as 
\begin{eqnarray*}
&& \int_{\sigma'}  d\vec{\rho'}  \int_{\sigma_0} d\vec{\rho}_0 \, \tilde{A}(\vec{\rho}_0) \, 
G(|\vec{\rho'} - \vec{\rho}_0|, \frac{\omega}{c d_1}) \,
G(|\vec{\rho} - \vec{\rho'}|, \frac{\omega}{c d_2}) \nonumber \\
&=&\int_{\sigma_0} d\vec{\rho}_0 \, \tilde{A}(\vec{\rho}_0) \,  
G(\vec{\rho}_0, \frac{\omega}{c d_1}) \, G(\vec{\rho}, \frac{\omega}{c d_2})
\\
&& \times \int_{\sigma'}  d\vec{\rho'} \, G(\vec{\rho'}, \frac{\omega}{c}(\frac{1}{d_1} + 
\frac{1}{d_2})) \,
e^{-i\frac{\omega}{c}(\frac{\vec{\rho}_{0}}{d_1} + 
\frac{\vec{\rho}}{d_2})\cdot \vec{\rho'}} \nonumber \\
&=& \frac{i 2\pi c}{\omega} \frac{d_1 d_2}{d_1 + d_2}
\int_{\sigma_0} d\vec{\rho}_0 \, \tilde{A}(\vec{\rho}_0) \,  
G(\vec{\rho}_0, \frac{\omega}{c d_1}) \, 
G(\vec{\rho}, \frac{\omega}{c d_2}) \, \\
&& \hspace{24mm} \times \ G(|\frac{\vec{\rho}_{0}}{d_1} + 
\frac{\vec{\rho}}{d_2}|, \frac{\omega}{c}(\frac{d_1 d_2}{d_1 + d_2})) 
\nonumber \\
&=& \frac{i 2\pi c}{\omega} \frac{d_1 d_2}{d_1 + d_2}
\int_{\sigma_0} d\vec{\rho}_0 \, \tilde{A}(\vec{\rho}_0) \,
G(|\vec{\rho} - \vec{\rho}_0|, \frac{\omega}{c (d_1 + d_2)})
\end{eqnarray*}
where we have applied Eq.~(\ref{Gaussian-10}), and the integral of
$d\vec{\rho'}$ has been taken to infinity.  Substituting this result into 
Eq.~(\ref{gg-11}) we obtain
\begin{eqnarray*}
&& g(\vec{\kappa}, \omega; \vec{\rho}, z) \nonumber \\
&=& C^2 \, \frac{i 2\pi c}{\omega} \frac{e^{i \frac{\omega}{c}( d_1+ d_2)}}{d_1 + d_2} 
\int_{\sigma_0} d\vec{\rho}_0 \, \tilde{A}(\vec{\rho}_0) \,
G(|\vec{\rho} - \vec{\rho}_0|, \frac{\omega}{c (d_1 + d_2)}) \nonumber \\
&=& C \, \frac{e^{i \frac{\omega}{c} z}}{z} 
\int_{\sigma_0} \, d\vec{\rho}_0 \, \tilde{A}(\vec{\rho}_0) \, 
G(|\vec{\rho} - \vec{\rho}_0|, \frac{\omega}{c z}).
\end{eqnarray*}
Therefore, the normalization constant $C$ must take the value of
$
C = - i \omega / 2 \pi c.
$
The normalized Green's function for free-space Fresnel propagation is thus
\begin{eqnarray}\label{gg-final}
g(\vec{\kappa}, \omega; \vec{\rho}, z) 
=  \frac{-i \omega}{ 2 \pi c} \ \frac{e^{i \frac{\omega}{c} z}}{z}
\int_{\sigma_0} \, d\vec{\rho}_0 \, \tilde{A}(\vec{\rho}_0) \, 
G(|\vec{\rho} - \vec{\rho}_0|, \frac{\omega}{c z}).
\end{eqnarray}

%\vspace{6mm}

\end{document}